\documentclass[aps,prl,twocolumn,superscriptaddress,showpacs,amsmath,amssymb,longbibliography,floatfix]{revtex4-2}

\usepackage{graphicx,bm,color}
\graphicspath{{Fig/}}
\usepackage[T1]{fontenc}
\usepackage{mathrsfs}
\usepackage{amsmath}
\usepackage{amssymb}
\usepackage{esint}
\usepackage{multirow}
\usepackage{float}
\usepackage{array}
\usepackage{makecell}
\usepackage{booktabs}
\usepackage{gensymb}
\usepackage{xspace}
\usepackage{hyperref}
\usepackage{setspace}
\usepackage[normalem]{ulem}

\hypersetup{
  colorlinks=true,
  citecolor=blue,
  linkcolor=black,
  urlcolor=black
}

\def\Eq#1{Eq.~(\ref{#1})}

\def\CMAO{CMAO\xspace}
\def\CMAOfull{CeMgAl$_{11}$O$_{19}$}

\newcommand{\be}{\begin{eqnarray}}
\newcommand{\ee}{\end{eqnarray}}

\renewcommand{\arraystretch}{1.5}

\begin{document}

\title{Hidden Triakontadipole Order in the Triangular-Lattice Magnet CeMgAl$_{11}$O$_{19}$}

\author{Yu-Bo Liu}
\thanks{These authors contributed equally to this work.}
\affiliation{Institute of Theoretical Physics, Chinese Academy of Sciences, Beijing 100190, China}

\author{Yuan Gao}
\thanks{These authors contributed equally to this work.}
\affiliation{Institute of Theoretical Physics, Chinese Academy of Sciences, Beijing 100190, China}

\author{Zhen Ma}
\thanks{These authors contributed equally to this work.}
\affiliation{Hubei Key Laboratory of Photoelectric Materials and Devices, School of Materials Science and Engineering, Hubei Normal University, Huangshi 435002, China}

\author{Shufan Cheng}
\thanks{These authors contributed equally to this work.}
\affiliation{National Laboratory of Solid State Microstructures and School of Physics, Nanjing University, Nanjing 210093, China}

\author{Xinyang Liu}
\affiliation{School of Physics, Beihang University, Beijing 100191, China}
\affiliation{Institute of Physics, Chinese Academy of Sciences, Beijing 100190, China}

\author{Jinsheng Wen}
\altaffiliation{jwen@nju.edu.cn}
\affiliation{National Laboratory of Solid State Microstructures and School of Physics, Nanjing University, Nanjing 210093, China}
\affiliation{Collaborative Innovation Center of Advanced Microstructures and Jiangsu Physical Science Research Center, Nanjing University, Nanjing 210093, China}
\affiliation{Jiangsu Key Laboratory of Quantum Information Science and Technology, Nanjing University, Suzhou 215163, China}

\author{Wei Li}
\email{wli@itp.ac.cn}
\affiliation{Institute of Theoretical Physics, Chinese Academy of Sciences, Beijing 100190, China}

\begin{abstract}
The absence of magnetic Bragg peaks and a broad excitation continuum are commonly recognized experimental features of a quantum spin liquid. The Ce-based triangular-lattice hexaaluminate \CMAOfull\ exhibits both signatures, yet its effective model and ground state remain under debate. Experiments have variously suggested a quantum spin liquid, proximate umbrella point, and disorder-suppressed magnetic order. Here we trace these contradictions to an overlooked multipolar degree of freedom, and establish a dipole--triakontadipole XXZ model with $J_{\perp} \simeq 0.65$~K and $J_c \simeq -0.13$~K via precision tensor-network calculations. The resulting planar triakontadipolar superfluid order is hidden from dipolar probes, explaining the absence of a magnetic Bragg peak in neutron scattering. The zero-field excitations separate into a low-energy triakontadipole wave and a broad higher-energy continuum. We further propose nuclear spin-lattice relaxation and spin-current experiments to probe the gapless excitations of this hidden triakontadipole superfluid. Our findings thus offer a hidden-order caveat for spin-liquid searches and provide a new platform for exploring high-rank multipolar physics in rare-earth triangular-lattice magnets.
\end{abstract}

\maketitle

\textit{Introduction.}---Frustrated triangular-lattice quantum magnets provide a rich platform for exploring exotic spin states, including the quantum spin liquid (QSL)~\cite{Balents2010, Mila2000, Savary2016, Broholm2020}, spin nematics~\cite{Chubukov1990, Bhattacharjee2006, Lauchli2006} and spin supersolids~\cite{Heidarian2010, Gao2022, Xiang2024, Shu2026EuCo2Al9}. In particular, rare-earth triangular-lattice magnets have attracted growing attention. Prominent examples include YbMgGaO$_4$, which exhibits no long-range magnetic order but hosts a broad spin-excitation continuum~\cite{Li2015, Paddison2016, Shen2016, Zhu2017}, and the $A$RECh$_2$ family ($A$ = alkali, RE = rare earth, Ch = chalcogen), which offers QSL candidates free from site-mixing disorder~\cite{Bordelon2019, Xing2019CsYbSe2, Dai2021, Scheie2023, Xie2024CsCeSe2}. The field continues to thrive, with new compounds and research advances emerging at a rapid pace.

\begin{figure*}[t!]
\centering
\includegraphics[width=0.85\linewidth]{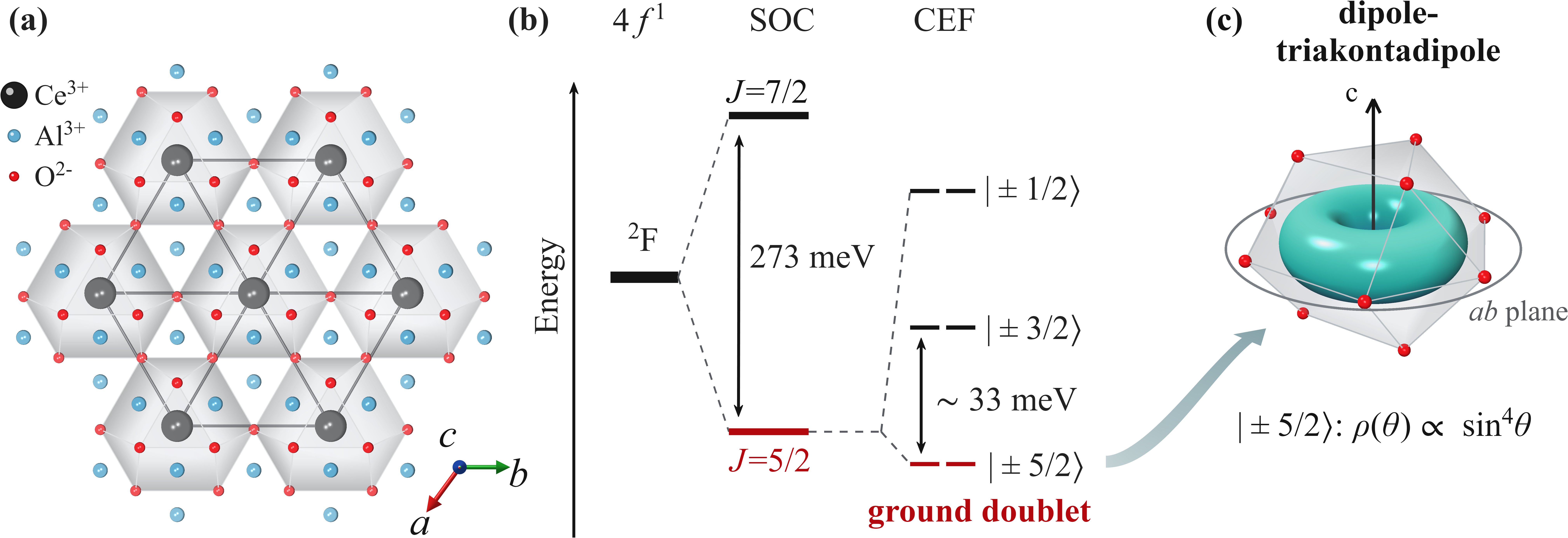}
\caption{(a) Ce$^{3+}$ ions form a triangular lattice within their CeO$_{12}$ coordination environment. (b) Spin--orbit coupling separates the $4f^1$ manifold into $J=5/2$ and $J=7/2$, while the crystal electric field selects the $|\pm5/2\rangle$ Kramers ground doublet. (c) The corresponding oblate $4f$ charge density, $\rho(\theta)\propto\sin^4\theta$, exhibits $U(1)$ rotational symmetry about the $c$ axis.}
\label{fig1}
\end{figure*}

Recently, effective spin-1/2 rare-earth hexaaluminates have gathered intense interest~\cite{Ashtar2019b, Gao2026, Cao2025, colman, Bippus2025CMAO, Kumar2025CMAOMagnetoelectric, Cao2024, Ma2024PrMgAl, Li2024PrMgAl, Tu2024PrMgAl, Kumar2025PrMgAl, Kumar2025PrMgAlME, Bu2022PrZnAl, Kumar2025NdMgAl, Cao2026NdMgAlMCE, Cao2025NdZnAl, Kumar2026SmMgAl}, where the rare-earth ions form triangular layers.
In this family, \CMAOfull\ (\CMAO) has a $|\pm5/2\rangle$ Kramers-doublet ground state well-isolated with the rest crystal electric field (CEF) levels~\cite{Gao2026, Cao2025, colman}. It exhibits a pronounced easy-axis anisotropy, with the $c$-axis susceptibility far exceeding the in-plane response~\cite{Cao2025, colman}. Specific-heat measurements show no singularity down to $30$~mK~\cite{Gao2026, Cao2025, colman}, while muon spin relaxation ($\mu$SR) detects persistent spin fluctuations down to $50$~mK~\cite{Cao2025}.
Neutron scattering bounds any static ordered moment below $0.01~\mu_B$/Ce down to $100$~mK, and resolves a broad excitation continuum extending to $\sim0.3$~meV, reminiscent of the tiny static dipole moment yet robust collective dynamics in URu$_2$Si$_2$~\cite{Broholm1987}. These findings pose a sharp tension: the static magnetic response is Ising-like, whereas inelastic neutron scattering (INS) data reveal dominant in-plane exchange couplings~\cite{Gao2026}. Moreover, the $120^\circ$ order expected from these antiferromagnetic in-plane couplings should produce a static magnetic Bragg peak at the $K$ point, yet no such peak has been observed down to the lowest measured temperatures~\cite{SM}.

Different experimental groups have proposed distinct pictures of the ground state. Cao \textit{et al.} proposed a gapless U(1) Dirac QSL state based on thermodynamic and $\mu$SR measurements~\cite{Cao2025}.
By fitting the field-polarized spin-wave spectrum, Gao \textit{et al.} constructed a microscopic spin model and placed \CMAO close to an exactly solvable umbrella point, attributing the excitation continuum to an ensemble average over the umbrella states~\cite{Gao2026}, a scenario tied to the fine-tuned degenerate point.
In contrast, Bastien \textit{et al.} obtained a substantially different exchange ratio from a density-matrix renormalization group (DMRG) analysis of the magnetization behaviors. They found \CMAO deviates from the umbrella point and invoked structural disorder to explain the absence of long-range order~\cite{colman}, while the origin of the broad continuum then leaves unexplained.
Thus, resolving the puzzle in \CMAO urgently calls for a coherent theoretical framework.

Beyond dipoles, strong spin--orbit coupling and crystal electric fields can endow rare-earth moments with higher-rank multipolar character~\cite{Santini2009, Kuramoto2009, Pourovskii2025}. Such multipoles can order without producing magnetic Bragg peaks, offering a paradigmatic route to hidden order, while their intrinsically anisotropic interactions can stabilize phases inaccessible to purely dipolar spin models~\cite{Rau2019, Li2016HiddenMultipolar, Liu2018NonKramers}, most prominently in dipolar--octupolar pyrochlores~\cite{Sibille2020Octupole, Tang2023Pr2Zr2O7, Gao2025QSI}. Experimental studies of multipolar physics on triangular lattices, by contrast, remain more limited. TmMgGaO$_4$ provides a notable example, exhibiting a 6-clock dipolar order assisted by the multipolar component~\cite{Shen2019, Li2020TmMgGaO, Hu2020BKT}, while spin nematics exemplify rank-2 quadrupolar hidden order~\cite{Nakatsuji2005, Lauchli2006, Sheng2025}. Yet multipolar order of rank $\ge3$ has remained essentially unexplored on this prototypical frustrated lattice.

In this work, we grow high-quality single crystals of \CMAO, perform magnetic specific heat and susceptibility measurements, and analyze the low-temperature magnetization and spin dynamics. We trace the origin of the prior contradictions to the $|\pm5/2\rangle$ doublet of Ce$^{3+}$ ion: its longitudinal component is a dipole, whereas its transverse pseudospin is a rank-5 \textit{triakontadipole}. This dual dipolar--multipolar nature reconciles at its root the Ising-like static response with the in-plane-dominated spin exchange and dynamics. Through precision many-body calculations, we establish a dipolar--triakontadipolar XXZ model with in-plane exchange $J_{\perp}\simeq0.65$~K, out-of-plane Ising coupling $J_c\simeq-0.13$~K, and Land\'e factor $g_c\simeq3.75$. With the microscopic model, we locate \CMAO in the $120^\circ$ ordered phase, a hidden triakontadipolar superfluid whose order parameter is invisible to dipolar probes. The calculated zero-field spectrum contains the low-energy triakontadipole wave and reproduces the measured higher-energy continuum, which we ascribe to precursor excitations of the nearby triakontadipolar stripe phase. We further propose dynamical probes of the multipolar physics, including the ultralow-temperature nuclear magnetic resonance (NMR) relaxation and spin Seebeck responses on the triangular-lattice compound.

\textit{Dipole–Triakontadipole magnet with XXZ couplings.}---The lattice structure is illustrated in Fig.~\ref{fig1}(a), where the magnetic Ce$^{3+}$ ($4f^1$) ions occupy the $2d$ sites of $D_{3h}$ point symmetry. Spin--orbit coupling separates the $4f^1$ manifold into the $J=5/2$ and $J=7/2$ multiplets separated by an energy gap $\Delta_{7/2}\simeq273$~meV, and the $D_{3h}$ CEF further splits the former into three Kramers doublets [Fig.~\ref{fig1}(b)]. The CEF Hamiltonian reads $H_{\rm CEF}=\sum_{k,q}B_k^q O_k^q$, and the only symmetry-allowed transverse term $O_6^6$ is inert within the $J=5/2$ manifold (see Appendix). The CEF ground-state doublet thus remains pure $|\pm5/2\rangle$ states, i.e., $|g,\pm\rangle=|J=5/2,m_J=\pm5/2\rangle$ in the $|J,m_J\rangle$ basis, whose $4f$ charge cloud is oblate [see Fig.~\ref{fig1}(c)]. Within this doublet, an in-plane flip requires $\Delta m_J=5$, and the transverse pseudospin is thus a rank-5 multipole---a triakontadipole. The small in-plane $g_{\perp}$ observed in experiments~\cite{Cao2025, Gao2026, colman} arises from a weak $J=7/2$ admixture that the sixfold crystal-field term $B_6^6O_6^6$ induces across the large spin--orbit gap $\Delta_{7/2}$. As it amounts to only $1\%$--$5\%$ of the ground-doublet weight (see Appendix), this admixture is neglected in our main discussion.

We therefore consider the effective XXZ model
\begin{equation}
\label{eq:XXZ}
H = \sum_{\langle ij\rangle} \Big[ J_{\perp}(S^x_iS^x_j+S^y_iS^y_j)+J_c S^z_iS^z_j \Big] -g_c \mu_B  B \sum_i S^z_i,
\end{equation}
where $\langle ij\rangle$ denotes nearest-neighbor bonds, and $B$ is the magnetic field applied along the $c$ axis, with $g_c$ the corresponding Land\'{e} factor. Moreover, off-diagonal interactions, including the out-of-plane Dzyaloshinskii--Moriya (DM) interaction and the bond-dependent anisotropic multipolar exchange $J_{\pm\pm}$, are symmetry-allowed couplings. The $J_{\pm\pm}$ interaction takes the form $2J_{\pm\pm}\sum_{\langle ij\rangle}[(S^x_iS^x_j-S^y_iS^y_j)\cos\phi_{ij}-(S^x_iS^y_j+S^y_iS^x_j)\sin\phi_{ij}]$, with the bond angles $\phi_{ij}=0,\pm 2\pi/3$ labeling the three bond directions. As anticipated from the very weak effective sixfold anisotropy of the ground doublet, and corroborated by the experimental analysis below, $J_{\pm\pm}$ is very small compared with $J_{\perp}$ and is omitted from \Eq{eq:XXZ}. The DM term is likewise neglected, as its expected spectral signatures~\cite{Shan2021, Hog2022} have not been observed experimentally~\cite{Gao2026}.

\begin{figure}[!t]
\centering
\includegraphics[width=\linewidth]{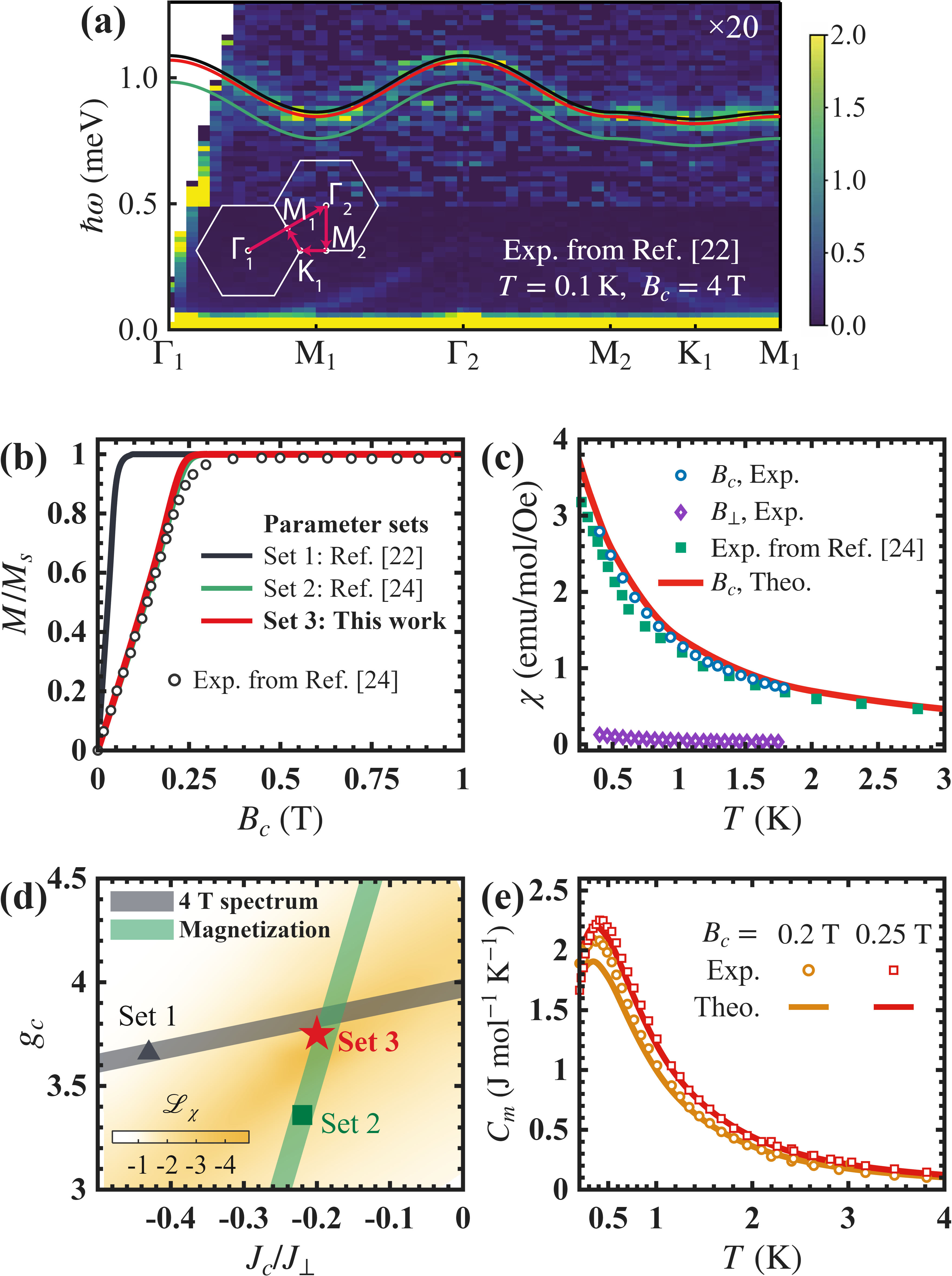}
\caption{Determination and validation of the model parameters. Sets 1 and 2 are taken from Refs.~\cite{Gao2026, colman}, respectively, and Set 3 is our model. (a) Field-polarized INS spectrum at $T=0.1$~K and $B=4$~T (adapted from Ref.~\cite{Gao2026}), with linear spin-wave theory (LSWT) dispersions of the transverse excitations contributing to $S^{+-}(\mathbf q,\omega)$ (see the Appendix). (b) Normalized $c$-axis magnetization at $T=0.03$~K (solid lines: tanTRG calculations). (c) Susceptibility for $B\parallel c$ and $B\perp c$, with red line denoting the calculated $\chi_c$ for Set 3. (d) Constraints in the $(J_c/J_\perp,g_c)$ plane. The gray and green bands from the $4$-T spectrum and the saturation field, respectively, and the susceptibility fitting loss (gold map, lower is better). The triangle, square, and red star mark Sets 1, 2, and 3. (e) Magnetic specific heat data are compared with the calculated results with Set 3.}
\label{fig2}
\end{figure}

We emphasize that, although \Eq{eq:XXZ} coincides in form with the Hamiltonians adopted in Refs.~\cite{Gao2026, colman}, its physical content is fundamentally different. Here the in-plane coupling $J_\perp$ is predominantly a triakontadipolar exchange~\cite{Santini2009} rather than a dipolar one. Below we compute the ground state using DMRG~\cite{White1992, Schollwock2011}, the real-time spin dynamics using the time-dependent variational principle (TDVP)~\cite{TDVP2011, Haegeman2016}, and the finite-temperature properties using the tangent-space tensor renormalization group (tanTRG) method~\cite{Li2023} (see Supplemental Notes~2 and~5~\cite{SM} for details and convergence tests).

\textit{Parameter determination for \CMAO.}---Given the XXZ Hamiltonian, we determine the parameters $J_{\perp}$, $J_c$, and $g_c$ for \CMAO by comparing model calculations with three independent observations, i.e., the low-temperature magnetization, the $c$-axis susceptibility, and the field-polarized INS spectrum. The $B=4$~T dispersion directly fixes $J_{\perp} \simeq 0.65$~K [Fig.~\ref{fig2}(a)], and further requires $g_c\mu_B B-3J_c=\mathrm{const.}$ [gray band in Fig.~\ref{fig2}(d)]. The saturation field in Fig.~\ref{fig2}(b) obeys $g_c\mu_B B_s=\frac32(J_\perp+2J_c)$ (see Appendix and Supplemental Note~3~\cite{SM}). The two bands intersect within the low-loss region of the susceptibility fit. In contrast, the earlier parameter sets, Sets 1 and 2~\cite{Gao2026, colman} (gray triangle and green square, respectively), lie on their own bands but outside the common intersection. The three constraints thus jointly fix the model: the intersection point $(J_c/J_\perp,g_c)\simeq(-0.2,3.75)$, marked by the red star, yields $J_{\perp}\simeq0.65$~K, $J_c\simeq-0.13$~K, and $g_c\simeq3.75$.

With this parameter set, the calculated spin-wave dispersion tracks the $4$-T spectrum [Fig.~\ref{fig2}(a)], the normalized $c$-axis magnetization at $0.03$~K follows the measured curve up to saturation [Fig.~\ref{fig2}(b)], and the out-of-plane susceptibility $\chi_c$ agrees with both our data~\cite{SM} and those of Bastien \textit{et al.}~\cite{colman} [Fig.~\ref{fig2}(c)]. Moreover, the computed magnetic specific heat at $0.20$~T and $0.25$~T shows good overall agreement with the measurements [Fig.~\ref{fig2}(e)], validating the parameter set.

\begin{figure*}[t!]
\centering
\includegraphics[width=0.95\linewidth]{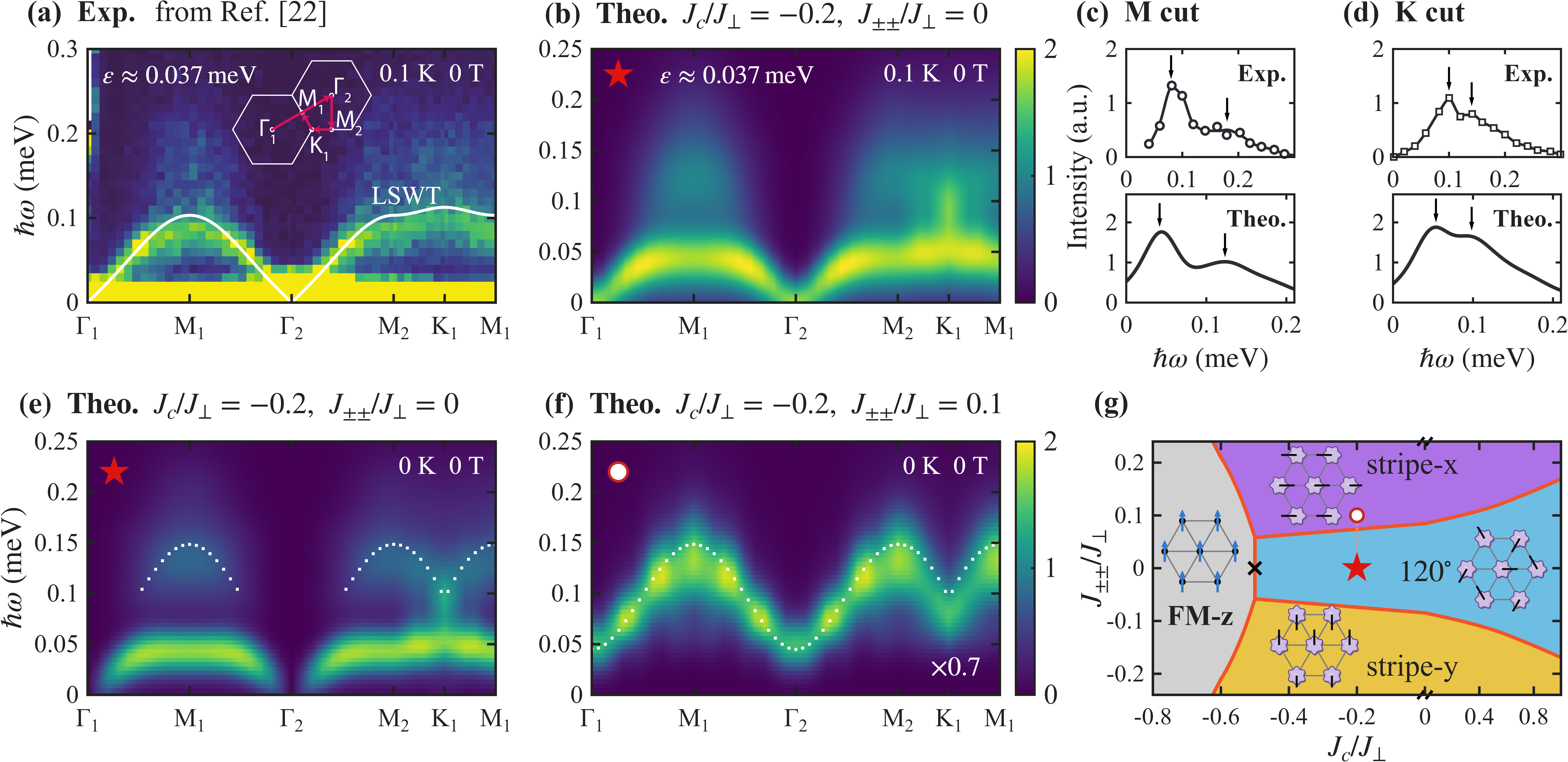}
\caption{(a) Zero-field INS spectrum at $T=0.1$~K adapted from Ref.~\cite{Gao2026}; the white solid curve is the LSWT dispersion of the $120^\circ$ state (TDSF) for our model (Set 3). (b) Calculated $S^{zz}(\mathbf q,\omega)$ at $T=0.1$~K, using bond dimension $D=2000$ in real-time evolution, with the same resolution $\varepsilon\approx0.037$~meV as experiments and the maximum intensity normalized to 2. (c,d) Energy cuts at the $M$ and $K$ points: experiment (upper panels) versus calculation (lower panels); arrows mark the low-energy magnon peak and the higher-energy shoulder. (e,f) Calculated $S^{zz}(\mathbf q,\omega)$ at $T=0$ for Set 3 with $J_{\pm\pm}/J_{\perp}=0$ and $0.1$, respectively (same resolution; intensity in (f) scaled by $0.7$); the white dotted curves show the stripe-magnon dispersion. (g) Ground-state phase diagram in the $(J_c/J_{\perp},J_{\pm\pm}/J_{\perp})$ plane; the red star and the red circle mark the parameters of \CMAO and the representative stripe-phase point used in (b,e) and (f), respectively, and the black cross is the exactly solvable umbrella point at $(-0.5,0)$~\cite{Miyashita1986, Momoi1992, Pal2021}.}
\label{fig3}
\end{figure*}

\textit{Triakontadipolar wave and excitation continuum.}---We now investigate the zero-field dynamics in comparison with the INS measurements~\cite{Gao2026}. Since the triakontadipole dynamical structure factor $S^{+-}(\mathbf q,\omega)$ is essentially invisible to neutron, with leaked dipole weight suppressed by $(g_{\perp}/g_c)^2\simeq0.01$, the measured INS response is dominated by the longitudinal channel $S^{zz}(\mathbf q,\omega)$. The linear spin-wave dispersion computed with our parameter set tracks the low-energy experimental band [white solid curve in Fig.~\ref{fig3}(a)], thereby identifying it as the triakontadipole wave, but captures none of the continuum features (see Appendix).

By contrast, the finite-temperature tanTRG calculation, augmented by real-time dynamics (see Supplemental Note~2~\cite{SM}), reproduces the measured spectrum at $0.1$~K [Fig.~\ref{fig3}(b)]. It includes three major features: gapless excitations toward the $\Gamma$ point, a relatively well-defined low-energy band, and a broad continuum extending to above $\sim0.2$~meV. The calculated energy cuts at the $M$ and $K$ points [lower panels of Figs.~\ref{fig3}(c,d)] agree with the measured ones (upper panels), resolving a low-energy peak together with a higher-energy shoulder. Although the $0.1$~K spectrum is further broadened by thermal effects, all these features persist down to zero temperature [Fig.~\ref{fig3}(e)] and are therefore intrinsic to the model rather than effects of finite cylinder width or limited energy resolution. Moreover, two-magnon excitations cannot account for the higher-energy continuum (see Supplemental Notes~6 and~9~\cite{SM}).

To understand the microscopic origin of the higher-energy continuum, we include the symmetry-allowed $J_{\pm\pm}$ interaction and map out the ground-state phase diagram in the $(J_c/J_{\perp},J_{\pm\pm}/J_{\perp})$ plane [Fig.~\ref{fig3}(g)]. Besides the $120^\circ$ (triakontadipole superfluid, TDSF) phase, it hosts a dipolar FM-$z$ phase for $J_c/J_{\perp}<-0.5$ and triakontadipolar stripe-$x$($y$) phases depending on the sign of $J_{\pm\pm}$. Our parameter set thus places \CMAO in close proximity to the stripe phase, as even a small $J_{\pm\pm}$ can drive the transition. As shown in Figs.~\ref{fig3}(e,f), the low-energy spectra differ substantially across the boundary, whereas the higher-energy features remain similar and closely follow the stripe-magnon dispersion (white dotted curves). Upon warming, the $K$-point correlations weaken while the $M$-point ones strengthen~\cite{SM}, further supporting the stripe character of the higher-energy excitations.

\textit{Discussion.}---Although the triakontadipolar order parameter itself evades neutron probes, its collective triakontadipole wave, a mode rarely observed in magnets, appears as the low-energy branch. A closer comparison with the experimental spectra nonetheless reveals two remaining differences, both concerning this branch. First, its bandwidth in the zero-temperature DMRG spectrum [Fig.~\ref{fig3}(e)] is narrower than in both LSWT and experiment. However, we observe that the band broadens with temperature, and at $0.1$~K it agrees better with experiment. We also note a residual sample dependence of the zero-field specific heat, in contrast to the close mutual agreement of the finite-field data~\cite{Gao2026, Cao2025, colman} (see Supplemental Note~4~\cite{SM}); further neutron measurements on independent samples will help settle this issue.

Second, the calculated low-energy branch is relatively well defined [Fig.~\ref{fig3}(b)], whereas the experimental data are more diffuse [Fig.~\ref{fig3}(a)]. Two effects omitted from our calculations may account for this: a weak but finite $J_{\pm\pm}$ can open additional decay channels for the low-energy branch (see Supplemental Note~7~\cite{SM}), and the Ce-site disorder of \CMAO is known to broaden the branch~\cite{colman, Li2017, Zhu2017, Ma2021}. It is also worth stressing that $g_\perp$ and $J_{\pm\pm}$ are small for the same microscopic reason: both directly reflect the sixfold anisotropy of the ground doublet, as filtered through the large spin--orbit gap (see Appendix).

Moreover, the low-energy triakontadipole excitations remain accessible through NMR relaxation and spin-current probes. Our model calculations predict a longitudinal nuclear spin-lattice relaxation rate $1/T_1^{zz}\propto T^2$ in the superfluid regime, and a normalized spin Seebeck current $\tilde I_2$ that turns negative upon cooling into the same regime. These calculations provide concrete NMR relaxation and transport signatures of the triakontadipole mode (see Fig.~\ref{fig4} in the Appendix).

\textit{Summary and outlook.}---Hidden order has long intrigued condensed matter physicists. In 3D compound URu$_2$Si$_2$, the pronounced thermodynamic transition~\cite{Palstra1985} contrasts with a tiny parasitic dipole moment~\cite{Broholm1987, Niklowitz2010}, motivating multipolar scenarios~\cite{Cricchio2009, Ikeda2012, Ressouche2012}.
Here we show that the transverse pseudospin of the $|\pm5/2\rangle$ Kramers doublet in CMAO is a rank-5 triakontadipole rather than an ordinary dipole. By analyzing the magnetization, neutron spectra, and thermodynamics on equal footing, we establish a dipole--triakontadipole model and identify it as the first triangular-lattice realization of rank-5 hidden order. The resulting model reconciles the $120^\circ$ order favored by the dominant in-plane exchange with the absence of magnetic Bragg peaks in neutron scattering. It attributes the higher-energy continuum to precursor excitations of the nearby triakontadipolar stripe phase without invoking fractionalized spinons~\cite{Shen2016, Paddison2016, Dai2021, Scheie2023, Ferrari2019}. Our results thus reveal how hidden multipolar order can mimic spin-liquid signatures.

The orbital origin of multipolar physics also favors strong magnetoelectric and magnetoelastic couplings~\cite{Kumar2025CMAOMagnetoelectric}, making ultrasound a promising bulk probe of the hidden order in \CMAO~\cite{Luthi1984CeB6, Patri2019}. Although external fields do not couple directly to the multipoles, the ferromagnetic dipole component efficiently links the entropy carried by gapless triakontadipolar excitations to magnetic fields. Together with the ultralow saturation field $B_c\approx0.23$~T, this coupling makes \CMAO\ a promising coolant for sub-Kelvin refrigeration~\cite{Xiang2024, Shu2026EuCo2Al9}.
Besides CMAO, within the REMgAl$_{11}$O$_{19}$ rare-earth hexaaluminate family, NdMgAl$_{11}$O$_{19}$~\cite{Kumar2025NdMgAl, Cao2026NdMgAlMCE} and PrMgAl$_{11}$O$_{19}$~\cite{Ma2024PrMgAl, Kumar2025PrMgAl, Cao2024} also exhibit pronounced magnetic anisotropy and small in-plane $g$ factors. Guided by the computed phase diagram, we call for concerted experimental and theoretical studies of multipolar physics in these rare-earth triangular-lattice quantum magnets.

\begin{acknowledgments}
\textit{Acknowledgments.---}We thank Tong Chen for providing the neutron data reported in Ref.~\cite{Gao2026}. W.L. is grateful to Zhentao Wang for helpful discussions. This work was supported by the National Key Projects for Research and Development of China (Grant Nos.~2024YFA1409200 and 2021YFA1400400), National Natural Science Foundation of China (Grant Nos.~12625408, 12534009, 12447101, 12225407 and 12434005), the Natural Science Foundation of Jiangsu Province (Grant No.~BK20233001), and the Postdoctoral Fellowship Program and China Postdoctoral Science Foundation (Grant No.~BX2026031). We thank the HPC cluster of ITP-CAS for technical support and generous allocation of CPU time.
\end{acknowledgments}

\appendix
\section{Appendix}
\textit{Sample preparation and thermodynamics measurements.}---Single crystals of \CMAO were grown in air by the optical floating-zone technique, with phase purity verified by powder x-ray diffraction and the orientation determined by Laue backscattering. Magnetic susceptibility was measured with a $^3$He-equipped MPMS down to $0.4$~K under a $0.1$~T field applied parallel and perpendicular to the $c$ axis, and specific heat was measured by the relaxation method in a PPMS equipped with a dilution refrigerator (see Supplemental Note~1~\cite{SM} for details).

\textit{Point-charge CEF calculations.}---We perform an all-ion point-charge calculation using PyCrystalField~\cite{Scheie2021PyCrystalField}, with its Ce$^{3+}$ radial integrals and shielding factors, with structures reconstructed from the published X-ray and neutron refinements~\cite{Gao2026}. The full $L=3$, $S=1/2$ basis with $\lambda=78$~meV is retained and the charge sum runs over a cluster of radius $R=80$~\AA, which is not strictly neutral and omits covalency and local disorder. The outcome in Table~\ref{tab:pointcharge_CEF} gives a first CEF gap $\Delta_1\simeq33$~meV, in agreement with Ref.~\cite{Cao2025} and far above the exchange scale.

The $D_{3h}$ site symmetry allows only four terms in $H_{\rm CEF}=\sum_{k,q}B_k^q O_k^q$ (even ranks $k\le6$ for $4f$ electrons): the threefold axis imposes $q\equiv0\pmod{3}$, and the horizontal mirror requires $k+q$ even and renders $B_6^6$ real, leaving $H_{\rm CEF}=B_2^0 O_2^0+B_4^0 O_4^0+B_6^0 O_6^0+B_6^6 O_6^6$. The axial terms are polynomials in $L_z$, hence diagonal in the $|m_L,m_S\rangle$ basis, and leave each $J=5/2$ doublet purely $|\pm m_J\rangle$. The transverse term $O_6^6=\frac12(L_+^6+L_-^6)$ is inert within the $J=5/2$ manifold because its rank $k=6$ exceeds $2J$.

Nevertheless, the $O_6^6$ term acts across the spin--orbit gap, however, coupling the dominant $|3,\downarrow\rangle$ component to the $J=7/2$ stretched state $|-3,\downarrow\rangle$ at $\Delta_{7/2}=7\lambda/2$, and  yielding $|g,+\rangle\approx0.901\,|3,\downarrow\rangle-0.375\,|2,\uparrow\rangle+0.218\,|-3,\downarrow\rangle$~\cite{Cao2025} with $0.218^2\simeq5\%$ $J=7/2$ weight. Moreover, our point-charge parameters give an even smaller admixture, $\sim1\%$, leading to a very small $g_\perp \simeq 0.19$ (see Tab.~\ref{tab:pointcharge_CEF}). This weak admixture may account for the small in-plane susceptibility observed in Fig.~\ref{fig2}(c)~\cite{Cao2025, colman}.

Applied to the published average structures of other family members~\cite{Cao2024,Kumar2025NdMgAl,Cao2025NdZnAl} within their respective Hund-term $LS$ manifolds, the same scheme yields the estimates (see the lower block of Table~\ref{tab:pointcharge_CEF}). PrMgAl$_{11}$O$_{19}$ hosts an Ising non-Kramers ground doublet with vanishing $g_\perp$ in the ideal $2d$ environment (local symmetry breaking and the associated quasi-doublet splitting not included), whereas NdMgAl$_{11}$O$_{19}$ and NdZnAl$_{11}$O$_{19}$ have in-plane $g$ factors smaller than their $c$-axis counterparts, though less strongly suppressed than in \CMAO.

\begin{table}[]
\caption{Single-ion CEF results for \CMAO (upper block) and other family members (lower block); the present-work entries are obtained from the all-ion point-charge scheme within $R=80$~\AA, and the range for \CMAO spans the two reconstructed structures. The lower-block estimates refer to the majority $2d$ site of each compound. $\Delta_1$ is the gap from the ground doublet to the next CEF level; $z\parallel c$.}
\label{tab:pointcharge_CEF}
\centering
\begingroup
\footnotesize
\setlength{\tabcolsep}{3pt}
\renewcommand{\arraystretch}{1.05}
\begin{tabular}{@{}llrrr@{}}
\toprule
Material & Model & $\Delta_1$ (meV) & $g_c$ & $g_\perp$ \\
\midrule
\multirow{3}{*}{\CMAO} & Bastien \textit{et al.}~\cite{colman} & 14.13 & 4.29 & 0 \\
 & Cao \textit{et al.}~\cite{Cao2025} & 36.21 & 3.71 & 0.39 \\
 & present work & 32.28--34.33 & 4.13 & 0.19 \\
\midrule
PrMgAl$_{11}$O$_{19}$ & \multirow{3}{*}{present work} & 15.36 & 5.86 & 0 \\
NdMgAl$_{11}$O$_{19}$ & & 11.83 & 3.75 & 2.06 \\
NdZnAl$_{11}$O$_{19}$ & & 8.11 & 3.83 & 2.01 \\
\bottomrule
\end{tabular}
\endgroup
\end{table}

\textit{Linear spin-wave calculations.}---For the $S=1/2$ triangular-lattice XXZ model with $J_{\pm\pm}=0$ and $B\parallel c$, the Holstein--Primakoff expansion about the fully polarized state ($S_i^z=S-a_i^\dagger a_i$, $S_i^+\simeq\sqrt{2S}\,a_i$) gives the transverse one-magnon dispersion $\varepsilon_{\bm q}^{\rm pol}=g_c\mu_BB-6SJ_c+6SJ_\perp g_{\bm q}$, where $g_{\bm q}=\frac13\sum_{\ell=1}^3\cos(\bm q\cdot\bm\delta_\ell)$ with $\bm\delta_\ell$ the three nearest-neighbor vectors. This dispersion underlies the $4$-T fit in Fig.~\ref{fig2}(a): its bandwidth fixes $J_\perp$, and its offset yields the gray constraint line in Fig.~\ref{fig2}(d). The saturation condition $\min_{\bm q}\varepsilon_{\bm q}^{\rm pol}=0$, with $\min_{\bm q}g_{\bm q}=-1/2$, gives $g_c\mu_BB_s=3S(J_\perp+2J_c)$, producing the green constraint line for the measured $B_s\approx0.25$~T~\cite{colman}.

At zero field, we expand the Hamiltonian about the coplanar $120^\circ$ and stripe states.  Obtained by diagonalizing the quadratic Holstein--Primakoff Hamiltonian via a standard Bogoliubov transformation, the single-magnon branch in $S^{zz}(\mathbf q,\omega)$ is shown as the white solid curve in Fig.~\ref{fig3}(a) for the $120^\circ$ state, for $S=1/2$, $J_{\perp}=0.65$~K, $J_c=-0.13$~K, and $J_{\pm\pm}=0$.
The magnon energy is $\varepsilon_{\bm q}=3SJ_{\perp}\sqrt{(1-g_{\bm q})(1+2\Delta g_{\bm q})}$,
where $\Delta=J_c/J_\perp$ and $g_{\bm q}$ is defined above.
Compared with the LSWT curve in Fig.~\ref{fig3}(a),
the low-energy branch obtained by DMRG in Fig.~\ref{fig3}(e) at the same exchange parameters
has a substantially reduced bandwidth, reflecting a strong downward renormalization.

The stripe dispersion overlaid in Figs.~\ref{fig3}(e,f) is calculated at $(J_c/J_\perp,J_{\pm\pm}/J_\perp)=(-0.2, 0.1)$. The two-sublattice LSWT calculation yields two magnon branches. The white dotted curves trace the branch with the larger $S^{zz}(\mathbf q,\omega)$ spectral weight at each momentum, rather than a fixed lower- or upper-energy branch. In Fig.~\ref{fig3}(e), portions in regions of weak DMRG intensity are omitted for clarity.

\begin{figure}[htbp]
\centering
\includegraphics[width=\columnwidth]{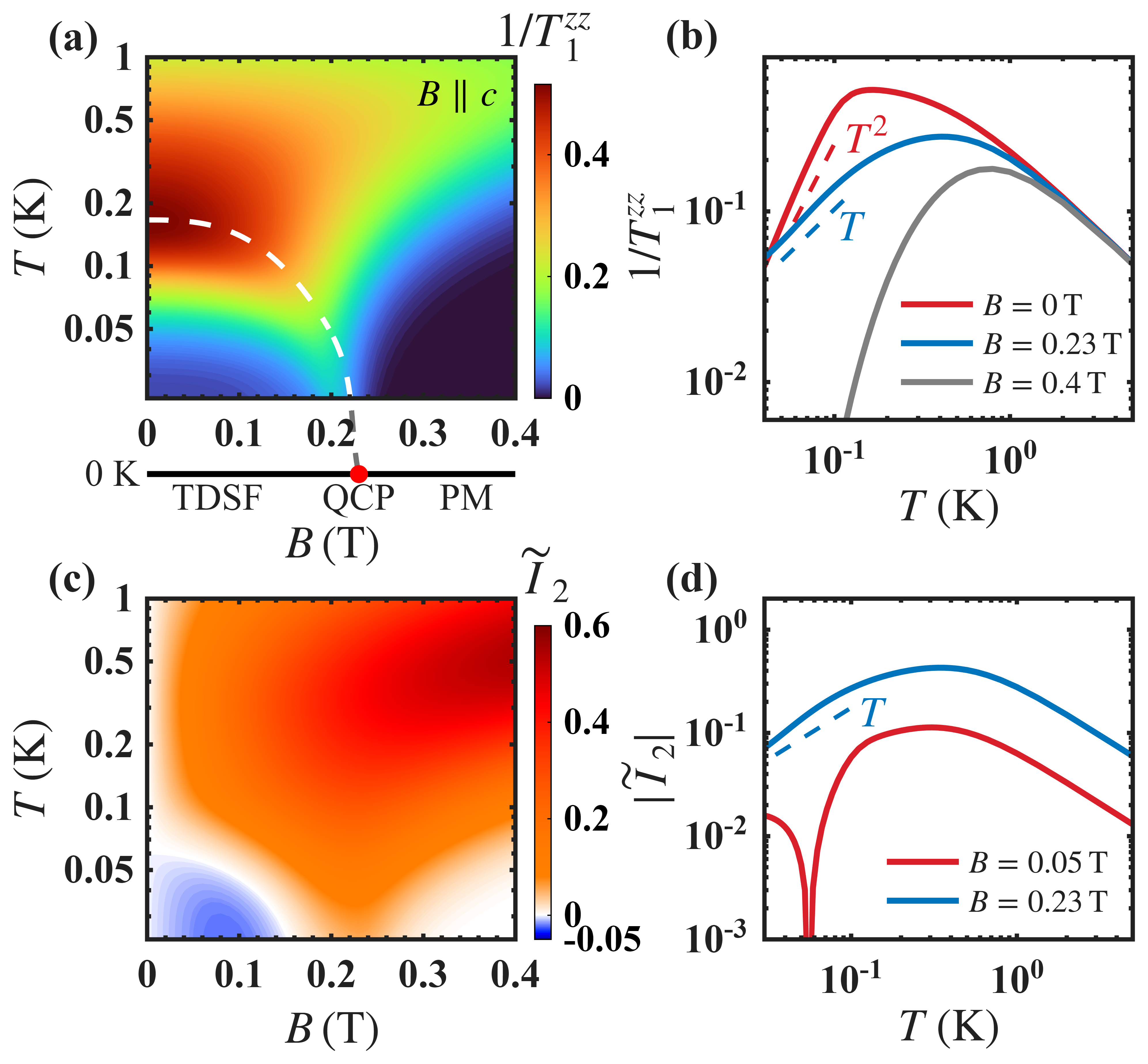}
\caption{(a) Field--temperature map of the longitudinal relaxation-rate proxy $1/T_1^{zz}$, where the schematic dashed curve connects to the quantum critical point (QCP, red dot) at $B_c\approx0.23$~T on the separate $T=0$ line. The QCP separates the TDSF and field-polarized phases on its low- and high-field sides, respectively. (b) Temperature dependence of $1/T_1^{zz}$ at $B=0$, $0.23$, and $0.4$~T. The red and blue dashed lines are $T^2$ and $T$ guides, respectively. (c) Field--temperature map of $\tilde I_2$, the normalized spin current, which changes sign with temperature and field. (d) Temperature dependence of $|\tilde I_2|$ at $B=0.05$ and $0.23$~T. The blue dashed line is a $T$ guide. The calculations involve model parameter $J_{\pm\pm}=0$ and $B\parallel c$.}
\label{fig4}
\end{figure}

\textit{NMR and spin-current probes.}---As in the INS analysis, we focus on the longitudinal relaxation channel $1/T_1^{zz}$, computed with tanTRG using an imaginary-time proxy (see Supplemental Note~2~\cite{SM}); since $1/T_1$ filters low-frequency spin fluctuations, it probes the gapless triakontadipole wave. We stress that the triakontadipolar order does not split NMR lines as it generates no static hyperfine field, so that the excitation is signaled through the relaxation rate rather than through spectral features. Figure~\ref{fig4}(a) shows stronger low-temperature relaxation in the TDSF regime and near the saturation QCP than in the gapped polarized regime: in the superfluid regime, $1/T_1^{zz}\propto T^2$ [Fig.~\ref{fig4}(b)]; at the saturation QCP ($z=2$, $d=2$), it crosses over to $1/T_1^{zz}\propto T$; and the gapped polarized phase implies thermally suppressed relaxation. Viewed from the polarized side, the field-induced TDSF is a Bose--Einstein condensate of pseudospin flips, each carrying five units of angular momentum---a rank-5 counterpart of the two-magnon spin-nematic condensate~\cite{Sheng2025}. Experimentally, the zero-field behavior can be approached with a weak in-plane probe field, following the NMR strategy employed for TmMgGaO$_4$~\cite{Li2020TmMgGaO, Hu2020BKT}: the field sets the nuclear quantization axis, allowing $c$-axis spin fluctuations to drive nuclear relaxation, while the small in-plane $g_\perp$ ensures that the electronic state is only weakly perturbed.

In the spin Seebeck effect (SSE) setup, a temperature gradient between a magnet and an adjacent metal induces a spin current through the interface~\cite{Uchida2010Insulator, Xiao2010PRB, Adachi2011, Bauer2012SpinCaloritronics}, which can be detected on the metal side via the inverse spin Hall effect; its sign reflects the contributing elementary excitations~\cite{Masuda2024, Tang2026SpinIceSSE}. Using the imaginary-time proxy~\cite{Gao2025SpinSeebeck}, we compute the normalized spin current $\tilde I_2$ with tanTRG. The calculated $\tilde I_2$ turns negative upon cooling into the TDSF regime [Fig.~\ref{fig4}(c)], with $|\tilde I_2|$ at $B=0.05$~T passing through a pronounced dip before rising again [Fig.~\ref{fig4}(d)]. This low-temperature response is consistent with Goldstone-mode-mediated spin transport, which can saturate rather than decay, as established for spin superfluids~\cite{Takei2014_ST, Qaiumzadeh2017_ST} and supersolids~\cite{Gao2025SpinSeebeck}. At the field-driven QCP, $B_c\approx0.23$~T, the leading response scales as $\tilde I_2\propto T^{d/z}=T$ with $d=z=2$ [Fig.~\ref{fig4}(d)].

\nocite{Gleis2023CBE,Li2024CBE,Xi2024_1oT1,Li2011,Chen2017,Chen2018}
\bibliography{references}

\begin{thebibliography}{89}%
\makeatletter
\providecommand \@ifxundefined [1]{%
 \@ifx{#1\undefined}
}%
\providecommand \@ifnum [1]{%
 \ifnum #1\expandafter \@firstoftwo
 \else \expandafter \@secondoftwo
 \fi
}%
\providecommand \@ifx [1]{%
 \ifx #1\expandafter \@firstoftwo
 \else \expandafter \@secondoftwo
 \fi
}%
\providecommand \natexlab [1]{#1}%
\providecommand \enquote  [1]{``#1''}%
\providecommand \bibnamefont  [1]{#1}%
\providecommand \bibfnamefont [1]{#1}%
\providecommand \citenamefont [1]{#1}%
\providecommand \href@noop [0]{\@secondoftwo}%
\providecommand \href [0]{\begingroup \@sanitize@url \@href}%
\providecommand \@href[1]{\@@startlink{#1}\@@href}%
\providecommand \@@href[1]{\endgroup#1\@@endlink}%
\providecommand \@sanitize@url [0]{\catcode `\\12\catcode `\$12\catcode
  `\&12\catcode `\#12\catcode `\^12\catcode `\_12\catcode `\%12\relax}%
\providecommand \@@startlink[1]{}%
\providecommand \@@endlink[0]{}%
\providecommand \url  [0]{\begingroup\@sanitize@url \@url }%
\providecommand \@url [1]{\endgroup\@href {#1}{\urlprefix }}%
\providecommand \urlprefix  [0]{URL }%
\providecommand \Eprint [0]{\href }%
\providecommand \doibase [0]{https://doi.org/}%
\providecommand \selectlanguage [0]{\@gobble}%
\providecommand \bibinfo  [0]{\@secondoftwo}%
\providecommand \bibfield  [0]{\@secondoftwo}%
\providecommand \translation [1]{[#1]}%
\providecommand \BibitemOpen [0]{}%
\providecommand \bibitemStop [0]{}%
\providecommand \bibitemNoStop [0]{.\EOS\space}%
\providecommand \EOS [0]{\spacefactor3000\relax}%
\providecommand \BibitemShut  [1]{\csname bibitem#1\endcsname}%
\let\auto@bib@innerbib\@empty
%</preamble>
\bibitem [{\citenamefont {Balents}(2010)}]{Balents2010}%
  \BibitemOpen
  \bibfield  {author} {\bibinfo {author} {\bibfnamefont {L.}~\bibnamefont
  {Balents}},\ }\bibfield  {title} {\bibinfo {title} {{Spin liquids in
  frustrated magnets}},\ }\href {https://doi.org/10.1038/nature08917}
  {\bibfield  {journal} {\bibinfo  {journal} {Nature}\ }\textbf {\bibinfo
  {volume} {464}},\ \bibinfo {pages} {199} (\bibinfo {year}
  {2010})}\BibitemShut {NoStop}%
\bibitem [{\citenamefont {Mila}(2000)}]{Mila2000}%
  \BibitemOpen
  \bibfield  {author} {\bibinfo {author} {\bibfnamefont {F.}~\bibnamefont
  {Mila}},\ }\bibfield  {title} {\bibinfo {title} {{Quantum spin liquids}},\
  }\href {https://doi.org/10.1088/0143-0807/21/6/302} {\bibfield  {journal}
  {\bibinfo  {journal} {Eur. J. Phys.}\ }\textbf {\bibinfo {volume} {21}},\
  \bibinfo {pages} {499} (\bibinfo {year} {2000})}\BibitemShut {NoStop}%
\bibitem [{\citenamefont {Savary}\ and\ \citenamefont
  {Balents}(2017)}]{Savary2016}%
  \BibitemOpen
  \bibfield  {author} {\bibinfo {author} {\bibfnamefont {L.}~\bibnamefont
  {Savary}}\ and\ \bibinfo {author} {\bibfnamefont {L.}~\bibnamefont
  {Balents}},\ }\bibfield  {title} {\bibinfo {title} {{Quantum spin liquids: a
  review}},\ }\href {https://doi.org/10.1088/0034-4885/80/1/016502} {\bibfield
  {journal} {\bibinfo  {journal} {Rep. Prog. Phys.}\ }\textbf {\bibinfo
  {volume} {80}},\ \bibinfo {pages} {016502} (\bibinfo {year}
  {2017})}\BibitemShut {NoStop}%
\bibitem [{\citenamefont {Broholm}\ \emph {et~al.}(2020)\citenamefont
  {Broholm}, \citenamefont {Cava}, \citenamefont {Kivelson}, \citenamefont
  {Nocera}, \citenamefont {Norman},\ and\ \citenamefont
  {Senthil}}]{Broholm2020}%
  \BibitemOpen
  \bibfield  {author} {\bibinfo {author} {\bibfnamefont {C.}~\bibnamefont
  {Broholm}}, \bibinfo {author} {\bibfnamefont {R.~J.}\ \bibnamefont {Cava}},
  \bibinfo {author} {\bibfnamefont {S.~A.}\ \bibnamefont {Kivelson}}, \bibinfo
  {author} {\bibfnamefont {D.~G.}\ \bibnamefont {Nocera}}, \bibinfo {author}
  {\bibfnamefont {M.~R.}\ \bibnamefont {Norman}},\ and\ \bibinfo {author}
  {\bibfnamefont {T.}~\bibnamefont {Senthil}},\ }\bibfield  {title} {\bibinfo
  {title} {{Quantum spin liquids}},\ }\href
  {https://doi.org/10.1126/science.aay0668} {\bibfield  {journal} {\bibinfo
  {journal} {Science}\ }\textbf {\bibinfo {volume} {367}},\ \bibinfo {pages}
  {eaay0668} (\bibinfo {year} {2020})}\BibitemShut {NoStop}%
\bibitem [{\citenamefont {Chubukov}(1990)}]{Chubukov1990}%
  \BibitemOpen
  \bibfield  {author} {\bibinfo {author} {\bibfnamefont {A.~V.}\ \bibnamefont
  {Chubukov}},\ }\bibfield  {title} {\bibinfo {title} {{Fluctuations in spin
  nematics}},\ }\href {https://doi.org/10.1088/0953-8984/2/6/018} {\bibfield
  {journal} {\bibinfo  {journal} {J. Phys.: Condens. Matter}\ }\textbf
  {\bibinfo {volume} {2}},\ \bibinfo {pages} {1593} (\bibinfo {year}
  {1990})}\BibitemShut {NoStop}%
\bibitem [{\citenamefont {Bhattacharjee}\ \emph {et~al.}(2006)\citenamefont
  {Bhattacharjee}, \citenamefont {Shenoy},\ and\ \citenamefont
  {Senthil}}]{Bhattacharjee2006}%
  \BibitemOpen
  \bibfield  {author} {\bibinfo {author} {\bibfnamefont {S.}~\bibnamefont
  {Bhattacharjee}}, \bibinfo {author} {\bibfnamefont {V.~B.}\ \bibnamefont
  {Shenoy}},\ and\ \bibinfo {author} {\bibfnamefont {T.}~\bibnamefont
  {Senthil}},\ }\bibfield  {title} {\bibinfo {title} {{Possible
  ferro-spin-nematic order in NiGa$_2$S$_4$}},\ }\href
  {https://doi.org/10.1103/physrevb.74.092406} {\bibfield  {journal} {\bibinfo
  {journal} {Phys. Rev. B}\ }\textbf {\bibinfo {volume} {74}},\ \bibinfo
  {pages} {092406} (\bibinfo {year} {2006})}\BibitemShut {NoStop}%
\bibitem [{\citenamefont {L{\"a}uchli}\ \emph {et~al.}(2006)\citenamefont
  {L{\"a}uchli}, \citenamefont {Mila},\ and\ \citenamefont
  {Penc}}]{Lauchli2006}%
  \BibitemOpen
  \bibfield  {author} {\bibinfo {author} {\bibfnamefont {A.}~\bibnamefont
  {L{\"a}uchli}}, \bibinfo {author} {\bibfnamefont {F.}~\bibnamefont {Mila}},\
  and\ \bibinfo {author} {\bibfnamefont {K.}~\bibnamefont {Penc}},\ }\bibfield
  {title} {\bibinfo {title} {{Quadrupolar phases of the $S=1$
  bilinear-biquadratic Heisenberg model on the triangular lattice}},\ }\href
  {https://doi.org/10.1103/physrevlett.97.087205} {\bibfield  {journal}
  {\bibinfo  {journal} {Phys. Rev. Lett.}\ }\textbf {\bibinfo {volume} {97}},\
  \bibinfo {pages} {087205} (\bibinfo {year} {2006})}\BibitemShut {NoStop}%
\bibitem [{\citenamefont {Heidarian}\ and\ \citenamefont
  {Paramekanti}(2010)}]{Heidarian2010}%
  \BibitemOpen
  \bibfield  {author} {\bibinfo {author} {\bibfnamefont {D.}~\bibnamefont
  {Heidarian}}\ and\ \bibinfo {author} {\bibfnamefont {A.}~\bibnamefont
  {Paramekanti}},\ }\bibfield  {title} {\bibinfo {title} {{Supersolidity in the
  triangular-lattice spin-$1/2$ XXZ model: A variational perspective}},\ }\href
  {https://doi.org/10.1103/physrevlett.104.015301} {\bibfield  {journal}
  {\bibinfo  {journal} {Phys. Rev. Lett.}\ }\textbf {\bibinfo {volume} {104}},\
  \bibinfo {pages} {015301} (\bibinfo {year} {2010})}\BibitemShut {NoStop}%
\bibitem [{\citenamefont {Gao}\ \emph {et~al.}(2022)\citenamefont {Gao},
  \citenamefont {Fan}, \citenamefont {Li}, \citenamefont {Yang}, \citenamefont
  {Zeng}, \citenamefont {Sheng} \emph {et~al.}}]{Gao2022}%
  \BibitemOpen
  \bibfield  {author} {\bibinfo {author} {\bibfnamefont {Y.}~\bibnamefont
  {Gao}}, \bibinfo {author} {\bibfnamefont {Y.-C.}\ \bibnamefont {Fan}},
  \bibinfo {author} {\bibfnamefont {H.}~\bibnamefont {Li}}, \bibinfo {author}
  {\bibfnamefont {F.}~\bibnamefont {Yang}}, \bibinfo {author} {\bibfnamefont
  {X.-T.}\ \bibnamefont {Zeng}}, \bibinfo {author} {\bibfnamefont {X.-L.}\
  \bibnamefont {Sheng}}, \emph {et~al.},\ }\bibfield  {title} {\bibinfo {title}
  {{Spin supersolidity in nearly ideal easy-axis triangular quantum
  antiferromagnet Na$_2$BaCo(PO$_4$)$_2$}},\ }\href
  {https://doi.org/10.1038/s41535-022-00500-3} {\bibfield  {journal} {\bibinfo
  {journal} {npj Quantum Mater.}\ }\textbf {\bibinfo {volume} {7}},\ \bibinfo
  {pages} {89} (\bibinfo {year} {2022})}\BibitemShut {NoStop}%
\bibitem [{\citenamefont {Xiang}\ \emph {et~al.}(2024)\citenamefont {Xiang},
  \citenamefont {Zhang}, \citenamefont {Gao}, \citenamefont {Schmidt},
  \citenamefont {Schmalzl}, \citenamefont {Wang} \emph {et~al.}}]{Xiang2024}%
  \BibitemOpen
  \bibfield  {author} {\bibinfo {author} {\bibfnamefont {J.}~\bibnamefont
  {Xiang}}, \bibinfo {author} {\bibfnamefont {C.}~\bibnamefont {Zhang}},
  \bibinfo {author} {\bibfnamefont {Y.}~\bibnamefont {Gao}}, \bibinfo {author}
  {\bibfnamefont {W.}~\bibnamefont {Schmidt}}, \bibinfo {author} {\bibfnamefont
  {K.}~\bibnamefont {Schmalzl}}, \bibinfo {author} {\bibfnamefont {C.-W.}\
  \bibnamefont {Wang}}, \emph {et~al.},\ }\bibfield  {title} {\bibinfo {title}
  {{Giant magnetocaloric effect in spin supersolid candidate
  Na$_2$BaCo(PO$_4$)$_2$}},\ }\href
  {https://doi.org/10.1038/s41586-023-06885-w} {\bibfield  {journal} {\bibinfo
  {journal} {Nature}\ }\textbf {\bibinfo {volume} {625}},\ \bibinfo {pages}
  {270} (\bibinfo {year} {2024})}\BibitemShut {NoStop}%
\bibitem [{\citenamefont {Shu}\ \emph {et~al.}(2026)\citenamefont {Shu},
  \citenamefont {Xu}, \citenamefont {Xi}, \citenamefont {He}, \citenamefont
  {Xiang}, \citenamefont {Qu} \emph {et~al.}}]{Shu2026EuCo2Al9}%
  \BibitemOpen
  \bibfield  {author} {\bibinfo {author} {\bibfnamefont {M.}~\bibnamefont
  {Shu}}, \bibinfo {author} {\bibfnamefont {X.}~\bibnamefont {Xu}}, \bibinfo
  {author} {\bibfnamefont {N.}~\bibnamefont {Xi}}, \bibinfo {author}
  {\bibfnamefont {M.}~\bibnamefont {He}}, \bibinfo {author} {\bibfnamefont
  {J.}~\bibnamefont {Xiang}}, \bibinfo {author} {\bibfnamefont
  {G.}~\bibnamefont {Qu}}, \emph {et~al.},\ }\bibfield  {title} {\bibinfo
  {title} {{Giant magnetocaloric effect and spin supersolid in a metallic
  dipolar magnet}},\ }\href {https://doi.org/10.1038/s41586-026-10144-z}
  {\bibfield  {journal} {\bibinfo  {journal} {Nature}\ }\textbf {\bibinfo
  {volume} {651}},\ \bibinfo {pages} {61} (\bibinfo {year} {2026})}\BibitemShut
  {NoStop}%
\bibitem [{\citenamefont {Li}\ \emph {et~al.}(2015)\citenamefont {Li},
  \citenamefont {Chen}, \citenamefont {Tong}, \citenamefont {Pi}, \citenamefont
  {Liu}, \citenamefont {Yang}, \citenamefont {Wang},\ and\ \citenamefont
  {Zhang}}]{Li2015}%
  \BibitemOpen
  \bibfield  {author} {\bibinfo {author} {\bibfnamefont {Y.}~\bibnamefont
  {Li}}, \bibinfo {author} {\bibfnamefont {G.}~\bibnamefont {Chen}}, \bibinfo
  {author} {\bibfnamefont {W.}~\bibnamefont {Tong}}, \bibinfo {author}
  {\bibfnamefont {L.}~\bibnamefont {Pi}}, \bibinfo {author} {\bibfnamefont
  {J.}~\bibnamefont {Liu}}, \bibinfo {author} {\bibfnamefont {Z.}~\bibnamefont
  {Yang}}, \bibinfo {author} {\bibfnamefont {X.}~\bibnamefont {Wang}},\ and\
  \bibinfo {author} {\bibfnamefont {Q.}~\bibnamefont {Zhang}},\ }\bibfield
  {title} {\bibinfo {title} {{Rare-earth triangular-lattice spin liquid: A
  single-crystal study of YbMgGaO$_4$}},\ }\href
  {https://doi.org/10.1103/physrevlett.115.167203} {\bibfield  {journal}
  {\bibinfo  {journal} {Phys. Rev. Lett.}\ }\textbf {\bibinfo {volume} {115}},\
  \bibinfo {pages} {167203} (\bibinfo {year} {2015})}\BibitemShut {NoStop}%
\bibitem [{\citenamefont {Paddison}\ \emph {et~al.}(2017)\citenamefont
  {Paddison}, \citenamefont {Daum}, \citenamefont {Dun}, \citenamefont
  {Ehlers}, \citenamefont {Liu}, \citenamefont {Stone}, \citenamefont {Zhou},\
  and\ \citenamefont {Mourigal}}]{Paddison2016}%
  \BibitemOpen
  \bibfield  {author} {\bibinfo {author} {\bibfnamefont {J.~A.~M.}\
  \bibnamefont {Paddison}}, \bibinfo {author} {\bibfnamefont {M.}~\bibnamefont
  {Daum}}, \bibinfo {author} {\bibfnamefont {Z.}~\bibnamefont {Dun}}, \bibinfo
  {author} {\bibfnamefont {G.}~\bibnamefont {Ehlers}}, \bibinfo {author}
  {\bibfnamefont {Y.}~\bibnamefont {Liu}}, \bibinfo {author} {\bibfnamefont
  {M.~B.}\ \bibnamefont {Stone}}, \bibinfo {author} {\bibfnamefont
  {H.}~\bibnamefont {Zhou}},\ and\ \bibinfo {author} {\bibfnamefont
  {M.}~\bibnamefont {Mourigal}},\ }\bibfield  {title} {\bibinfo {title}
  {{Continuous excitations of the triangular-lattice quantum spin liquid
  YbMgGaO$_4$}},\ }\href {https://doi.org/10.1038/nphys3971} {\bibfield
  {journal} {\bibinfo  {journal} {Nat. Phys.}\ }\textbf {\bibinfo {volume}
  {13}},\ \bibinfo {pages} {117} (\bibinfo {year} {2017})}\BibitemShut
  {NoStop}%
\bibitem [{\citenamefont {Shen}\ \emph {et~al.}(2016)\citenamefont {Shen},
  \citenamefont {Li}, \citenamefont {Wo}, \citenamefont {Li}, \citenamefont
  {Shen}, \citenamefont {Pan} \emph {et~al.}}]{Shen2016}%
  \BibitemOpen
  \bibfield  {author} {\bibinfo {author} {\bibfnamefont {Y.}~\bibnamefont
  {Shen}}, \bibinfo {author} {\bibfnamefont {Y.-D.}\ \bibnamefont {Li}},
  \bibinfo {author} {\bibfnamefont {H.}~\bibnamefont {Wo}}, \bibinfo {author}
  {\bibfnamefont {Y.}~\bibnamefont {Li}}, \bibinfo {author} {\bibfnamefont
  {S.}~\bibnamefont {Shen}}, \bibinfo {author} {\bibfnamefont {B.}~\bibnamefont
  {Pan}}, \emph {et~al.},\ }\bibfield  {title} {\bibinfo {title} {{Evidence for
  a spinon Fermi surface in a triangular-lattice quantum-spin-liquid
  candidate}},\ }\href {https://doi.org/10.1038/nature20614} {\bibfield
  {journal} {\bibinfo  {journal} {Nature}\ }\textbf {\bibinfo {volume} {540}},\
  \bibinfo {pages} {559} (\bibinfo {year} {2016})}\BibitemShut {NoStop}%
\bibitem [{\citenamefont {Zhu}\ \emph {et~al.}(2017)\citenamefont {Zhu},
  \citenamefont {Maksimov}, \citenamefont {White},\ and\ \citenamefont
  {Chernyshev}}]{Zhu2017}%
  \BibitemOpen
  \bibfield  {author} {\bibinfo {author} {\bibfnamefont {Z.}~\bibnamefont
  {Zhu}}, \bibinfo {author} {\bibfnamefont {P.~A.}\ \bibnamefont {Maksimov}},
  \bibinfo {author} {\bibfnamefont {S.~R.}\ \bibnamefont {White}},\ and\
  \bibinfo {author} {\bibfnamefont {A.~L.}\ \bibnamefont {Chernyshev}},\
  }\bibfield  {title} {\bibinfo {title} {{Disorder-induced mimicry of a spin
  liquid in YbMgGaO$_4$}},\ }\href
  {https://doi.org/10.1103/physrevlett.119.157201} {\bibfield  {journal}
  {\bibinfo  {journal} {Phys. Rev. Lett.}\ }\textbf {\bibinfo {volume} {119}},\
  \bibinfo {pages} {157201} (\bibinfo {year} {2017})}\BibitemShut {NoStop}%
\bibitem [{\citenamefont {Bordelon}\ \emph {et~al.}(2019)\citenamefont
  {Bordelon}, \citenamefont {Kenney}, \citenamefont {Liu}, \citenamefont
  {Hogan}, \citenamefont {Posthuma}, \citenamefont {Kavand} \emph
  {et~al.}}]{Bordelon2019}%
  \BibitemOpen
  \bibfield  {author} {\bibinfo {author} {\bibfnamefont {M.~M.}\ \bibnamefont
  {Bordelon}}, \bibinfo {author} {\bibfnamefont {E.}~\bibnamefont {Kenney}},
  \bibinfo {author} {\bibfnamefont {C.}~\bibnamefont {Liu}}, \bibinfo {author}
  {\bibfnamefont {T.}~\bibnamefont {Hogan}}, \bibinfo {author} {\bibfnamefont
  {L.}~\bibnamefont {Posthuma}}, \bibinfo {author} {\bibfnamefont
  {M.}~\bibnamefont {Kavand}}, \emph {et~al.},\ }\bibfield  {title} {\bibinfo
  {title} {{Field-tunable quantum disordered ground state in the
  triangular-lattice antiferromagnet NaYbO$_2$}},\ }\href
  {https://doi.org/10.1038/s41567-019-0594-5} {\bibfield  {journal} {\bibinfo
  {journal} {Nat. Phys.}\ }\textbf {\bibinfo {volume} {15}},\ \bibinfo {pages}
  {1058} (\bibinfo {year} {2019})}\BibitemShut {NoStop}%
\bibitem [{\citenamefont {Xing}\ \emph {et~al.}(2020)\citenamefont {Xing},
  \citenamefont {Sanjeewa}, \citenamefont {Kim}, \citenamefont {Stewart},
  \citenamefont {Du}, \citenamefont {Reboredo}, \citenamefont {Custelcean},\
  and\ \citenamefont {Sefat}}]{Xing2019CsYbSe2}%
  \BibitemOpen
  \bibfield  {author} {\bibinfo {author} {\bibfnamefont {J.}~\bibnamefont
  {Xing}}, \bibinfo {author} {\bibfnamefont {L.~D.}\ \bibnamefont {Sanjeewa}},
  \bibinfo {author} {\bibfnamefont {J.}~\bibnamefont {Kim}}, \bibinfo {author}
  {\bibfnamefont {G.~R.}\ \bibnamefont {Stewart}}, \bibinfo {author}
  {\bibfnamefont {M.-H.}\ \bibnamefont {Du}}, \bibinfo {author} {\bibfnamefont
  {F.~A.}\ \bibnamefont {Reboredo}}, \bibinfo {author} {\bibfnamefont
  {R.}~\bibnamefont {Custelcean}},\ and\ \bibinfo {author} {\bibfnamefont
  {A.~S.}\ \bibnamefont {Sefat}},\ }\bibfield  {title} {\bibinfo {title}
  {{Crystal Synthesis and Frustrated Magnetism in Triangular Lattice CsRESe$_2$
  (RE = La--Lu): Quantum Spin Liquid Candidates CsCeSe$_2$ and CsYbSe$_2$}},\
  }\href {https://doi.org/10.1021/acsmaterialslett.9b00464} {\bibfield
  {journal} {\bibinfo  {journal} {ACS Mater. Lett.}\ }\textbf {\bibinfo
  {volume} {2}},\ \bibinfo {pages} {71} (\bibinfo {year} {2020})}\BibitemShut
  {NoStop}%
\bibitem [{\citenamefont {Dai}\ \emph {et~al.}(2021)\citenamefont {Dai},
  \citenamefont {Zhang}, \citenamefont {Xie}, \citenamefont {Duan},
  \citenamefont {Gao}, \citenamefont {Zhu} \emph {et~al.}}]{Dai2021}%
  \BibitemOpen
  \bibfield  {author} {\bibinfo {author} {\bibfnamefont {P.-L.}\ \bibnamefont
  {Dai}}, \bibinfo {author} {\bibfnamefont {G.}~\bibnamefont {Zhang}}, \bibinfo
  {author} {\bibfnamefont {Y.}~\bibnamefont {Xie}}, \bibinfo {author}
  {\bibfnamefont {C.}~\bibnamefont {Duan}}, \bibinfo {author} {\bibfnamefont
  {Y.}~\bibnamefont {Gao}}, \bibinfo {author} {\bibfnamefont {Z.}~\bibnamefont
  {Zhu}}, \emph {et~al.},\ }\bibfield  {title} {\bibinfo {title} {{Spinon
  Fermi-surface spin liquid in a triangular-lattice antiferromagnet
  NaYbSe$_2$}},\ }\href {https://doi.org/10.1103/physrevx.11.021044} {\bibfield
   {journal} {\bibinfo  {journal} {Phys. Rev. X}\ }\textbf {\bibinfo {volume}
  {11}},\ \bibinfo {pages} {021044} (\bibinfo {year} {2021})}\BibitemShut
  {NoStop}%
\bibitem [{\citenamefont {Scheie}\ \emph {et~al.}(2024)\citenamefont {Scheie},
  \citenamefont {Ghioldi}, \citenamefont {Xing}, \citenamefont {Paddison},
  \citenamefont {Sherman}, \citenamefont {Dupont} \emph {et~al.}}]{Scheie2023}%
  \BibitemOpen
  \bibfield  {author} {\bibinfo {author} {\bibfnamefont {A.~O.}\ \bibnamefont
  {Scheie}}, \bibinfo {author} {\bibfnamefont {E.~A.}\ \bibnamefont {Ghioldi}},
  \bibinfo {author} {\bibfnamefont {J.}~\bibnamefont {Xing}}, \bibinfo {author}
  {\bibfnamefont {J.~A.~M.}\ \bibnamefont {Paddison}}, \bibinfo {author}
  {\bibfnamefont {N.~E.}\ \bibnamefont {Sherman}}, \bibinfo {author}
  {\bibfnamefont {M.}~\bibnamefont {Dupont}}, \emph {et~al.},\ }\bibfield
  {title} {\bibinfo {title} {{Proximate spin liquid and fractionalization in
  the triangular antiferromagnet KYbSe$_2$}},\ }\href
  {https://doi.org/10.1038/s41567-023-02259-1} {\bibfield  {journal} {\bibinfo
  {journal} {Nat. Phys.}\ }\textbf {\bibinfo {volume} {20}},\ \bibinfo {pages}
  {74} (\bibinfo {year} {2024})}\BibitemShut {NoStop}%
\bibitem [{\citenamefont {Xie}\ \emph {et~al.}(2024)\citenamefont {Xie},
  \citenamefont {Gozel}, \citenamefont {Xing}, \citenamefont {Zhao},
  \citenamefont {Avdoshenko}, \citenamefont {Wu} \emph
  {et~al.}}]{Xie2024CsCeSe2}%
  \BibitemOpen
  \bibfield  {author} {\bibinfo {author} {\bibfnamefont {T.}~\bibnamefont
  {Xie}}, \bibinfo {author} {\bibfnamefont {S.}~\bibnamefont {Gozel}}, \bibinfo
  {author} {\bibfnamefont {J.}~\bibnamefont {Xing}}, \bibinfo {author}
  {\bibfnamefont {N.}~\bibnamefont {Zhao}}, \bibinfo {author} {\bibfnamefont
  {S.~M.}\ \bibnamefont {Avdoshenko}}, \bibinfo {author} {\bibfnamefont
  {L.}~\bibnamefont {Wu}}, \emph {et~al.},\ }\bibfield  {title} {\bibinfo
  {title} {{Quantum Spin Dynamics Due to Strong Kitaev Interactions in the
  Triangular-Lattice Antiferromagnet CsCeSe$_2$}},\ }\href
  {https://doi.org/10.1103/physrevlett.133.096703} {\bibfield  {journal}
  {\bibinfo  {journal} {Phys. Rev. Lett.}\ }\textbf {\bibinfo {volume} {133}},\
  \bibinfo {pages} {096703} (\bibinfo {year} {2024})}\BibitemShut {NoStop}%
\bibitem [{\citenamefont {Ashtar}\ \emph {et~al.}(2019)\citenamefont {Ashtar},
  \citenamefont {Marwat}, \citenamefont {Gao}, \citenamefont {Zhang},
  \citenamefont {Pi}, \citenamefont {Yuan},\ and\ \citenamefont
  {Tian}}]{Ashtar2019b}%
  \BibitemOpen
  \bibfield  {author} {\bibinfo {author} {\bibfnamefont {M.}~\bibnamefont
  {Ashtar}}, \bibinfo {author} {\bibfnamefont {M.~A.}\ \bibnamefont {Marwat}},
  \bibinfo {author} {\bibfnamefont {Y.~X.}\ \bibnamefont {Gao}}, \bibinfo
  {author} {\bibfnamefont {Z.~T.}\ \bibnamefont {Zhang}}, \bibinfo {author}
  {\bibfnamefont {L.}~\bibnamefont {Pi}}, \bibinfo {author} {\bibfnamefont
  {S.~L.}\ \bibnamefont {Yuan}},\ and\ \bibinfo {author} {\bibfnamefont
  {Z.~M.}\ \bibnamefont {Tian}},\ }\bibfield  {title} {\bibinfo {title}
  {{REZnAl$_{11}$O$_{19}$ (RE = Pr, Nd, Sm--Tb): a new family of ideal 2D
  triangular lattice frustrated magnets}},\ }\href
  {https://doi.org/10.1039/c9tc02643f} {\bibfield  {journal} {\bibinfo
  {journal} {J. Mater. Chem. C}\ }\textbf {\bibinfo {volume} {7}},\ \bibinfo
  {pages} {10073} (\bibinfo {year} {2019})}\BibitemShut {NoStop}%
\bibitem [{\citenamefont {Gao}\ \emph {et~al.}(2026)\citenamefont {Gao},
  \citenamefont {Chen}, \citenamefont {Liu}, \citenamefont {Klemm},
  \citenamefont {Zhang}, \citenamefont {Ma} \emph {et~al.}}]{Gao2026}%
  \BibitemOpen
  \bibfield  {author} {\bibinfo {author} {\bibfnamefont {B.}~\bibnamefont
  {Gao}}, \bibinfo {author} {\bibfnamefont {T.}~\bibnamefont {Chen}}, \bibinfo
  {author} {\bibfnamefont {C.}~\bibnamefont {Liu}}, \bibinfo {author}
  {\bibfnamefont {M.~L.}\ \bibnamefont {Klemm}}, \bibinfo {author}
  {\bibfnamefont {S.}~\bibnamefont {Zhang}}, \bibinfo {author} {\bibfnamefont
  {Z.}~\bibnamefont {Ma}}, \emph {et~al.},\ }\bibfield  {title} {\bibinfo
  {title} {{Spin excitation continuum from degenerate states in the mixed
  ferro-antiferromagnetic exchange system CeMgAl$_{11}$O$_{19}$}},\ }\href
  {https://doi.org/10.1126/sciadv.aed7778} {\bibfield  {journal} {\bibinfo
  {journal} {Sci. Adv.}\ }\textbf {\bibinfo {volume} {12}},\ \bibinfo {pages}
  {eaed7778} (\bibinfo {year} {2026})}\BibitemShut {NoStop}%
\bibitem [{\citenamefont {Cao}\ \emph {et~al.}(2025{\natexlab{a}})\citenamefont
  {Cao}, \citenamefont {Koda}, \citenamefont {Le}, \citenamefont {Pomjakushin},
  \citenamefont {Liu}, \citenamefont {Fu} \emph {et~al.}}]{Cao2025}%
  \BibitemOpen
  \bibfield  {author} {\bibinfo {author} {\bibfnamefont {Y.}~\bibnamefont
  {Cao}}, \bibinfo {author} {\bibfnamefont {A.}~\bibnamefont {Koda}}, \bibinfo
  {author} {\bibfnamefont {M.~D.}\ \bibnamefont {Le}}, \bibinfo {author}
  {\bibfnamefont {V.}~\bibnamefont {Pomjakushin}}, \bibinfo {author}
  {\bibfnamefont {B.}~\bibnamefont {Liu}}, \bibinfo {author} {\bibfnamefont
  {Z.}~\bibnamefont {Fu}}, \emph {et~al.},\ }\bibfield  {title} {\bibinfo
  {title} {{U(1) Dirac quantum spin liquid candidate in triangular-lattice
  antiferromagnet CeMgAl$_{11}$O$_{19}$}},\ }\href
  {https://doi.org/10.1007/s11433-024-2634-9} {\bibfield  {journal} {\bibinfo
  {journal} {Sci. China Phys. Mech. Astron.}\ }\textbf {\bibinfo {volume}
  {68}},\ \bibinfo {pages} {267011} (\bibinfo {year}
  {2025}{\natexlab{a}})}\BibitemShut {NoStop}%
\bibitem [{\citenamefont {Bastien}\ \emph {et~al.}(2026)\citenamefont
  {Bastien}, \citenamefont {Eli{\'a}{\v{s}}}, \citenamefont {Anderle},
  \citenamefont {Kancko}, \citenamefont {Corr{\^e}a}, \citenamefont {Kumar}
  \emph {et~al.}}]{colman}%
  \BibitemOpen
  \bibfield  {author} {\bibinfo {author} {\bibfnamefont {G.}~\bibnamefont
  {Bastien}}, \bibinfo {author} {\bibfnamefont {A.}~\bibnamefont
  {Eli{\'a}{\v{s}}}}, \bibinfo {author} {\bibfnamefont {V.}~\bibnamefont
  {Anderle}}, \bibinfo {author} {\bibfnamefont {A.}~\bibnamefont {Kancko}},
  \bibinfo {author} {\bibfnamefont {C.~A.}\ \bibnamefont {Corr{\^e}a}},
  \bibinfo {author} {\bibfnamefont {S.}~\bibnamefont {Kumar}}, \emph {et~al.},\
  }\bibfield  {title} {\bibinfo {title} {{Quantum disordered ground state and
  relative proximity to an exactly solvable model in the frustrated magnet
  CeMgAl$_{11}$O$_{19}$}},\ }\href {https://doi.org/10.1103/kqxq-g723}
  {\bibfield  {journal} {\bibinfo  {journal} {Phys. Rev. Res.}\ }\textbf
  {\bibinfo {volume} {8}},\ \bibinfo {pages} {L042007} (\bibinfo {year}
  {2026})}\BibitemShut {NoStop}%
\bibitem [{\citenamefont {Bippus}\ \emph {et~al.}(2025)\citenamefont {Bippus},
  \citenamefont {Schneider},\ and\ \citenamefont {Sbierski}}]{Bippus2025CMAO}%
  \BibitemOpen
  \bibfield  {author} {\bibinfo {author} {\bibfnamefont {F.}~\bibnamefont
  {Bippus}}, \bibinfo {author} {\bibfnamefont {B.}~\bibnamefont {Schneider}},\
  and\ \bibinfo {author} {\bibfnamefont {B.}~\bibnamefont {Sbierski}},\
  }\bibfield  {title} {\bibinfo {title} {{Pseudo-Majorana functional
  renormalization for frustrated XXZ spin-$1/2$ models with field or
  magnetization along the spin-$Z$ direction at finite temperature}},\ }\href
  {https://doi.org/10.1103/PhysRevB.111.054420} {\bibfield  {journal} {\bibinfo
   {journal} {Phys. Rev. B}\ }\textbf {\bibinfo {volume} {111}},\ \bibinfo
  {pages} {054420} (\bibinfo {year} {2025})}\BibitemShut {NoStop}%
\bibitem [{\citenamefont {Kumar}\ \emph
  {et~al.}(2025{\natexlab{a}})\citenamefont {Kumar}, \citenamefont {Bastien},
  \citenamefont {Savinov}, \citenamefont {Proschek}, \citenamefont
  {Eli{\'a}{\v{s}}}, \citenamefont {Za{\l}{\k{e}}ski}, \citenamefont
  {{\'S}liwi{\'n}ska-Bartkowiak}, \citenamefont {Colman},\ and\ \citenamefont
  {Kamba}}]{Kumar2025CMAOMagnetoelectric}%
  \BibitemOpen
  \bibfield  {author} {\bibinfo {author} {\bibfnamefont {S.}~\bibnamefont
  {Kumar}}, \bibinfo {author} {\bibfnamefont {G.}~\bibnamefont {Bastien}},
  \bibinfo {author} {\bibfnamefont {M.}~\bibnamefont {Savinov}}, \bibinfo
  {author} {\bibfnamefont {P.}~\bibnamefont {Proschek}}, \bibinfo {author}
  {\bibfnamefont {A.}~\bibnamefont {Eli{\'a}{\v{s}}}}, \bibinfo {author}
  {\bibfnamefont {K.}~\bibnamefont {Za{\l}{\k{e}}ski}}, \bibinfo {author}
  {\bibfnamefont {M.}~\bibnamefont {{\'S}liwi{\'n}ska-Bartkowiak}}, \bibinfo
  {author} {\bibfnamefont {R.~H.}\ \bibnamefont {Colman}},\ and\ \bibinfo
  {author} {\bibfnamefont {S.}~\bibnamefont {Kamba}},\ }\bibfield  {title}
  {\bibinfo {title} {{Crystal-Field--Driven Magnetoelectricity in the
  Triangular Quantum Magnet CeMgAl$_{11}$O$_{19}$}},\ }\href
  {https://doi.org/10.1103/c3hd-pwby} {\bibfield  {journal} {\bibinfo
  {journal} {Phys. Rev. B}\ }\textbf {\bibinfo {volume} {112}},\ \bibinfo
  {pages} {224431} (\bibinfo {year} {2025}{\natexlab{a}})}\BibitemShut
  {NoStop}%
\bibitem [{\citenamefont {Cao}\ \emph {et~al.}(2024)\citenamefont {Cao},
  \citenamefont {Bu}, \citenamefont {Fu}, \citenamefont {Zhao}, \citenamefont
  {Gardner}, \citenamefont {Ouyang} \emph {et~al.}}]{Cao2024}%
  \BibitemOpen
  \bibfield  {author} {\bibinfo {author} {\bibfnamefont {Y.}~\bibnamefont
  {Cao}}, \bibinfo {author} {\bibfnamefont {H.}~\bibnamefont {Bu}}, \bibinfo
  {author} {\bibfnamefont {Z.}~\bibnamefont {Fu}}, \bibinfo {author}
  {\bibfnamefont {J.}~\bibnamefont {Zhao}}, \bibinfo {author} {\bibfnamefont
  {J.~S.}\ \bibnamefont {Gardner}}, \bibinfo {author} {\bibfnamefont
  {Z.}~\bibnamefont {Ouyang}}, \emph {et~al.},\ }\bibfield  {title} {\bibinfo
  {title} {{Synthesis, disorder and Ising anisotropy in a new spin liquid
  candidate PrMgAl$_{11}$O$_{19}$}},\ }\href
  {https://doi.org/10.1088/2752-5724/ad4a93} {\bibfield  {journal} {\bibinfo
  {journal} {Mater. Futures}\ }\textbf {\bibinfo {volume} {3}},\ \bibinfo
  {pages} {035201} (\bibinfo {year} {2024})}\BibitemShut {NoStop}%
\bibitem [{\citenamefont {Ma}\ \emph {et~al.}(2024)\citenamefont {Ma},
  \citenamefont {Zheng}, \citenamefont {Chen}, \citenamefont {Xu},
  \citenamefont {Dong}, \citenamefont {Wang} \emph {et~al.}}]{Ma2024PrMgAl}%
  \BibitemOpen
  \bibfield  {author} {\bibinfo {author} {\bibfnamefont {Z.}~\bibnamefont
  {Ma}}, \bibinfo {author} {\bibfnamefont {S.}~\bibnamefont {Zheng}}, \bibinfo
  {author} {\bibfnamefont {Y.}~\bibnamefont {Chen}}, \bibinfo {author}
  {\bibfnamefont {R.}~\bibnamefont {Xu}}, \bibinfo {author} {\bibfnamefont
  {Z.-Y.}\ \bibnamefont {Dong}}, \bibinfo {author} {\bibfnamefont
  {J.}~\bibnamefont {Wang}}, \emph {et~al.},\ }\bibfield  {title} {\bibinfo
  {title} {{Possible gapless quantum spin liquid behavior in the
  triangular-lattice Ising antiferromagnet PrMgAl$_{11}$O$_{19}$}},\ }\href
  {https://doi.org/10.1103/physrevb.109.165143} {\bibfield  {journal} {\bibinfo
   {journal} {Phys. Rev. B}\ }\textbf {\bibinfo {volume} {109}},\ \bibinfo
  {pages} {165143} (\bibinfo {year} {2024})}\BibitemShut {NoStop}%
\bibitem [{\citenamefont {Li}\ \emph {et~al.}(2024{\natexlab{a}})\citenamefont
  {Li}, \citenamefont {Rutherford}, \citenamefont {Wang}, \citenamefont
  {Liang}, \citenamefont {Li}, \citenamefont {Zhang} \emph
  {et~al.}}]{Li2024PrMgAl}%
  \BibitemOpen
  \bibfield  {author} {\bibinfo {author} {\bibfnamefont {N.}~\bibnamefont
  {Li}}, \bibinfo {author} {\bibfnamefont {A.}~\bibnamefont {Rutherford}},
  \bibinfo {author} {\bibfnamefont {Y.~Y.}\ \bibnamefont {Wang}}, \bibinfo
  {author} {\bibfnamefont {H.}~\bibnamefont {Liang}}, \bibinfo {author}
  {\bibfnamefont {Q.~J.}\ \bibnamefont {Li}}, \bibinfo {author} {\bibfnamefont
  {Z.~J.}\ \bibnamefont {Zhang}}, \emph {et~al.},\ }\bibfield  {title}
  {\bibinfo {title} {{Ising-type quantum spin liquid state in
  PrMgAl$_{11}$O$_{19}$}},\ }\href
  {https://doi.org/10.1103/physrevb.110.134401} {\bibfield  {journal} {\bibinfo
   {journal} {Phys. Rev. B}\ }\textbf {\bibinfo {volume} {110}},\ \bibinfo
  {pages} {134401} (\bibinfo {year} {2024}{\natexlab{a}})}\BibitemShut
  {NoStop}%
\bibitem [{\citenamefont {Tu}\ \emph {et~al.}(2024)\citenamefont {Tu},
  \citenamefont {Ma}, \citenamefont {Wang}, \citenamefont {Jiao}, \citenamefont
  {Dai},\ and\ \citenamefont {Li}}]{Tu2024PrMgAl}%
  \BibitemOpen
  \bibfield  {author} {\bibinfo {author} {\bibfnamefont {C.~P.}\ \bibnamefont
  {Tu}}, \bibinfo {author} {\bibfnamefont {Z.}~\bibnamefont {Ma}}, \bibinfo
  {author} {\bibfnamefont {H.~R.}\ \bibnamefont {Wang}}, \bibinfo {author}
  {\bibfnamefont {Y.~H.}\ \bibnamefont {Jiao}}, \bibinfo {author}
  {\bibfnamefont {D.~Z.}\ \bibnamefont {Dai}},\ and\ \bibinfo {author}
  {\bibfnamefont {S.~Y.}\ \bibnamefont {Li}},\ }\bibfield  {title} {\bibinfo
  {title} {{Gapped quantum spin liquid in a triangular-lattice Ising-type
  antiferromagnet PrMgAl$_{11}$O$_{19}$}},\ }\href
  {https://doi.org/10.1103/PhysRevResearch.6.043147} {\bibfield  {journal}
  {\bibinfo  {journal} {Phys. Rev. Res.}\ }\textbf {\bibinfo {volume} {6}},\
  \bibinfo {pages} {043147} (\bibinfo {year} {2024})}\BibitemShut {NoStop}%
\bibitem [{\citenamefont {Kumar}\ \emph
  {et~al.}(2025{\natexlab{b}})\citenamefont {Kumar}, \citenamefont {Klicpera},
  \citenamefont {Eli{\'a}{\v{s}}}, \citenamefont {Kratochv{\'i}lov{\'a}},
  \citenamefont {Kancko}, \citenamefont {Correa} \emph
  {et~al.}}]{Kumar2025PrMgAl}%
  \BibitemOpen
  \bibfield  {author} {\bibinfo {author} {\bibfnamefont {S.}~\bibnamefont
  {Kumar}}, \bibinfo {author} {\bibfnamefont {M.}~\bibnamefont {Klicpera}},
  \bibinfo {author} {\bibfnamefont {A.}~\bibnamefont {Eli{\'a}{\v{s}}}},
  \bibinfo {author} {\bibfnamefont {M.}~\bibnamefont {Kratochv{\'i}lov{\'a}}},
  \bibinfo {author} {\bibfnamefont {A.}~\bibnamefont {Kancko}}, \bibinfo
  {author} {\bibfnamefont {C.}~\bibnamefont {Correa}}, \emph {et~al.},\
  }\bibfield  {title} {\bibinfo {title} {{Induced quantum magnetism on a
  triangular lattice of non-Kramers ions in PrMgAl$_{11}$O$_{19}$}},\ }\href
  {https://doi.org/10.1103/physrevb.111.174444} {\bibfield  {journal} {\bibinfo
   {journal} {Phys. Rev. B}\ }\textbf {\bibinfo {volume} {111}},\ \bibinfo
  {pages} {174444} (\bibinfo {year} {2025}{\natexlab{b}})}\BibitemShut
  {NoStop}%
\bibitem [{\citenamefont {Kumar}\ \emph
  {et~al.}(2026{\natexlab{a}})\citenamefont {Kumar}, \citenamefont {Bastien},
  \citenamefont {Colman}, \citenamefont {Savinov}, \citenamefont {Proschek},
  \citenamefont {Vali{\v{s}}ka}, \citenamefont {Kempi{\'n}ski}, \citenamefont
  {Kempi{\'n}ski}, \citenamefont {{\'S}liwi{\'n}ska-Bartkowiak},\ and\
  \citenamefont {Kamba}}]{Kumar2025PrMgAlME}%
  \BibitemOpen
  \bibfield  {author} {\bibinfo {author} {\bibfnamefont {S.}~\bibnamefont
  {Kumar}}, \bibinfo {author} {\bibfnamefont {G.}~\bibnamefont {Bastien}},
  \bibinfo {author} {\bibfnamefont {R.~H.}\ \bibnamefont {Colman}}, \bibinfo
  {author} {\bibfnamefont {M.}~\bibnamefont {Savinov}}, \bibinfo {author}
  {\bibfnamefont {P.}~\bibnamefont {Proschek}}, \bibinfo {author}
  {\bibfnamefont {M.}~\bibnamefont {Vali{\v{s}}ka}}, \bibinfo {author}
  {\bibfnamefont {M.}~\bibnamefont {Kempi{\'n}ski}}, \bibinfo {author}
  {\bibfnamefont {W.}~\bibnamefont {Kempi{\'n}ski}}, \bibinfo {author}
  {\bibfnamefont {M.}~\bibnamefont {{\'S}liwi{\'n}ska-Bartkowiak}},\ and\
  \bibinfo {author} {\bibfnamefont {S.}~\bibnamefont {Kamba}},\ }\bibfield
  {title} {\bibinfo {title} {{Crystal-field--driven magnetoelectric coupling in
  the non-Kramers hexaaluminate PrMgAl$_{11}$O$_{19}$}},\ }\href
  {https://doi.org/10.1016/j.jallcom.2026.187535} {\bibfield  {journal}
  {\bibinfo  {journal} {J. Alloys Compd.}\ }\textbf {\bibinfo {volume}
  {1061}},\ \bibinfo {pages} {187535} (\bibinfo {year}
  {2026}{\natexlab{a}})}\BibitemShut {NoStop}%
\bibitem [{\citenamefont {Bu}\ \emph {et~al.}(2022)\citenamefont {Bu},
  \citenamefont {Ashtar}, \citenamefont {Shiroka}, \citenamefont {Walker},
  \citenamefont {Fu}, \citenamefont {Zhao} \emph {et~al.}}]{Bu2022PrZnAl}%
  \BibitemOpen
  \bibfield  {author} {\bibinfo {author} {\bibfnamefont {H.}~\bibnamefont
  {Bu}}, \bibinfo {author} {\bibfnamefont {M.}~\bibnamefont {Ashtar}}, \bibinfo
  {author} {\bibfnamefont {T.}~\bibnamefont {Shiroka}}, \bibinfo {author}
  {\bibfnamefont {H.~C.}\ \bibnamefont {Walker}}, \bibinfo {author}
  {\bibfnamefont {Z.}~\bibnamefont {Fu}}, \bibinfo {author} {\bibfnamefont
  {J.}~\bibnamefont {Zhao}}, \emph {et~al.},\ }\bibfield  {title} {\bibinfo
  {title} {{Gapless triangular-lattice spin-liquid candidate
  PrZnAl$_{11}$O$_{19}$}},\ }\href
  {https://doi.org/10.1103/PhysRevB.106.134428} {\bibfield  {journal} {\bibinfo
   {journal} {Phys. Rev. B}\ }\textbf {\bibinfo {volume} {106}},\ \bibinfo
  {pages} {134428} (\bibinfo {year} {2022})}\BibitemShut {NoStop}%
\bibitem [{\citenamefont {Kumar}\ \emph
  {et~al.}(2026{\natexlab{b}})\citenamefont {Kumar}, \citenamefont {Bastien},
  \citenamefont {Prokle{\v{s}}ka}, \citenamefont {Kempi{\'n}ski}, \citenamefont
  {Kempi{\'n}ski}, \citenamefont {Za{\l}{\k{e}}ski}, \citenamefont {Kancko},
  \citenamefont {Corr{\^e}a}, \citenamefont {Treu}, \citenamefont {Gegenwart},
  \citenamefont {{\'S}liwi{\'n}ska-Bartkowiak},\ and\ \citenamefont
  {Colman}}]{Kumar2025NdMgAl}%
  \BibitemOpen
  \bibfield  {author} {\bibinfo {author} {\bibfnamefont {S.}~\bibnamefont
  {Kumar}}, \bibinfo {author} {\bibfnamefont {G.}~\bibnamefont {Bastien}},
  \bibinfo {author} {\bibfnamefont {J.}~\bibnamefont {Prokle{\v{s}}ka}},
  \bibinfo {author} {\bibfnamefont {M.}~\bibnamefont {Kempi{\'n}ski}}, \bibinfo
  {author} {\bibfnamefont {W.}~\bibnamefont {Kempi{\'n}ski}}, \bibinfo {author}
  {\bibfnamefont {K.}~\bibnamefont {Za{\l}{\k{e}}ski}}, \bibinfo {author}
  {\bibfnamefont {A.}~\bibnamefont {Kancko}}, \bibinfo {author} {\bibfnamefont
  {C.~A.}\ \bibnamefont {Corr{\^e}a}}, \bibinfo {author} {\bibfnamefont
  {T.}~\bibnamefont {Treu}}, \bibinfo {author} {\bibfnamefont {P.}~\bibnamefont
  {Gegenwart}}, \bibinfo {author} {\bibfnamefont {M.}~\bibnamefont
  {{\'S}liwi{\'n}ska-Bartkowiak}},\ and\ \bibinfo {author} {\bibfnamefont
  {R.~H.}\ \bibnamefont {Colman}},\ }\bibfield  {title} {\bibinfo {title}
  {{Absence of Long-Range Order and Magnetic Anisotropy in the Triangular
  Magnet NdMgAl$_{11}$O$_{19}$}},\ }\href {https://doi.org/10.1103/x1rf-xmrd}
  {\bibfield  {journal} {\bibinfo  {journal} {Phys. Rev. B}\ }\textbf {\bibinfo
  {volume} {114}},\ \bibinfo {pages} {134403} (\bibinfo {year}
  {2026}{\natexlab{b}})}\BibitemShut {NoStop}%
\bibitem [{\citenamefont {Cao}\ \emph {et~al.}(2026)\citenamefont {Cao},
  \citenamefont {Sun}, \citenamefont {Fu}, \citenamefont {Tian}, \citenamefont
  {Luo}, \citenamefont {Xiang}, \citenamefont {Sun}, \citenamefont {Zhao},\
  and\ \citenamefont {Guo}}]{Cao2026NdMgAlMCE}%
  \BibitemOpen
  \bibfield  {author} {\bibinfo {author} {\bibfnamefont {Y.}~\bibnamefont
  {Cao}}, \bibinfo {author} {\bibfnamefont {H.}~\bibnamefont {Sun}}, \bibinfo
  {author} {\bibfnamefont {Z.}~\bibnamefont {Fu}}, \bibinfo {author}
  {\bibfnamefont {Z.}~\bibnamefont {Tian}}, \bibinfo {author} {\bibfnamefont
  {H.}~\bibnamefont {Luo}}, \bibinfo {author} {\bibfnamefont {J.}~\bibnamefont
  {Xiang}}, \bibinfo {author} {\bibfnamefont {P.}~\bibnamefont {Sun}}, \bibinfo
  {author} {\bibfnamefont {J.}~\bibnamefont {Zhao}},\ and\ \bibinfo {author}
  {\bibfnamefont {H.}~\bibnamefont {Guo}},\ }\href
  {https://arxiv.org/abs/2607.12262} {\bibinfo {title} {{Giant magnetocaloric
  effect at low fields in triangular-lattice NdMgAl$_{11}$O$_{19}$}}} (\bibinfo
  {year} {2026}),\ \Eprint {https://arxiv.org/abs/2607.12262} {arXiv:2607.12262
  [cond-mat.str-el]} \BibitemShut {NoStop}%
\bibitem [{\citenamefont {Cao}\ \emph {et~al.}(2025{\natexlab{b}})\citenamefont
  {Cao}, \citenamefont {Bu}, \citenamefont {Shiroka}, \citenamefont {Walker},
  \citenamefont {Fu}, \citenamefont {Tian} \emph {et~al.}}]{Cao2025NdZnAl}%
  \BibitemOpen
  \bibfield  {author} {\bibinfo {author} {\bibfnamefont {Y.}~\bibnamefont
  {Cao}}, \bibinfo {author} {\bibfnamefont {H.}~\bibnamefont {Bu}}, \bibinfo
  {author} {\bibfnamefont {T.}~\bibnamefont {Shiroka}}, \bibinfo {author}
  {\bibfnamefont {H.~C.}\ \bibnamefont {Walker}}, \bibinfo {author}
  {\bibfnamefont {Z.}~\bibnamefont {Fu}}, \bibinfo {author} {\bibfnamefont
  {Z.}~\bibnamefont {Tian}}, \emph {et~al.},\ }\bibfield  {title} {\bibinfo
  {title} {{Magnetic ground state and persistent spin fluctuations in the
  triangular-lattice antiferromagnet NdZnAl$_{11}$O$_{19}$}},\ }\href
  {https://doi.org/10.1103/tnwb-9hv8} {\bibfield  {journal} {\bibinfo
  {journal} {Phys. Rev. B}\ }\textbf {\bibinfo {volume} {112}},\ \bibinfo
  {pages} {144409} (\bibinfo {year} {2025}{\natexlab{b}})}\BibitemShut
  {NoStop}%
\bibitem [{\citenamefont {Kumar}\ \emph
  {et~al.}(2026{\natexlab{c}})\citenamefont {Kumar}, \citenamefont
  {Salajov{\'a}}, \citenamefont {Kancko}, \citenamefont {Corr{\^e}a},
  \citenamefont {Halder},\ and\ \citenamefont {Colman}}]{Kumar2026SmMgAl}%
  \BibitemOpen
  \bibfield  {author} {\bibinfo {author} {\bibfnamefont {S.}~\bibnamefont
  {Kumar}}, \bibinfo {author} {\bibfnamefont {B.}~\bibnamefont {Salajov{\'a}}},
  \bibinfo {author} {\bibfnamefont {A.}~\bibnamefont {Kancko}}, \bibinfo
  {author} {\bibfnamefont {C.~A.}\ \bibnamefont {Corr{\^e}a}}, \bibinfo
  {author} {\bibfnamefont {S.}~\bibnamefont {Halder}},\ and\ \bibinfo {author}
  {\bibfnamefont {R.~H.}\ \bibnamefont {Colman}},\ }\bibfield  {title}
  {\bibinfo {title} {{Strongly Quenched Kramers Doublet Magnetism in
  SmMgAl$_{11}$O$_{19}$}},\ }\href {https://doi.org/10.1103/gt2r-3mn7}
  {\bibfield  {journal} {\bibinfo  {journal} {Phys. Rev. B}\ }\textbf {\bibinfo
  {volume} {113}},\ \bibinfo {pages} {224443} (\bibinfo {year}
  {2026}{\natexlab{c}})}\BibitemShut {NoStop}%
\bibitem [{\citenamefont {Broholm}\ \emph {et~al.}(1987)\citenamefont
  {Broholm}, \citenamefont {Kjems}, \citenamefont {Buyers}, \citenamefont
  {Matthews}, \citenamefont {Palstra}, \citenamefont {Menovsky},\ and\
  \citenamefont {Mydosh}}]{Broholm1987}%
  \BibitemOpen
  \bibfield  {author} {\bibinfo {author} {\bibfnamefont {C.}~\bibnamefont
  {Broholm}}, \bibinfo {author} {\bibfnamefont {J.~K.}\ \bibnamefont {Kjems}},
  \bibinfo {author} {\bibfnamefont {W.~J.~L.}\ \bibnamefont {Buyers}}, \bibinfo
  {author} {\bibfnamefont {P.}~\bibnamefont {Matthews}}, \bibinfo {author}
  {\bibfnamefont {T.~T.~M.}\ \bibnamefont {Palstra}}, \bibinfo {author}
  {\bibfnamefont {A.~A.}\ \bibnamefont {Menovsky}},\ and\ \bibinfo {author}
  {\bibfnamefont {J.~A.}\ \bibnamefont {Mydosh}},\ }\bibfield  {title}
  {\bibinfo {title} {{Magnetic excitations and ordering in the heavy-electron
  superconductor URu$_2$Si$_2$}},\ }\href
  {https://doi.org/10.1103/PhysRevLett.58.1467} {\bibfield  {journal} {\bibinfo
   {journal} {Phys. Rev. Lett.}\ }\textbf {\bibinfo {volume} {58}},\ \bibinfo
  {pages} {1467} (\bibinfo {year} {1987})}\BibitemShut {NoStop}%
\bibitem [{SM()}]{SM}%
  \BibitemOpen
  \href@noop {} {\bibinfo {title} {See supplemental material for experimental
  and numerical methods and supporting analyses, which includes
  refs.~\cite{Gleis2023CBE, Li2024CBE, Xi2024_1oT1, Li2011, Chen2017,
  Chen2018}}}\BibitemShut {NoStop}%
\bibitem [{\citenamefont {Santini}\ \emph {et~al.}(2009)\citenamefont
  {Santini}, \citenamefont {Carretta}, \citenamefont {Amoretti}, \citenamefont
  {Caciuffo}, \citenamefont {Magnani},\ and\ \citenamefont
  {Lander}}]{Santini2009}%
  \BibitemOpen
  \bibfield  {author} {\bibinfo {author} {\bibfnamefont {P.}~\bibnamefont
  {Santini}}, \bibinfo {author} {\bibfnamefont {S.}~\bibnamefont {Carretta}},
  \bibinfo {author} {\bibfnamefont {G.}~\bibnamefont {Amoretti}}, \bibinfo
  {author} {\bibfnamefont {R.}~\bibnamefont {Caciuffo}}, \bibinfo {author}
  {\bibfnamefont {N.}~\bibnamefont {Magnani}},\ and\ \bibinfo {author}
  {\bibfnamefont {G.~H.}\ \bibnamefont {Lander}},\ }\bibfield  {title}
  {\bibinfo {title} {{Multipolar interactions in $f$-electron systems: The
  paradigm of actinide dioxides}},\ }\href
  {https://doi.org/10.1103/revmodphys.81.807} {\bibfield  {journal} {\bibinfo
  {journal} {Rev. Mod. Phys.}\ }\textbf {\bibinfo {volume} {81}},\ \bibinfo
  {pages} {807} (\bibinfo {year} {2009})}\BibitemShut {NoStop}%
\bibitem [{\citenamefont {Kuramoto}\ \emph {et~al.}(2009)\citenamefont
  {Kuramoto}, \citenamefont {Kusunose},\ and\ \citenamefont
  {Kiss}}]{Kuramoto2009}%
  \BibitemOpen
  \bibfield  {author} {\bibinfo {author} {\bibfnamefont {Y.}~\bibnamefont
  {Kuramoto}}, \bibinfo {author} {\bibfnamefont {H.}~\bibnamefont {Kusunose}},\
  and\ \bibinfo {author} {\bibfnamefont {A.}~\bibnamefont {Kiss}},\ }\bibfield
  {title} {\bibinfo {title} {{Multipole Orders and Fluctuations in Strongly
  Correlated Electron Systems}},\ }\href
  {https://doi.org/10.1143/jpsj.78.072001} {\bibfield  {journal} {\bibinfo
  {journal} {J. Phys. Soc. Jpn.}\ }\textbf {\bibinfo {volume} {78}},\ \bibinfo
  {pages} {072001} (\bibinfo {year} {2009})}\BibitemShut {NoStop}%
\bibitem [{\citenamefont {Pourovskii}\ \emph {et~al.}(2025)\citenamefont
  {Pourovskii}, \citenamefont {Fiore~Mosca}, \citenamefont {Celiberti},
  \citenamefont {Khmelevskyi}, \citenamefont {Paramekanti},\ and\ \citenamefont
  {Franchini}}]{Pourovskii2025}%
  \BibitemOpen
  \bibfield  {author} {\bibinfo {author} {\bibfnamefont {L.~V.}\ \bibnamefont
  {Pourovskii}}, \bibinfo {author} {\bibfnamefont {D.}~\bibnamefont
  {Fiore~Mosca}}, \bibinfo {author} {\bibfnamefont {L.}~\bibnamefont
  {Celiberti}}, \bibinfo {author} {\bibfnamefont {S.}~\bibnamefont
  {Khmelevskyi}}, \bibinfo {author} {\bibfnamefont {A.}~\bibnamefont
  {Paramekanti}},\ and\ \bibinfo {author} {\bibfnamefont {C.}~\bibnamefont
  {Franchini}},\ }\bibfield  {title} {\bibinfo {title} {{Hidden orders in
  spin-orbit entangled correlated insulators}},\ }\href
  {https://doi.org/10.1038/s41578-025-00824-z} {\bibfield  {journal} {\bibinfo
  {journal} {Nat. Rev. Mater.}\ }\textbf {\bibinfo {volume} {10}},\ \bibinfo
  {pages} {674} (\bibinfo {year} {2025})}\BibitemShut {NoStop}%
\bibitem [{\citenamefont {Rau}\ and\ \citenamefont {Gingras}(2019)}]{Rau2019}%
  \BibitemOpen
  \bibfield  {author} {\bibinfo {author} {\bibfnamefont {J.~G.}\ \bibnamefont
  {Rau}}\ and\ \bibinfo {author} {\bibfnamefont {M.~J.~P.}\ \bibnamefont
  {Gingras}},\ }\bibfield  {title} {\bibinfo {title} {{Frustrated quantum
  rare-earth pyrochlores}},\ }\href
  {https://doi.org/10.1146/annurev-conmatphys-022317-110520} {\bibfield
  {journal} {\bibinfo  {journal} {Annu. Rev. Condens. Matter Phys.}\ }\textbf
  {\bibinfo {volume} {10}},\ \bibinfo {pages} {357} (\bibinfo {year}
  {2019})}\BibitemShut {NoStop}%
\bibitem [{\citenamefont {Li}\ \emph {et~al.}(2016)\citenamefont {Li},
  \citenamefont {Wang},\ and\ \citenamefont {Chen}}]{Li2016HiddenMultipolar}%
  \BibitemOpen
  \bibfield  {author} {\bibinfo {author} {\bibfnamefont {Y.-D.}\ \bibnamefont
  {Li}}, \bibinfo {author} {\bibfnamefont {X.}~\bibnamefont {Wang}},\ and\
  \bibinfo {author} {\bibfnamefont {G.}~\bibnamefont {Chen}},\ }\bibfield
  {title} {\bibinfo {title} {{Hidden multipolar orders of dipole-octupole
  doublets on a triangular lattice}},\ }\href
  {https://doi.org/10.1103/PhysRevB.94.201114} {\bibfield  {journal} {\bibinfo
  {journal} {Phys. Rev. B}\ }\textbf {\bibinfo {volume} {94}},\ \bibinfo
  {pages} {201114} (\bibinfo {year} {2016})}\BibitemShut {NoStop}%
\bibitem [{\citenamefont {Liu}\ \emph {et~al.}(2018)\citenamefont {Liu},
  \citenamefont {Li},\ and\ \citenamefont {Chen}}]{Liu2018NonKramers}%
  \BibitemOpen
  \bibfield  {author} {\bibinfo {author} {\bibfnamefont {C.}~\bibnamefont
  {Liu}}, \bibinfo {author} {\bibfnamefont {Y.-D.}\ \bibnamefont {Li}},\ and\
  \bibinfo {author} {\bibfnamefont {G.}~\bibnamefont {Chen}},\ }\bibfield
  {title} {\bibinfo {title} {{Selective measurements of intertwined multipolar
  orders: Non-Kramers doublets on a triangular lattice}},\ }\href
  {https://doi.org/10.1103/PhysRevB.98.045119} {\bibfield  {journal} {\bibinfo
  {journal} {Phys. Rev. B}\ }\textbf {\bibinfo {volume} {98}},\ \bibinfo
  {pages} {045119} (\bibinfo {year} {2018})}\BibitemShut {NoStop}%
\bibitem [{\citenamefont {Sibille}\ \emph {et~al.}(2020)\citenamefont
  {Sibille}, \citenamefont {Gauthier}, \citenamefont {Lhotel}, \citenamefont
  {Por{\'e}e}, \citenamefont {Pomjakushin}, \citenamefont {Ewings} \emph
  {et~al.}}]{Sibille2020Octupole}%
  \BibitemOpen
  \bibfield  {author} {\bibinfo {author} {\bibfnamefont {R.}~\bibnamefont
  {Sibille}}, \bibinfo {author} {\bibfnamefont {N.}~\bibnamefont {Gauthier}},
  \bibinfo {author} {\bibfnamefont {E.}~\bibnamefont {Lhotel}}, \bibinfo
  {author} {\bibfnamefont {V.}~\bibnamefont {Por{\'e}e}}, \bibinfo {author}
  {\bibfnamefont {V.}~\bibnamefont {Pomjakushin}}, \bibinfo {author}
  {\bibfnamefont {R.~A.}\ \bibnamefont {Ewings}}, \emph {et~al.},\ }\bibfield
  {title} {\bibinfo {title} {{A quantum liquid of magnetic octupoles on the
  pyrochlore lattice}},\ }\href {https://doi.org/10.1038/s41567-020-0827-7}
  {\bibfield  {journal} {\bibinfo  {journal} {Nat. Phys.}\ }\textbf {\bibinfo
  {volume} {16}},\ \bibinfo {pages} {546} (\bibinfo {year} {2020})}\BibitemShut
  {NoStop}%
\bibitem [{\citenamefont {Tang}\ \emph {et~al.}(2023)\citenamefont {Tang},
  \citenamefont {Gritsenko}, \citenamefont {Kimura}, \citenamefont
  {Bhattacharjee}, \citenamefont {Sakai}, \citenamefont {Fu} \emph
  {et~al.}}]{Tang2023Pr2Zr2O7}%
  \BibitemOpen
  \bibfield  {author} {\bibinfo {author} {\bibfnamefont {N.}~\bibnamefont
  {Tang}}, \bibinfo {author} {\bibfnamefont {Y.}~\bibnamefont {Gritsenko}},
  \bibinfo {author} {\bibfnamefont {K.}~\bibnamefont {Kimura}}, \bibinfo
  {author} {\bibfnamefont {S.}~\bibnamefont {Bhattacharjee}}, \bibinfo {author}
  {\bibfnamefont {A.}~\bibnamefont {Sakai}}, \bibinfo {author} {\bibfnamefont
  {M.}~\bibnamefont {Fu}}, \emph {et~al.},\ }\bibfield  {title} {\bibinfo
  {title} {{Spin--orbital liquid state and liquid--gas metamagnetic transition
  on a pyrochlore lattice}},\ }\href
  {https://doi.org/10.1038/s41567-022-01816-4} {\bibfield  {journal} {\bibinfo
  {journal} {Nat. Phys.}\ }\textbf {\bibinfo {volume} {19}},\ \bibinfo {pages}
  {92} (\bibinfo {year} {2023})}\BibitemShut {NoStop}%
\bibitem [{\citenamefont {Gao}\ \emph {et~al.}(2025{\natexlab{a}})\citenamefont
  {Gao}, \citenamefont {Desrochers}, \citenamefont {Tam}, \citenamefont
  {Kirschbaum}, \citenamefont {Steffens}, \citenamefont {Hiess} \emph
  {et~al.}}]{Gao2025QSI}%
  \BibitemOpen
  \bibfield  {author} {\bibinfo {author} {\bibfnamefont {B.}~\bibnamefont
  {Gao}}, \bibinfo {author} {\bibfnamefont {F.}~\bibnamefont {Desrochers}},
  \bibinfo {author} {\bibfnamefont {D.~W.}\ \bibnamefont {Tam}}, \bibinfo
  {author} {\bibfnamefont {D.~M.}\ \bibnamefont {Kirschbaum}}, \bibinfo
  {author} {\bibfnamefont {P.}~\bibnamefont {Steffens}}, \bibinfo {author}
  {\bibfnamefont {A.}~\bibnamefont {Hiess}}, \emph {et~al.},\ }\bibfield
  {title} {\bibinfo {title} {{Neutron scattering and thermodynamic evidence for
  emergent photons and fractionalization in a pyrochlore spin ice}},\ }\href
  {https://doi.org/10.1038/s41567-025-02922-9} {\bibfield  {journal} {\bibinfo
  {journal} {Nat. Phys.}\ }\textbf {\bibinfo {volume} {21}},\ \bibinfo {pages}
  {1203} (\bibinfo {year} {2025}{\natexlab{a}})}\BibitemShut {NoStop}%
\bibitem [{\citenamefont {Shen}\ \emph {et~al.}(2019)\citenamefont {Shen},
  \citenamefont {Liu}, \citenamefont {Qin}, \citenamefont {Shen}, \citenamefont
  {Li}, \citenamefont {Bewley} \emph {et~al.}}]{Shen2019}%
  \BibitemOpen
  \bibfield  {author} {\bibinfo {author} {\bibfnamefont {Y.}~\bibnamefont
  {Shen}}, \bibinfo {author} {\bibfnamefont {C.}~\bibnamefont {Liu}}, \bibinfo
  {author} {\bibfnamefont {Y.}~\bibnamefont {Qin}}, \bibinfo {author}
  {\bibfnamefont {S.}~\bibnamefont {Shen}}, \bibinfo {author} {\bibfnamefont
  {Y.-D.}\ \bibnamefont {Li}}, \bibinfo {author} {\bibfnamefont
  {R.}~\bibnamefont {Bewley}}, \emph {et~al.},\ }\bibfield  {title} {\bibinfo
  {title} {{Intertwined dipolar and multipolar order in the triangular-lattice
  magnet TmMgGaO$_4$}},\ }\href {https://doi.org/10.1038/s41467-019-12410-3}
  {\bibfield  {journal} {\bibinfo  {journal} {Nat. Commun.}\ }\textbf {\bibinfo
  {volume} {10}},\ \bibinfo {pages} {4530} (\bibinfo {year}
  {2019})}\BibitemShut {NoStop}%
\bibitem [{\citenamefont {Li}\ \emph {et~al.}(2020)\citenamefont {Li},
  \citenamefont {Liao}, \citenamefont {Chen}, \citenamefont {Zeng},
  \citenamefont {Sheng}, \citenamefont {Qi}, \citenamefont {Meng},\ and\
  \citenamefont {Li}}]{Li2020TmMgGaO}%
  \BibitemOpen
  \bibfield  {author} {\bibinfo {author} {\bibfnamefont {H.}~\bibnamefont
  {Li}}, \bibinfo {author} {\bibfnamefont {Y.~D.}\ \bibnamefont {Liao}},
  \bibinfo {author} {\bibfnamefont {B.-B.}\ \bibnamefont {Chen}}, \bibinfo
  {author} {\bibfnamefont {X.-T.}\ \bibnamefont {Zeng}}, \bibinfo {author}
  {\bibfnamefont {X.-L.}\ \bibnamefont {Sheng}}, \bibinfo {author}
  {\bibfnamefont {Y.}~\bibnamefont {Qi}}, \bibinfo {author} {\bibfnamefont
  {Z.~Y.}\ \bibnamefont {Meng}},\ and\ \bibinfo {author} {\bibfnamefont
  {W.}~\bibnamefont {Li}},\ }\bibfield  {title} {\bibinfo {title}
  {{Kosterlitz-Thouless melting of magnetic order in the triangular quantum
  Ising material TmMgGaO$_4$}},\ }\href
  {https://doi.org/10.1038/s41467-020-14907-8} {\bibfield  {journal} {\bibinfo
  {journal} {Nat. Commun.}\ }\textbf {\bibinfo {volume} {11}},\ \bibinfo
  {pages} {1111} (\bibinfo {year} {2020})}\BibitemShut {NoStop}%
\bibitem [{\citenamefont {Hu}\ \emph {et~al.}(2020)\citenamefont {Hu},
  \citenamefont {Ma}, \citenamefont {Liao}, \citenamefont {Li}, \citenamefont
  {Ma}, \citenamefont {Cui} \emph {et~al.}}]{Hu2020BKT}%
  \BibitemOpen
  \bibfield  {author} {\bibinfo {author} {\bibfnamefont {Z.}~\bibnamefont
  {Hu}}, \bibinfo {author} {\bibfnamefont {Z.}~\bibnamefont {Ma}}, \bibinfo
  {author} {\bibfnamefont {Y.-D.}\ \bibnamefont {Liao}}, \bibinfo {author}
  {\bibfnamefont {H.}~\bibnamefont {Li}}, \bibinfo {author} {\bibfnamefont
  {C.}~\bibnamefont {Ma}}, \bibinfo {author} {\bibfnamefont {Y.}~\bibnamefont
  {Cui}}, \emph {et~al.},\ }\bibfield  {title} {\bibinfo {title} {{Evidence of
  the Berezinskii-Kosterlitz-Thouless phase in a frustrated magnet}},\ }\href
  {https://doi.org/10.1038/s41467-020-19380-x} {\bibfield  {journal} {\bibinfo
  {journal} {Nat. Commun.}\ }\textbf {\bibinfo {volume} {11}},\ \bibinfo
  {pages} {5631} (\bibinfo {year} {2020})}\BibitemShut {NoStop}%
\bibitem [{\citenamefont {Nakatsuji}\ \emph {et~al.}(2005)\citenamefont
  {Nakatsuji}, \citenamefont {Nambu}, \citenamefont {Tonomura}, \citenamefont
  {Sakai}, \citenamefont {Jonas}, \citenamefont {Broholm}, \citenamefont
  {Tsunetsugu}, \citenamefont {Qiu},\ and\ \citenamefont
  {Maeno}}]{Nakatsuji2005}%
  \BibitemOpen
  \bibfield  {author} {\bibinfo {author} {\bibfnamefont {S.}~\bibnamefont
  {Nakatsuji}}, \bibinfo {author} {\bibfnamefont {Y.}~\bibnamefont {Nambu}},
  \bibinfo {author} {\bibfnamefont {H.}~\bibnamefont {Tonomura}}, \bibinfo
  {author} {\bibfnamefont {O.}~\bibnamefont {Sakai}}, \bibinfo {author}
  {\bibfnamefont {S.}~\bibnamefont {Jonas}}, \bibinfo {author} {\bibfnamefont
  {C.}~\bibnamefont {Broholm}}, \bibinfo {author} {\bibfnamefont
  {H.}~\bibnamefont {Tsunetsugu}}, \bibinfo {author} {\bibfnamefont
  {Y.}~\bibnamefont {Qiu}},\ and\ \bibinfo {author} {\bibfnamefont
  {Y.}~\bibnamefont {Maeno}},\ }\bibfield  {title} {\bibinfo {title} {{Spin
  disorder on a triangular lattice}},\ }\href
  {https://doi.org/10.1126/science.1114727} {\bibfield  {journal} {\bibinfo
  {journal} {Science}\ }\textbf {\bibinfo {volume} {309}},\ \bibinfo {pages}
  {1697} (\bibinfo {year} {2005})}\BibitemShut {NoStop}%
\bibitem [{\citenamefont {Sheng}\ \emph {et~al.}(2025)\citenamefont {Sheng},
  \citenamefont {Mei}, \citenamefont {Wang}, \citenamefont {Xu}, \citenamefont
  {Jiang}, \citenamefont {Xu} \emph {et~al.}}]{Sheng2025}%
  \BibitemOpen
  \bibfield  {author} {\bibinfo {author} {\bibfnamefont {J.}~\bibnamefont
  {Sheng}}, \bibinfo {author} {\bibfnamefont {J.-W.}\ \bibnamefont {Mei}},
  \bibinfo {author} {\bibfnamefont {L.}~\bibnamefont {Wang}}, \bibinfo {author}
  {\bibfnamefont {X.}~\bibnamefont {Xu}}, \bibinfo {author} {\bibfnamefont
  {W.}~\bibnamefont {Jiang}}, \bibinfo {author} {\bibfnamefont
  {L.}~\bibnamefont {Xu}}, \emph {et~al.},\ }\bibfield  {title} {\bibinfo
  {title} {{Bose--Einstein condensation of a two-magnon bound state in a spin-1
  triangular lattice}},\ }\href {https://doi.org/10.1038/s41563-024-02071-z}
  {\bibfield  {journal} {\bibinfo  {journal} {Nat. Mater.}\ }\textbf {\bibinfo
  {volume} {24}},\ \bibinfo {pages} {544} (\bibinfo {year} {2025})}\BibitemShut
  {NoStop}%
\bibitem [{\citenamefont {Shan}\ \emph {et~al.}(2021)\citenamefont {Shan},
  \citenamefont {Jin}, \citenamefont {Datta},\ and\ \citenamefont
  {Yao}}]{Shan2021}%
  \BibitemOpen
  \bibfield  {author} {\bibinfo {author} {\bibfnamefont {C.}~\bibnamefont
  {Shan}}, \bibinfo {author} {\bibfnamefont {S.}~\bibnamefont {Jin}}, \bibinfo
  {author} {\bibfnamefont {T.}~\bibnamefont {Datta}},\ and\ \bibinfo {author}
  {\bibfnamefont {D.-X.}\ \bibnamefont {Yao}},\ }\bibfield  {title} {\bibinfo
  {title} {{Torque equilibrium spin wave theory of Raman scattering in an
  anisotropic triangular lattice antiferromagnet with Dzyaloshinskii-Moriya
  interaction}},\ }\href {https://doi.org/10.1103/physrevb.103.024417}
  {\bibfield  {journal} {\bibinfo  {journal} {Phys. Rev. B}\ }\textbf {\bibinfo
  {volume} {103}},\ \bibinfo {pages} {024417} (\bibinfo {year}
  {2021})}\BibitemShut {NoStop}%
\bibitem [{\citenamefont {El~Hog}\ \emph {et~al.}(2022)\citenamefont {El~Hog},
  \citenamefont {Sharafullin}, \citenamefont {Diep}, \citenamefont {Garbouj},
  \citenamefont {Debbichi},\ and\ \citenamefont {Said}}]{Hog2022}%
  \BibitemOpen
  \bibfield  {author} {\bibinfo {author} {\bibfnamefont {S.}~\bibnamefont
  {El~Hog}}, \bibinfo {author} {\bibfnamefont {I.~F.}\ \bibnamefont
  {Sharafullin}}, \bibinfo {author} {\bibfnamefont {H.}~\bibnamefont {Diep}},
  \bibinfo {author} {\bibfnamefont {H.}~\bibnamefont {Garbouj}}, \bibinfo
  {author} {\bibfnamefont {M.}~\bibnamefont {Debbichi}},\ and\ \bibinfo
  {author} {\bibfnamefont {M.}~\bibnamefont {Said}},\ }\bibfield  {title}
  {\bibinfo {title} {{Frustrated antiferromagnetic triangular lattice with
  Dzyaloshinskii-Moriya interaction: Ground states, spin waves, skyrmion
  crystal, phase transition}},\ }\href
  {https://doi.org/10.1016/j.jmmm.2022.169920} {\bibfield  {journal} {\bibinfo
  {journal} {J. Magn. Magn. Mater.}\ }\textbf {\bibinfo {volume} {563}},\
  \bibinfo {pages} {169920} (\bibinfo {year} {2022})}\BibitemShut {NoStop}%
\bibitem [{\citenamefont {White}(1992)}]{White1992}%
  \BibitemOpen
  \bibfield  {author} {\bibinfo {author} {\bibfnamefont {S.~R.}\ \bibnamefont
  {White}},\ }\bibfield  {title} {\bibinfo {title} {{Density matrix formulation
  for quantum renormalization groups}},\ }\href
  {https://doi.org/10.1103/physrevlett.69.2863} {\bibfield  {journal} {\bibinfo
   {journal} {Phys. Rev. Lett.}\ }\textbf {\bibinfo {volume} {69}},\ \bibinfo
  {pages} {2863} (\bibinfo {year} {1992})}\BibitemShut {NoStop}%
\bibitem [{\citenamefont {Schollw{\"o}ck}(2011)}]{Schollwock2011}%
  \BibitemOpen
  \bibfield  {author} {\bibinfo {author} {\bibfnamefont {U.}~\bibnamefont
  {Schollw{\"o}ck}},\ }\bibfield  {title} {\bibinfo {title} {{The
  density-matrix renormalization group in the age of matrix product states}},\
  }\href {https://doi.org/10.1016/j.aop.2010.09.012} {\bibfield  {journal}
  {\bibinfo  {journal} {Ann. Phys.}\ }\textbf {\bibinfo {volume} {326}},\
  \bibinfo {pages} {96} (\bibinfo {year} {2011})}\BibitemShut {NoStop}%
\bibitem [{\citenamefont {Haegeman}\ \emph {et~al.}(2011)\citenamefont
  {Haegeman}, \citenamefont {Cirac}, \citenamefont {Osborne}, \citenamefont
  {Pi{\v{z}}orn}, \citenamefont {Verschelde},\ and\ \citenamefont
  {Verstraete}}]{TDVP2011}%
  \BibitemOpen
  \bibfield  {author} {\bibinfo {author} {\bibfnamefont {J.}~\bibnamefont
  {Haegeman}}, \bibinfo {author} {\bibfnamefont {J.~I.}\ \bibnamefont {Cirac}},
  \bibinfo {author} {\bibfnamefont {T.~J.}\ \bibnamefont {Osborne}}, \bibinfo
  {author} {\bibfnamefont {I.}~\bibnamefont {Pi{\v{z}}orn}}, \bibinfo {author}
  {\bibfnamefont {H.}~\bibnamefont {Verschelde}},\ and\ \bibinfo {author}
  {\bibfnamefont {F.}~\bibnamefont {Verstraete}},\ }\bibfield  {title}
  {\bibinfo {title} {{Time-Dependent Variational Principle for Quantum
  Lattices}},\ }\href {https://doi.org/10.1103/PhysRevLett.107.070601}
  {\bibfield  {journal} {\bibinfo  {journal} {Phys. Rev. Lett.}\ }\textbf
  {\bibinfo {volume} {107}},\ \bibinfo {pages} {070601} (\bibinfo {year}
  {2011})}\BibitemShut {NoStop}%
\bibitem [{\citenamefont {Haegeman}\ \emph {et~al.}(2016)\citenamefont
  {Haegeman}, \citenamefont {Lubich}, \citenamefont {Oseledets}, \citenamefont
  {Vandereycken},\ and\ \citenamefont {Verstraete}}]{Haegeman2016}%
  \BibitemOpen
  \bibfield  {author} {\bibinfo {author} {\bibfnamefont {J.}~\bibnamefont
  {Haegeman}}, \bibinfo {author} {\bibfnamefont {C.}~\bibnamefont {Lubich}},
  \bibinfo {author} {\bibfnamefont {I.}~\bibnamefont {Oseledets}}, \bibinfo
  {author} {\bibfnamefont {B.}~\bibnamefont {Vandereycken}},\ and\ \bibinfo
  {author} {\bibfnamefont {F.}~\bibnamefont {Verstraete}},\ }\bibfield  {title}
  {\bibinfo {title} {{Unifying time evolution and optimization with matrix
  product states}},\ }\href {https://doi.org/10.1103/physrevb.94.165116}
  {\bibfield  {journal} {\bibinfo  {journal} {Phys. Rev. B}\ }\textbf {\bibinfo
  {volume} {94}},\ \bibinfo {pages} {165116} (\bibinfo {year}
  {2016})}\BibitemShut {NoStop}%
\bibitem [{\citenamefont {Li}\ \emph {et~al.}(2023)\citenamefont {Li},
  \citenamefont {Gao}, \citenamefont {He}, \citenamefont {Qi}, \citenamefont
  {Chen},\ and\ \citenamefont {Li}}]{Li2023}%
  \BibitemOpen
  \bibfield  {author} {\bibinfo {author} {\bibfnamefont {Q.}~\bibnamefont
  {Li}}, \bibinfo {author} {\bibfnamefont {Y.}~\bibnamefont {Gao}}, \bibinfo
  {author} {\bibfnamefont {Y.-Y.}\ \bibnamefont {He}}, \bibinfo {author}
  {\bibfnamefont {Y.}~\bibnamefont {Qi}}, \bibinfo {author} {\bibfnamefont
  {B.-B.}\ \bibnamefont {Chen}},\ and\ \bibinfo {author} {\bibfnamefont
  {W.}~\bibnamefont {Li}},\ }\bibfield  {title} {\bibinfo {title} {{Tangent
  Space Approach for Thermal Tensor Network Simulations of the 2D Hubbard
  Model}},\ }\href {https://doi.org/10.1103/physrevlett.130.226502} {\bibfield
  {journal} {\bibinfo  {journal} {Phys. Rev. Lett.}\ }\textbf {\bibinfo
  {volume} {130}},\ \bibinfo {pages} {226502} (\bibinfo {year}
  {2023})}\BibitemShut {NoStop}%
\bibitem [{\citenamefont {Miyashita}(1986)}]{Miyashita1986}%
  \BibitemOpen
  \bibfield  {author} {\bibinfo {author} {\bibfnamefont {S.}~\bibnamefont
  {Miyashita}},\ }\bibfield  {title} {\bibinfo {title} {{The Ground State and
  Thermodynamic Properties of Generalized Heisenberg Models on the Triangular
  Lattice}},\ }\href {https://doi.org/10.1143/ptps.87.112} {\bibfield
  {journal} {\bibinfo  {journal} {Prog. Theor. Phys. Suppl.}\ }\textbf
  {\bibinfo {volume} {87}},\ \bibinfo {pages} {112} (\bibinfo {year}
  {1986})}\BibitemShut {NoStop}%
\bibitem [{\citenamefont {Momoi}\ and\ \citenamefont
  {Suzuki}(1992)}]{Momoi1992}%
  \BibitemOpen
  \bibfield  {author} {\bibinfo {author} {\bibfnamefont {T.}~\bibnamefont
  {Momoi}}\ and\ \bibinfo {author} {\bibfnamefont {M.}~\bibnamefont {Suzuki}},\
  }\bibfield  {title} {\bibinfo {title} {{Ground-state properties and phase
  diagram of the quantum $XXZ$ antiferromagnet on a triangular lattice}},\
  }\href {https://doi.org/10.1143/jpsj.61.3732} {\bibfield  {journal} {\bibinfo
   {journal} {J. Phys. Soc. Jpn.}\ }\textbf {\bibinfo {volume} {61}},\ \bibinfo
  {pages} {3732} (\bibinfo {year} {1992})}\BibitemShut {NoStop}%
\bibitem [{\citenamefont {Pal}\ \emph {et~al.}(2021)\citenamefont {Pal},
  \citenamefont {Sharma}, \citenamefont {Changlani},\ and\ \citenamefont
  {Pujari}}]{Pal2021}%
  \BibitemOpen
  \bibfield  {author} {\bibinfo {author} {\bibfnamefont {S.}~\bibnamefont
  {Pal}}, \bibinfo {author} {\bibfnamefont {P.}~\bibnamefont {Sharma}},
  \bibinfo {author} {\bibfnamefont {H.~J.}\ \bibnamefont {Changlani}},\ and\
  \bibinfo {author} {\bibfnamefont {S.}~\bibnamefont {Pujari}},\ }\bibfield
  {title} {\bibinfo {title} {{Colorful points in the XY regime of XXZ quantum
  magnets}},\ }\href {https://doi.org/10.1103/physrevb.103.144414} {\bibfield
  {journal} {\bibinfo  {journal} {Phys. Rev. B}\ }\textbf {\bibinfo {volume}
  {103}},\ \bibinfo {pages} {144414} (\bibinfo {year} {2021})}\BibitemShut
  {NoStop}%
\bibitem [{\citenamefont {Li}\ \emph {et~al.}(2017)\citenamefont {Li},
  \citenamefont {Adroja}, \citenamefont {Bewley}, \citenamefont {Voneshen},
  \citenamefont {Tsirlin}, \citenamefont {Gegenwart},\ and\ \citenamefont
  {Zhang}}]{Li2017}%
  \BibitemOpen
  \bibfield  {author} {\bibinfo {author} {\bibfnamefont {Y.}~\bibnamefont
  {Li}}, \bibinfo {author} {\bibfnamefont {D.}~\bibnamefont {Adroja}}, \bibinfo
  {author} {\bibfnamefont {R.~I.}\ \bibnamefont {Bewley}}, \bibinfo {author}
  {\bibfnamefont {D.}~\bibnamefont {Voneshen}}, \bibinfo {author}
  {\bibfnamefont {A.~A.}\ \bibnamefont {Tsirlin}}, \bibinfo {author}
  {\bibfnamefont {P.}~\bibnamefont {Gegenwart}},\ and\ \bibinfo {author}
  {\bibfnamefont {Q.}~\bibnamefont {Zhang}},\ }\bibfield  {title} {\bibinfo
  {title} {{Crystalline electric-field randomness in the triangular-lattice
  spin-liquid YbMgGaO$_4$}},\ }\href
  {https://doi.org/10.1103/physrevlett.118.107202} {\bibfield  {journal}
  {\bibinfo  {journal} {Phys. Rev. Lett.}\ }\textbf {\bibinfo {volume} {118}},\
  \bibinfo {pages} {107202} (\bibinfo {year} {2017})}\BibitemShut {NoStop}%
\bibitem [{\citenamefont {Ma}\ \emph {et~al.}(2021)\citenamefont {Ma},
  \citenamefont {Dong}, \citenamefont {Wang}, \citenamefont {Zheng},
  \citenamefont {Ran}, \citenamefont {Bao} \emph {et~al.}}]{Ma2021}%
  \BibitemOpen
  \bibfield  {author} {\bibinfo {author} {\bibfnamefont {Z.}~\bibnamefont
  {Ma}}, \bibinfo {author} {\bibfnamefont {Z.-Y.}\ \bibnamefont {Dong}},
  \bibinfo {author} {\bibfnamefont {J.}~\bibnamefont {Wang}}, \bibinfo {author}
  {\bibfnamefont {S.}~\bibnamefont {Zheng}}, \bibinfo {author} {\bibfnamefont
  {K.}~\bibnamefont {Ran}}, \bibinfo {author} {\bibfnamefont {S.}~\bibnamefont
  {Bao}}, \emph {et~al.},\ }\bibfield  {title} {\bibinfo {title}
  {{Disorder-induced broadening of the spin waves in the triangular-lattice
  quantum spin liquid candidate YbZnGaO$_4$}},\ }\href
  {https://doi.org/10.1103/physrevb.104.224433} {\bibfield  {journal} {\bibinfo
   {journal} {Phys. Rev. B}\ }\textbf {\bibinfo {volume} {104}},\ \bibinfo
  {pages} {224433} (\bibinfo {year} {2021})}\BibitemShut {NoStop}%
\bibitem [{\citenamefont {Palstra}\ \emph {et~al.}(1985)\citenamefont
  {Palstra}, \citenamefont {Menovsky}, \citenamefont {van~den Berg},
  \citenamefont {Dirkmaat}, \citenamefont {Kes}, \citenamefont {Nieuwenhuys},\
  and\ \citenamefont {Mydosh}}]{Palstra1985}%
  \BibitemOpen
  \bibfield  {author} {\bibinfo {author} {\bibfnamefont {T.~T.~M.}\
  \bibnamefont {Palstra}}, \bibinfo {author} {\bibfnamefont {A.~A.}\
  \bibnamefont {Menovsky}}, \bibinfo {author} {\bibfnamefont {J.}~\bibnamefont
  {van~den Berg}}, \bibinfo {author} {\bibfnamefont {A.~J.}\ \bibnamefont
  {Dirkmaat}}, \bibinfo {author} {\bibfnamefont {P.~H.}\ \bibnamefont {Kes}},
  \bibinfo {author} {\bibfnamefont {G.~J.}\ \bibnamefont {Nieuwenhuys}},\ and\
  \bibinfo {author} {\bibfnamefont {J.~A.}\ \bibnamefont {Mydosh}},\ }\bibfield
   {title} {\bibinfo {title} {{Superconducting and magnetic transitions in the
  heavy-fermion system URu$_2$Si$_2$}},\ }\href
  {https://doi.org/10.1103/PhysRevLett.55.2727} {\bibfield  {journal} {\bibinfo
   {journal} {Phys. Rev. Lett.}\ }\textbf {\bibinfo {volume} {55}},\ \bibinfo
  {pages} {2727} (\bibinfo {year} {1985})}\BibitemShut {NoStop}%
\bibitem [{\citenamefont {Niklowitz}\ \emph {et~al.}(2010)\citenamefont
  {Niklowitz}, \citenamefont {Pfleiderer}, \citenamefont {Keller},
  \citenamefont {Vojta}, \citenamefont {Huang},\ and\ \citenamefont
  {Mydosh}}]{Niklowitz2010}%
  \BibitemOpen
  \bibfield  {author} {\bibinfo {author} {\bibfnamefont {P.~G.}\ \bibnamefont
  {Niklowitz}}, \bibinfo {author} {\bibfnamefont {C.}~\bibnamefont
  {Pfleiderer}}, \bibinfo {author} {\bibfnamefont {T.}~\bibnamefont {Keller}},
  \bibinfo {author} {\bibfnamefont {M.}~\bibnamefont {Vojta}}, \bibinfo
  {author} {\bibfnamefont {Y.-K.}\ \bibnamefont {Huang}},\ and\ \bibinfo
  {author} {\bibfnamefont {J.~A.}\ \bibnamefont {Mydosh}},\ }\bibfield  {title}
  {\bibinfo {title} {{Parasitic small-moment antiferromagnetism and nonlinear
  coupling of hidden order and antiferromagnetism in URu$_2$Si$_2$ observed by
  Larmor diffraction}},\ }\href
  {https://doi.org/10.1103/PhysRevLett.104.106406} {\bibfield  {journal}
  {\bibinfo  {journal} {Phys. Rev. Lett.}\ }\textbf {\bibinfo {volume} {104}},\
  \bibinfo {pages} {106406} (\bibinfo {year} {2010})}\BibitemShut {NoStop}%
\bibitem [{\citenamefont {Cricchio}\ \emph {et~al.}(2009)\citenamefont
  {Cricchio}, \citenamefont {Bultmark}, \citenamefont {Gr{\aa}n{\"a}s},\ and\
  \citenamefont {Nordstr{\"o}m}}]{Cricchio2009}%
  \BibitemOpen
  \bibfield  {author} {\bibinfo {author} {\bibfnamefont {F.}~\bibnamefont
  {Cricchio}}, \bibinfo {author} {\bibfnamefont {F.}~\bibnamefont {Bultmark}},
  \bibinfo {author} {\bibfnamefont {O.}~\bibnamefont {Gr{\aa}n{\"a}s}},\ and\
  \bibinfo {author} {\bibfnamefont {L.}~\bibnamefont {Nordstr{\"o}m}},\
  }\bibfield  {title} {\bibinfo {title} {{Itinerant magnetic multipole moments
  of rank five as the hidden order in URu$_2$Si$_2$}},\ }\href
  {https://doi.org/10.1103/PhysRevLett.103.107202} {\bibfield  {journal}
  {\bibinfo  {journal} {Phys. Rev. Lett.}\ }\textbf {\bibinfo {volume} {103}},\
  \bibinfo {pages} {107202} (\bibinfo {year} {2009})}\BibitemShut {NoStop}%
\bibitem [{\citenamefont {Ikeda}\ \emph {et~al.}(2012)\citenamefont {Ikeda},
  \citenamefont {Suzuki}, \citenamefont {Arita}, \citenamefont {Takimoto},
  \citenamefont {Shibauchi},\ and\ \citenamefont {Matsuda}}]{Ikeda2012}%
  \BibitemOpen
  \bibfield  {author} {\bibinfo {author} {\bibfnamefont {H.}~\bibnamefont
  {Ikeda}}, \bibinfo {author} {\bibfnamefont {M.-T.}\ \bibnamefont {Suzuki}},
  \bibinfo {author} {\bibfnamefont {R.}~\bibnamefont {Arita}}, \bibinfo
  {author} {\bibfnamefont {T.}~\bibnamefont {Takimoto}}, \bibinfo {author}
  {\bibfnamefont {T.}~\bibnamefont {Shibauchi}},\ and\ \bibinfo {author}
  {\bibfnamefont {Y.}~\bibnamefont {Matsuda}},\ }\bibfield  {title} {\bibinfo
  {title} {{Emergent rank-5 nematic order in URu$_2$Si$_2$}},\ }\href
  {https://doi.org/10.1038/nphys2330} {\bibfield  {journal} {\bibinfo
  {journal} {Nat. Phys.}\ }\textbf {\bibinfo {volume} {8}},\ \bibinfo {pages}
  {528} (\bibinfo {year} {2012})}\BibitemShut {NoStop}%
\bibitem [{\citenamefont {Ressouche}\ \emph {et~al.}(2012)\citenamefont
  {Ressouche}, \citenamefont {Ballou}, \citenamefont {Bourdarot}, \citenamefont
  {Aoki}, \citenamefont {Simonet}, \citenamefont {Fernandez-Diaz},
  \citenamefont {Stunault},\ and\ \citenamefont {Flouquet}}]{Ressouche2012}%
  \BibitemOpen
  \bibfield  {author} {\bibinfo {author} {\bibfnamefont {E.}~\bibnamefont
  {Ressouche}}, \bibinfo {author} {\bibfnamefont {R.}~\bibnamefont {Ballou}},
  \bibinfo {author} {\bibfnamefont {F.}~\bibnamefont {Bourdarot}}, \bibinfo
  {author} {\bibfnamefont {D.}~\bibnamefont {Aoki}}, \bibinfo {author}
  {\bibfnamefont {V.}~\bibnamefont {Simonet}}, \bibinfo {author} {\bibfnamefont
  {M.~T.}\ \bibnamefont {Fernandez-Diaz}}, \bibinfo {author} {\bibfnamefont
  {A.}~\bibnamefont {Stunault}},\ and\ \bibinfo {author} {\bibfnamefont
  {J.}~\bibnamefont {Flouquet}},\ }\bibfield  {title} {\bibinfo {title}
  {{Hidden order in URu$_2$Si$_2$ unveiled}},\ }\href
  {https://doi.org/10.1103/PhysRevLett.109.067202} {\bibfield  {journal}
  {\bibinfo  {journal} {Phys. Rev. Lett.}\ }\textbf {\bibinfo {volume} {109}},\
  \bibinfo {pages} {067202} (\bibinfo {year} {2012})}\BibitemShut {NoStop}%
\bibitem [{\citenamefont {Ferrari}\ and\ \citenamefont
  {Becca}(2019)}]{Ferrari2019}%
  \BibitemOpen
  \bibfield  {author} {\bibinfo {author} {\bibfnamefont {F.}~\bibnamefont
  {Ferrari}}\ and\ \bibinfo {author} {\bibfnamefont {F.}~\bibnamefont
  {Becca}},\ }\bibfield  {title} {\bibinfo {title} {{Dynamical structure factor
  of the $J_1-J_2$ Heisenberg model on the triangular lattice: Magnons,
  spinons, and gauge fields}},\ }\href
  {https://doi.org/10.1103/physrevx.9.031026} {\bibfield  {journal} {\bibinfo
  {journal} {Phys. Rev. X}\ }\textbf {\bibinfo {volume} {9}},\ \bibinfo {pages}
  {031026} (\bibinfo {year} {2019})}\BibitemShut {NoStop}%
\bibitem [{\citenamefont {L{\"u}thi}\ \emph {et~al.}(1984)\citenamefont
  {L{\"u}thi}, \citenamefont {Blumenr{\"o}der}, \citenamefont {Hillebrands},
  \citenamefont {Zirngiebl}, \citenamefont {G{\"u}ntherodt},\ and\
  \citenamefont {Winzer}}]{Luthi1984CeB6}%
  \BibitemOpen
  \bibfield  {author} {\bibinfo {author} {\bibfnamefont {B.}~\bibnamefont
  {L{\"u}thi}}, \bibinfo {author} {\bibfnamefont {S.}~\bibnamefont
  {Blumenr{\"o}der}}, \bibinfo {author} {\bibfnamefont {B.}~\bibnamefont
  {Hillebrands}}, \bibinfo {author} {\bibfnamefont {E.}~\bibnamefont
  {Zirngiebl}}, \bibinfo {author} {\bibfnamefont {G.}~\bibnamefont
  {G{\"u}ntherodt}},\ and\ \bibinfo {author} {\bibfnamefont {K.}~\bibnamefont
  {Winzer}},\ }\bibfield  {title} {\bibinfo {title} {{Elastic and
  magnetoelastic effects in CeB$_6$}},\ }\href@noop {} {\bibfield  {journal}
  {\bibinfo  {journal} {Z. Phys. B: Condens. Matter}\ }\textbf {\bibinfo
  {volume} {58}},\ \bibinfo {pages} {31} (\bibinfo {year} {1984})}\BibitemShut
  {NoStop}%
\bibitem [{\citenamefont {Patri}\ \emph {et~al.}(2019)\citenamefont {Patri},
  \citenamefont {Sakai}, \citenamefont {Lee}, \citenamefont {Paramekanti},
  \citenamefont {Nakatsuji},\ and\ \citenamefont {Kim}}]{Patri2019}%
  \BibitemOpen
  \bibfield  {author} {\bibinfo {author} {\bibfnamefont {A.~S.}\ \bibnamefont
  {Patri}}, \bibinfo {author} {\bibfnamefont {A.}~\bibnamefont {Sakai}},
  \bibinfo {author} {\bibfnamefont {S.}~\bibnamefont {Lee}}, \bibinfo {author}
  {\bibfnamefont {A.}~\bibnamefont {Paramekanti}}, \bibinfo {author}
  {\bibfnamefont {S.}~\bibnamefont {Nakatsuji}},\ and\ \bibinfo {author}
  {\bibfnamefont {Y.}~\bibnamefont {Kim}},\ }\bibfield  {title} {\bibinfo
  {title} {{Unveiling hidden multipolar orders with magnetostriction}},\ }\href
  {https://doi.org/10.1038/s41467-019-11913-3} {\bibfield  {journal} {\bibinfo
  {journal} {Nat. Commun.}\ }\textbf {\bibinfo {volume} {10}},\ \bibinfo
  {pages} {4092} (\bibinfo {year} {2019})}\BibitemShut {NoStop}%
\bibitem [{\citenamefont {Scheie}(2021)}]{Scheie2021PyCrystalField}%
  \BibitemOpen
  \bibfield  {author} {\bibinfo {author} {\bibfnamefont {A.}~\bibnamefont
  {Scheie}},\ }\bibfield  {title} {\bibinfo {title} {{PyCrystalField: software
  for calculation, analysis and fitting of crystal electric field
  Hamiltonians}},\ }\href {https://doi.org/10.1107/S160057672001554X}
  {\bibfield  {journal} {\bibinfo  {journal} {J. Appl. Cryst.}\ }\textbf
  {\bibinfo {volume} {54}},\ \bibinfo {pages} {356} (\bibinfo {year}
  {2021})}\BibitemShut {NoStop}%
\bibitem [{\citenamefont {Uchida}\ \emph {et~al.}(2010)\citenamefont {Uchida},
  \citenamefont {Xiao}, \citenamefont {Adachi}, \citenamefont {Ohe},
  \citenamefont {Takahashi}, \citenamefont {Ieda}, \citenamefont {Ota},
  \citenamefont {Kajiwara}, \citenamefont {Umezawa}, \citenamefont {Kawai},
  \citenamefont {Bauer}, \citenamefont {Maekawa},\ and\ \citenamefont
  {Saitoh}}]{Uchida2010Insulator}%
  \BibitemOpen
  \bibfield  {author} {\bibinfo {author} {\bibfnamefont {K.}~\bibnamefont
  {Uchida}}, \bibinfo {author} {\bibfnamefont {J.}~\bibnamefont {Xiao}},
  \bibinfo {author} {\bibfnamefont {H.}~\bibnamefont {Adachi}}, \bibinfo
  {author} {\bibfnamefont {J.}~\bibnamefont {Ohe}}, \bibinfo {author}
  {\bibfnamefont {S.}~\bibnamefont {Takahashi}}, \bibinfo {author}
  {\bibfnamefont {J.}~\bibnamefont {Ieda}}, \bibinfo {author} {\bibfnamefont
  {T.}~\bibnamefont {Ota}}, \bibinfo {author} {\bibfnamefont {Y.}~\bibnamefont
  {Kajiwara}}, \bibinfo {author} {\bibfnamefont {H.}~\bibnamefont {Umezawa}},
  \bibinfo {author} {\bibfnamefont {H.}~\bibnamefont {Kawai}}, \bibinfo
  {author} {\bibfnamefont {G.~E.~W.}\ \bibnamefont {Bauer}}, \bibinfo {author}
  {\bibfnamefont {S.}~\bibnamefont {Maekawa}},\ and\ \bibinfo {author}
  {\bibfnamefont {E.}~\bibnamefont {Saitoh}},\ }\bibfield  {title} {\bibinfo
  {title} {Spin {Seebeck} insulator},\ }\href
  {https://doi.org/10.1038/nmat2856} {\bibfield  {journal} {\bibinfo  {journal}
  {Nat. Mater.}\ }\textbf {\bibinfo {volume} {9}},\ \bibinfo {pages} {894}
  (\bibinfo {year} {2010})}\BibitemShut {NoStop}%
\bibitem [{\citenamefont {Xiao}\ \emph {et~al.}(2010)\citenamefont {Xiao},
  \citenamefont {Bauer}, \citenamefont {Uchida}, \citenamefont {Saitoh},\ and\
  \citenamefont {Maekawa}}]{Xiao2010PRB}%
  \BibitemOpen
  \bibfield  {author} {\bibinfo {author} {\bibfnamefont {J.}~\bibnamefont
  {Xiao}}, \bibinfo {author} {\bibfnamefont {G.~E.~W.}\ \bibnamefont {Bauer}},
  \bibinfo {author} {\bibfnamefont {K.-c.}\ \bibnamefont {Uchida}}, \bibinfo
  {author} {\bibfnamefont {E.}~\bibnamefont {Saitoh}},\ and\ \bibinfo {author}
  {\bibfnamefont {S.}~\bibnamefont {Maekawa}},\ }\bibfield  {title} {\bibinfo
  {title} {Theory of magnon-driven spin {Seebeck} effect},\ }\href
  {https://doi.org/10.1103/PhysRevB.81.214418} {\bibfield  {journal} {\bibinfo
  {journal} {Phys. Rev. B}\ }\textbf {\bibinfo {volume} {81}},\ \bibinfo
  {pages} {214418} (\bibinfo {year} {2010})}\BibitemShut {NoStop}%
\bibitem [{\citenamefont {Adachi}\ \emph {et~al.}(2011)\citenamefont {Adachi},
  \citenamefont {Ohe}, \citenamefont {Takahashi},\ and\ \citenamefont
  {Maekawa}}]{Adachi2011}%
  \BibitemOpen
  \bibfield  {author} {\bibinfo {author} {\bibfnamefont {H.}~\bibnamefont
  {Adachi}}, \bibinfo {author} {\bibfnamefont {J.-i.}\ \bibnamefont {Ohe}},
  \bibinfo {author} {\bibfnamefont {S.}~\bibnamefont {Takahashi}},\ and\
  \bibinfo {author} {\bibfnamefont {S.}~\bibnamefont {Maekawa}},\ }\bibfield
  {title} {\bibinfo {title} {Linear-response theory of spin {Seebeck} effect in
  ferromagnetic insulators},\ }\href
  {https://doi.org/10.1103/PhysRevB.83.094410} {\bibfield  {journal} {\bibinfo
  {journal} {Phys. Rev. B}\ }\textbf {\bibinfo {volume} {83}},\ \bibinfo
  {pages} {094410} (\bibinfo {year} {2011})}\BibitemShut {NoStop}%
\bibitem [{\citenamefont {Bauer}\ \emph {et~al.}(2012)\citenamefont {Bauer},
  \citenamefont {Saitoh},\ and\ \citenamefont {van
  Wees}}]{Bauer2012SpinCaloritronics}%
  \BibitemOpen
  \bibfield  {author} {\bibinfo {author} {\bibfnamefont {G.~E.~W.}\
  \bibnamefont {Bauer}}, \bibinfo {author} {\bibfnamefont {E.}~\bibnamefont
  {Saitoh}},\ and\ \bibinfo {author} {\bibfnamefont {B.~J.}\ \bibnamefont {van
  Wees}},\ }\bibfield  {title} {\bibinfo {title} {Spin caloritronics},\ }\href
  {https://doi.org/10.1038/nmat3301} {\bibfield  {journal} {\bibinfo  {journal}
  {Nat. Mater.}\ }\textbf {\bibinfo {volume} {11}},\ \bibinfo {pages} {391}
  (\bibinfo {year} {2012})}\BibitemShut {NoStop}%
\bibitem [{\citenamefont {Masuda}\ and\ \citenamefont
  {Sato}(2024)}]{Masuda2024}%
  \BibitemOpen
  \bibfield  {author} {\bibinfo {author} {\bibfnamefont {K.}~\bibnamefont
  {Masuda}}\ and\ \bibinfo {author} {\bibfnamefont {M.}~\bibnamefont {Sato}},\
  }\bibfield  {title} {\bibinfo {title} {Microscopic theory of spin {Seebeck}
  effect in antiferromagnets},\ }\href {https://doi.org/10.7566/JPSJ.93.034702}
  {\bibfield  {journal} {\bibinfo  {journal} {J. Phys. Soc. Jpn.}\ }\textbf
  {\bibinfo {volume} {93}},\ \bibinfo {pages} {034702} (\bibinfo {year}
  {2024})}\BibitemShut {NoStop}%
\bibitem [{\citenamefont {Tang}\ \emph {et~al.}(2026)\citenamefont {Tang},
  \citenamefont {Willsher}, \citenamefont {Glamsch}, \citenamefont {Aqeel},
  \citenamefont {Scheuchenpflug}, \citenamefont {Schulze}, \citenamefont
  {Liebald}, \citenamefont {Rytz}, \citenamefont {Guguschev}, \citenamefont
  {Albrecht}, \citenamefont {Moessner},\ and\ \citenamefont
  {Gegenwart}}]{Tang2026SpinIceSSE}%
  \BibitemOpen
  \bibfield  {author} {\bibinfo {author} {\bibfnamefont {N.}~\bibnamefont
  {Tang}}, \bibinfo {author} {\bibfnamefont {J.}~\bibnamefont {Willsher}},
  \bibinfo {author} {\bibfnamefont {S.}~\bibnamefont {Glamsch}}, \bibinfo
  {author} {\bibfnamefont {A.}~\bibnamefont {Aqeel}}, \bibinfo {author}
  {\bibfnamefont {L.}~\bibnamefont {Scheuchenpflug}}, \bibinfo {author}
  {\bibfnamefont {M.}~\bibnamefont {Schulze}}, \bibinfo {author} {\bibfnamefont
  {C.}~\bibnamefont {Liebald}}, \bibinfo {author} {\bibfnamefont
  {D.}~\bibnamefont {Rytz}}, \bibinfo {author} {\bibfnamefont {C.}~\bibnamefont
  {Guguschev}}, \bibinfo {author} {\bibfnamefont {M.}~\bibnamefont {Albrecht}},
  \bibinfo {author} {\bibfnamefont {R.}~\bibnamefont {Moessner}},\ and\
  \bibinfo {author} {\bibfnamefont {P.}~\bibnamefont {Gegenwart}},\ }\href
  {https://arxiv.org/abs/2606.02501} {\bibinfo {title} {Electrical observation
  via spin {Seebeck} effect of fractionalized excitations in a magnetic
  insulator}} (\bibinfo {year} {2026}),\ \Eprint
  {https://arxiv.org/abs/2606.02501} {arXiv:2606.02501 [cond-mat.str-el]}
  \BibitemShut {NoStop}%
\bibitem [{\citenamefont {Gao}\ \emph {et~al.}(2025{\natexlab{b}})\citenamefont
  {Gao}, \citenamefont {Huang}, \citenamefont {Maekawa},\ and\ \citenamefont
  {Li}}]{Gao2025SpinSeebeck}%
  \BibitemOpen
  \bibfield  {author} {\bibinfo {author} {\bibfnamefont {Y.}~\bibnamefont
  {Gao}}, \bibinfo {author} {\bibfnamefont {Y.}~\bibnamefont {Huang}}, \bibinfo
  {author} {\bibfnamefont {S.}~\bibnamefont {Maekawa}},\ and\ \bibinfo {author}
  {\bibfnamefont {W.}~\bibnamefont {Li}},\ }\bibfield  {title} {\bibinfo
  {title} {Spin seebeck effect of triangular lattice spin supersolid},\ }\href
  {https://doi.org/10.1103/t25p-x319} {\bibfield  {journal} {\bibinfo
  {journal} {Phys. Rev. Lett.}\ }\textbf {\bibinfo {volume} {135}},\ \bibinfo
  {pages} {236504} (\bibinfo {year} {2025}{\natexlab{b}})}\BibitemShut
  {NoStop}%
\bibitem [{\citenamefont {Takei}\ and\ \citenamefont
  {Tserkovnyak}(2014)}]{Takei2014_ST}%
  \BibitemOpen
  \bibfield  {author} {\bibinfo {author} {\bibfnamefont {S.}~\bibnamefont
  {Takei}}\ and\ \bibinfo {author} {\bibfnamefont {Y.}~\bibnamefont
  {Tserkovnyak}},\ }\bibfield  {title} {\bibinfo {title} {Superfluid spin
  transport through easy-plane ferromagnetic insulators},\ }\href
  {https://doi.org/10.1103/PhysRevLett.112.227201} {\bibfield  {journal}
  {\bibinfo  {journal} {Phys. Rev. Lett.}\ }\textbf {\bibinfo {volume} {112}},\
  \bibinfo {pages} {227201} (\bibinfo {year} {2014})}\BibitemShut {NoStop}%
\bibitem [{\citenamefont {Qaiumzadeh}\ \emph {et~al.}(2017)\citenamefont
  {Qaiumzadeh}, \citenamefont {Skarsv{\aa}g}, \citenamefont {Holmqvist},\ and\
  \citenamefont {Brataas}}]{Qaiumzadeh2017_ST}%
  \BibitemOpen
  \bibfield  {author} {\bibinfo {author} {\bibfnamefont {A.}~\bibnamefont
  {Qaiumzadeh}}, \bibinfo {author} {\bibfnamefont {H.}~\bibnamefont
  {Skarsv{\aa}g}}, \bibinfo {author} {\bibfnamefont {C.}~\bibnamefont
  {Holmqvist}},\ and\ \bibinfo {author} {\bibfnamefont {A.}~\bibnamefont
  {Brataas}},\ }\bibfield  {title} {\bibinfo {title} {Spin superfluidity in
  biaxial antiferromagnetic insulators},\ }\href
  {https://doi.org/10.1103/PhysRevLett.118.137201} {\bibfield  {journal}
  {\bibinfo  {journal} {Phys. Rev. Lett.}\ }\textbf {\bibinfo {volume} {118}},\
  \bibinfo {pages} {137201} (\bibinfo {year} {2017})}\BibitemShut {NoStop}%
\bibitem [{\citenamefont {Gleis}\ \emph {et~al.}(2023)\citenamefont {Gleis},
  \citenamefont {Li},\ and\ \citenamefont {von Delft}}]{Gleis2023CBE}%
  \BibitemOpen
  \bibfield  {author} {\bibinfo {author} {\bibfnamefont {A.}~\bibnamefont
  {Gleis}}, \bibinfo {author} {\bibfnamefont {J.-W.}\ \bibnamefont {Li}},\ and\
  \bibinfo {author} {\bibfnamefont {J.}~\bibnamefont {von Delft}},\ }\bibfield
  {title} {\bibinfo {title} {{Controlled Bond Expansion for Density Matrix
  Renormalization Group Ground State Search at Single-Site Costs}},\ }\href
  {https://doi.org/10.1103/PhysRevLett.130.246402} {\bibfield  {journal}
  {\bibinfo  {journal} {Phys. Rev. Lett.}\ }\textbf {\bibinfo {volume} {130}},\
  \bibinfo {pages} {246402} (\bibinfo {year} {2023})}\BibitemShut {NoStop}%
\bibitem [{\citenamefont {Li}\ \emph {et~al.}(2024{\natexlab{b}})\citenamefont
  {Li}, \citenamefont {Gleis},\ and\ \citenamefont {von Delft}}]{Li2024CBE}%
  \BibitemOpen
  \bibfield  {author} {\bibinfo {author} {\bibfnamefont {J.-W.}\ \bibnamefont
  {Li}}, \bibinfo {author} {\bibfnamefont {A.}~\bibnamefont {Gleis}},\ and\
  \bibinfo {author} {\bibfnamefont {J.}~\bibnamefont {von Delft}},\ }\bibfield
  {title} {\bibinfo {title} {{Time-Dependent Variational Principle with
  Controlled Bond Expansion for Matrix Product States}},\ }\href
  {https://doi.org/10.1103/PhysRevLett.133.026401} {\bibfield  {journal}
  {\bibinfo  {journal} {Phys. Rev. Lett.}\ }\textbf {\bibinfo {volume} {133}},\
  \bibinfo {pages} {026401} (\bibinfo {year} {2024}{\natexlab{b}})}\BibitemShut
  {NoStop}%
\bibitem [{\citenamefont {Xi}\ \emph {et~al.}(2026)\citenamefont {Xi},
  \citenamefont {Gao}, \citenamefont {Lin}, \citenamefont {Li}, \citenamefont
  {Li}, \citenamefont {Lv}, \citenamefont {Liang}, \citenamefont {Wang},
  \citenamefont {Yu}, \citenamefont {Wang},\ and\ \citenamefont
  {Li}}]{Xi2024_1oT1}%
  \BibitemOpen
  \bibfield  {author} {\bibinfo {author} {\bibfnamefont {N.}~\bibnamefont
  {Xi}}, \bibinfo {author} {\bibfnamefont {Y.}~\bibnamefont {Gao}}, \bibinfo
  {author} {\bibfnamefont {H.}~\bibnamefont {Lin}}, \bibinfo {author}
  {\bibfnamefont {Q.}~\bibnamefont {Li}}, \bibinfo {author} {\bibfnamefont
  {C.}~\bibnamefont {Li}}, \bibinfo {author} {\bibfnamefont {E.}~\bibnamefont
  {Lv}}, \bibinfo {author} {\bibfnamefont {S.}~\bibnamefont {Liang}}, \bibinfo
  {author} {\bibfnamefont {L.}~\bibnamefont {Wang}}, \bibinfo {author}
  {\bibfnamefont {R.}~\bibnamefont {Yu}}, \bibinfo {author} {\bibfnamefont
  {X.}~\bibnamefont {Wang}},\ and\ \bibinfo {author} {\bibfnamefont
  {W.}~\bibnamefont {Li}},\ }\bibfield  {title} {\bibinfo {title} {{Thermal
  tensor network approach for spin-lattice relaxation in quantum magnets}},\
  }\href {https://doi.org/10.1103/q6dt-kzgc} {\bibfield  {journal} {\bibinfo
  {journal} {Phys. Rev. B}\ }\textbf {\bibinfo {volume} {113}},\ \bibinfo
  {pages} {165105} (\bibinfo {year} {2026})}\BibitemShut {NoStop}%
\bibitem [{\citenamefont {Li}\ \emph {et~al.}(2011)\citenamefont {Li},
  \citenamefont {Ran}, \citenamefont {Gong}, \citenamefont {Zhao},
  \citenamefont {Xi}, \citenamefont {Ye},\ and\ \citenamefont {Su}}]{Li2011}%
  \BibitemOpen
  \bibfield  {author} {\bibinfo {author} {\bibfnamefont {W.}~\bibnamefont
  {Li}}, \bibinfo {author} {\bibfnamefont {S.-J.}\ \bibnamefont {Ran}},
  \bibinfo {author} {\bibfnamefont {S.-S.}\ \bibnamefont {Gong}}, \bibinfo
  {author} {\bibfnamefont {Y.}~\bibnamefont {Zhao}}, \bibinfo {author}
  {\bibfnamefont {B.}~\bibnamefont {Xi}}, \bibinfo {author} {\bibfnamefont
  {F.}~\bibnamefont {Ye}},\ and\ \bibinfo {author} {\bibfnamefont
  {G.}~\bibnamefont {Su}},\ }\bibfield  {title} {\bibinfo {title} {Linearized
  tensor renormalization group algorithm for the calculation of thermodynamic
  properties of quantum lattice models},\ }\href
  {https://doi.org/10.1103/PhysRevLett.106.127202} {\bibfield  {journal}
  {\bibinfo  {journal} {Phys. Rev. Lett.}\ }\textbf {\bibinfo {volume} {106}},\
  \bibinfo {pages} {127202} (\bibinfo {year} {2011})}\BibitemShut {NoStop}%
\bibitem [{\citenamefont {Chen}\ \emph {et~al.}(2017)\citenamefont {Chen},
  \citenamefont {Liu}, \citenamefont {Chen},\ and\ \citenamefont
  {Li}}]{Chen2017}%
  \BibitemOpen
  \bibfield  {author} {\bibinfo {author} {\bibfnamefont {B.-B.}\ \bibnamefont
  {Chen}}, \bibinfo {author} {\bibfnamefont {Y.-J.}\ \bibnamefont {Liu}},
  \bibinfo {author} {\bibfnamefont {Z.}~\bibnamefont {Chen}},\ and\ \bibinfo
  {author} {\bibfnamefont {W.}~\bibnamefont {Li}},\ }\bibfield  {title}
  {\bibinfo {title} {{Series-expansion thermal tensor network approach for
  quantum lattice models}},\ }\href
  {https://doi.org/10.1103/PhysRevB.95.161104} {\bibfield  {journal} {\bibinfo
  {journal} {Phys. Rev. B}\ }\textbf {\bibinfo {volume} {95}},\ \bibinfo
  {pages} {161104(R)} (\bibinfo {year} {2017})}\BibitemShut {NoStop}%
\bibitem [{\citenamefont {Chen}\ \emph {et~al.}(2018)\citenamefont {Chen},
  \citenamefont {Chen}, \citenamefont {Chen}, \citenamefont {Li},\ and\
  \citenamefont {Weichselbaum}}]{Chen2018}%
  \BibitemOpen
  \bibfield  {author} {\bibinfo {author} {\bibfnamefont {B.-B.}\ \bibnamefont
  {Chen}}, \bibinfo {author} {\bibfnamefont {L.}~\bibnamefont {Chen}}, \bibinfo
  {author} {\bibfnamefont {Z.}~\bibnamefont {Chen}}, \bibinfo {author}
  {\bibfnamefont {W.}~\bibnamefont {Li}},\ and\ \bibinfo {author}
  {\bibfnamefont {A.}~\bibnamefont {Weichselbaum}},\ }\bibfield  {title}
  {\bibinfo {title} {Exponential thermal tensor network approach for quantum
  lattice models},\ }\href {https://doi.org/10.1103/PhysRevX.8.031082}
  {\bibfield  {journal} {\bibinfo  {journal} {Phys. Rev. X}\ }\textbf {\bibinfo
  {volume} {8}},\ \bibinfo {pages} {031082} (\bibinfo {year}
  {2018})}\BibitemShut {NoStop}%
\end{thebibliography}%

% Supplemental Material is included below as editable source.
\clearpage
\onecolumngrid
\begingroup
\setstretch{1.1}
\raggedbottom
\graphicspath{{SM/}}
\setcounter{section}{0}
\setcounter{equation}{0}
\setcounter{figure}{0}
\setcounter{table}{0}
\setcounter{secnumdepth}{1}
\renewcommand{\thesection}{\arabic{section}}
\renewcommand{\theequation}{S\arabic{equation}}
\renewcommand{\thefigure}{S\arabic{figure}}
\renewcommand{\thetable}{S\arabic{table}}
\renewcommand{\theHsection}{SM.\arabic{section}}
\renewcommand{\theHequation}{SM.\arabic{equation}}
\renewcommand{\theHfigure}{SM.\arabic{figure}}
\renewcommand{\theHtable}{SM.\arabic{table}}
\makeatletter
\@removefromreset{equation}{section}
\def\theequation@prefix{}
\def\@sectioncntformat#1{Supplemental Note \thesection:\quad}
\def\@hangfrom@section#1#2#3{\@hangfrom{#1#2}\MakeTextUppercase{#3}}
\renewcommand{\@seccntformat}[1]{Supplemental Note \csname the#1\endcsname:\quad}
\makeatother
\begin{center}
\bfseries Supplemental Material for\\
Hidden Triakontadipole Order in the Triangular-Lattice Magnet CeMgAl$_{11}$O$_{19}$\\[0.6em]
\normalfont\normalsize
Yu-Bo Liu, \textit{et. al.}
\end{center}
\onecolumngrid
This Supplemental Material provides experimental and numerical methods, additional experimental comparisons, and numerical analyses supporting the main text.
\hyperref[sec:experimental-methods]{Supplemental Note~\ref*{sec:experimental-methods}} describes sample preparation and magnetic susceptibility and specific-heat measurements.
\hyperref[sec:numerical-methods]{Supplemental Note~\ref*{sec:numerical-methods}} presents the quantum many-body methods and the real- and imaginary-time approaches for spin dynamics.
In \hyperref[sec:parameter-constraints]{Supplemental Note~\ref*{sec:parameter-constraints}}, we derive the two parameter-constraint lines in main-text Fig.~2(d).
In \hyperref[sec:experimental-specific-heat]{Supplemental Note~\ref*{sec:experimental-specific-heat}}, we compare specific-heat measurements from different samples and explain the choice of finite-field data for validating the model parameters.
Subsequently, \hyperref[sec:D-convergence]{Supplemental Note~\ref*{sec:D-convergence}} demonstrates convergence of the zero- and finite-temperature tensor-network data in the main text with bond dimension $D$.
In \hyperref[sec:time-dependence]{Supplemental Note~\ref*{sec:time-dependence}}, the resolution dependence of the spectra is examined to distinguish the low-energy magnon branch from the broad higher-energy continuum.
\hyperref[sec:three-magnon]{Supplemental Note~\ref*{sec:three-magnon}} provides a detailed derivation of the three-magnon decay vertex and damping rate.
\hyperref[sec:dipole-triakontadipole-sf]{Supplemental Note~\ref*{sec:dipole-triakontadipole-sf}} presents the temperature dependence of the dipole--triakontadipole structure factor and a comparison with elastic neutron scattering.
Finally, \hyperref[sec:continuum-analysis]{Supplemental Note~\ref*{sec:continuum-analysis}} presents further analyses of the excitation continuum, including cylinder-size dependence and a two-magnon continuum comparison.

\section{Sample preparation and magnetothermal measurements}
\label{sec:experimental-methods}

\textit{Crystal growth and structural characterizations.}---Polycrystalline CeMgAl$_{11}$O$_{19}$ (CMAO) was synthesized by a solid-state reaction: stoichiometric CeO$_2$ (99.99\%), MgO (99.99\%), and Al$_2$O$_3$ (99.99\%) were thoroughly ground in an agate mortar and sintered in an alumina crucible at 1550~$^\circ$C for 60 hours with intermediate regrinding. High-quality single crystals were grown in air in a two-mirror optical floating-zone furnace (MF-2400, Cyberstar), with the seed and feed rods counter-rotating at 20~rpm and translated downward at 5~mm/h. Phase purity was verified by powder x-ray diffraction (SmartLab SE, Rigaku), and the crystal orientation was determined by Laue backscattering (Photonic Science).

\textit{Magnetic susceptibility and specific heat measurements.}---Magnetic susceptibility was measured in an MPMS (Quantum Design) with a $^3$He insert covering $0.4$--$1.8$~K, at a field of $0.1$~T applied parallel and perpendicular to the $c$ axis. Specific heat was measured via the relaxation method in a PPMS (Quantum Design), equipped with a dilution refrigerator, reaching millikelvin temperatures.

\section{Equilibrium and dynamical simulations}
\label{sec:numerical-methods}

\textit{Tensor network approaches.}---We perform quantum many-body calculations with tensor networks, representing pure states as matrix product states and thermal density operators as matrix product operators~\cite{Li2011, Chen2017, Chen2018, Li2023}. Ground states are obtained using the density-matrix renormalization group (DMRG)~\cite{White1992, Schollwock2011}. Imaginary- and real-time evolution are implemented using tangent-space methods based on the time-dependent variational principle (TDVP)~\cite{TDVP2011, Haegeman2016}. The high-temperature initialization is obtained using the series-expansion thermal tensor-network method~\cite{Chen2017}. At lower temperatures, the tangent-space tensor renormalization group (tanTRG)~\cite{Li2023} with controlled bond expansion~\cite{Gleis2023CBE, Li2024CBE} evolves the thermal density matrix $\rho(\beta)\equiv e^{-\beta H}$. The tensor-network calculations in the main text use $L_y\times L_x=6\times15$ cylinders with periodic boundary conditions along $y$ and open boundary conditions along $x$. We use a bond dimension of $D=1000$ for most calculations and $D=2000$ for the finite-temperature dynamical spectra.

For spin dynamics, we evaluate the real-time correlation function $C^{zz}(\mathbf q,t)\equiv\langle S_{\mathbf q}^z(t)S_{-\mathbf q}^z\rangle$, with $S_{\mathbf q}^z\equiv N^{-1/2}\sum_j e^{-i\mathbf q\cdot\mathbf r_j}S_j^z$ and $N=L_xL_y$. The zero- and finite-temperature expressions are
\begin{equation}
\begin{aligned}
C_0^{zz}(\mathbf q,t)&=e^{iE_0t}\langle\psi_0|S_{\mathbf q}^z e^{-iHt}S_{-\mathbf q}^z|\psi_0\rangle,\\
C_\beta^{zz}(\mathbf q,t)&=\frac{1}{Z}\operatorname{Tr}[e^{-\beta H/2}e^{iHt}S_{\mathbf q}^z e^{-iHt}S_{-\mathbf q}^ze^{-\beta H/2}],
\end{aligned}
\label{eq:real_time_correlations}
\end{equation}
where $|\psi_0\rangle$ is the normalized ground state with energy $E_0$, and $Z\equiv\operatorname{Tr}(e^{-\beta H})$ is the partition function. Numerically, TDVP constructs $e^{-iHt}S_{-\mathbf q}^z|\psi_0\rangle$ for the ground-state calculation and $e^{-iHt}S_{-\mathbf q}^z e^{-\beta H/2} e^{iHt}$ for the thermal calculation. The spectrum is then evaluated by trapezoidal integration with a Parzen window function, which sets an energy resolution of $\varepsilon\approx8/t_{\max}$, with $t_{\max}$ the longest evolution time.

\textit{Imaginary-time approach for spin dynamics.---} At low temperatures, we use an imaginary-time proxy, i.e., evaluating the midpoint correlations
\begin{equation}
G_{AB}\equiv\langle A(\beta/2)B\rangle_\beta
=\frac{1}{Z}\operatorname{Tr}\!\left[e^{-\beta H/2}A e^{-\beta H/2}B\right],
\label{eq:midpoint_contraction}
\end{equation}
to avoid long real-time evolution or analytic continuation. For a local, momentum-independent hyperfine form factor, the nuclear magnetic resonance (NMR) Kubo relation reads $\frac{1}{T_1}\propto \lim_{\omega\to0}T\sum_{\mathbf q}\sum_{\alpha=x,y,z}\frac{1}{\omega} \cdot \chi_{\alpha\alpha}^{\prime\prime}(\mathbf q,\omega)$,
where $\chi_{\alpha\alpha}^{\prime\prime}$ is the imaginary part of the dynamical spin susceptibility. Since the transverse pseudospins of CMAO are predominantly rank-5 multipoles, we retain only the $zz$ channel. Its first-order imaginary-time proxy~\cite{Xi2024_1oT1} is $\mathcal S_1^{zz}(0)\equiv \frac{\beta}{N_{\mathcal B}}\sum_{i\in\mathcal B} \left[G_{S_i^zS_i^z}-\langle S_i^z\rangle_\beta^2\right]$,
which subtracts the static magnetization contribution and captures the leading regular low-frequency term at low temperatures.

For the spin Seebeck effect (SSE), the low-temperature approximation to the normalized spin current follows from the midpoint derivative of the transverse correlation~\cite{Gao2025SpinSeebeck}. Using $\partial_\tau S_j^-(\tau)=[H,S_j^-](\tau)$, we evaluate 
$$\tilde I_2\equiv\frac{\beta}{N_{\mathcal B}}\sum_{j\in\mathcal B} G_{O_jS_j^+},$$ with $O_j\equiv[H,S_j^-]$.
For both proxies, the sums and normalization are restricted to the bulk region $\mathcal B$, containing $N_{\mathcal B}$ sites.
Equation~\eqref{eq:midpoint_contraction} thus yields both the NMR and SSE proxies without reconstructing real-frequency spectra.

\section{Experimental constraints on the model parameters}
\label{sec:parameter-constraints}

The two linear parameter constraints in Fig.~2(d) of the main text are determined by the field-polarized excitation spectrum and the saturation field. Here we derive these constraints in the XXZ model for CMAO and test the spin-wave description of the magnetization against zero-temperature DMRG results.

\begin{figure}[H]
\centering
\includegraphics[width=0.85\textwidth]{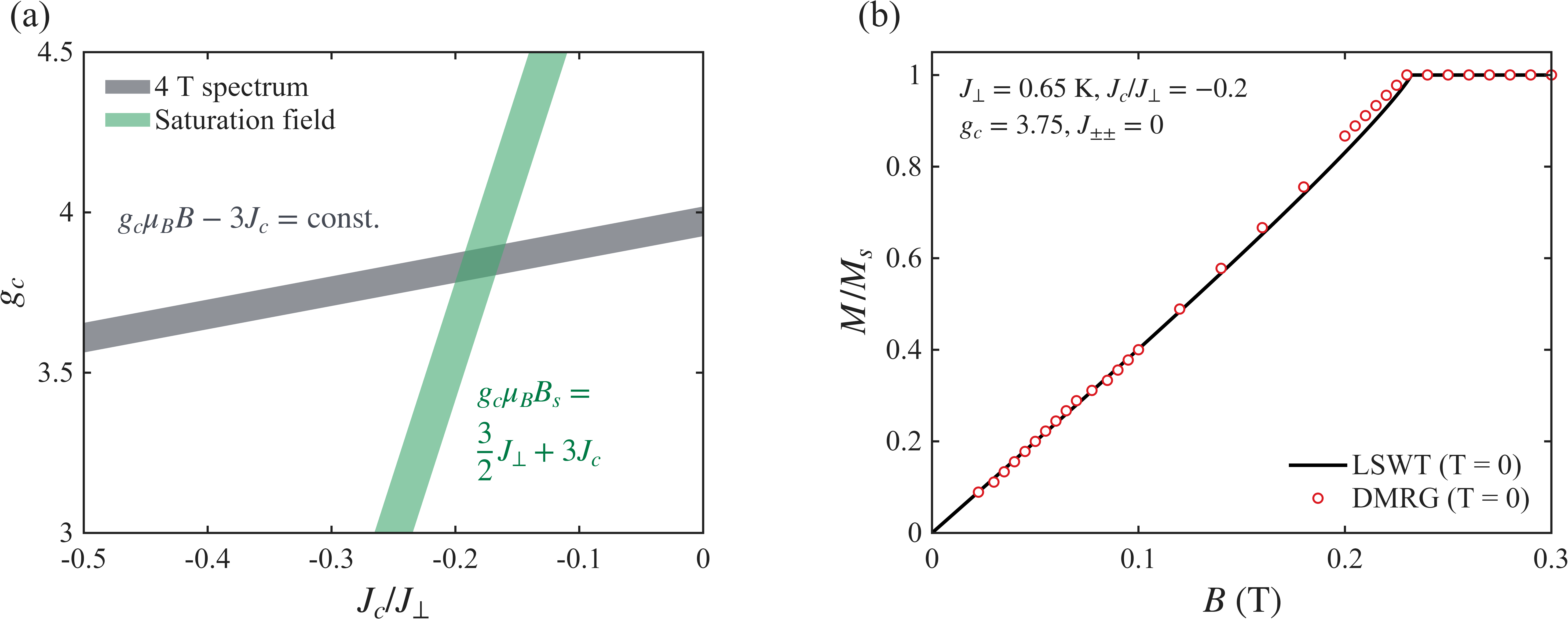}
\caption{Experimental constraints on the model parameters and comparison of the calculated magnetization curves.
(a) The gray and green bands show the constraints from the field-polarized excitation spectrum measured at $B=4$~T and the magnetization saturation field, respectively, as in main-text Fig.~2(d).
The equations beside the bands give the corresponding constraints for $S=1/2$, with fixed $J_{\perp}\simeq 0.65$~K and $B_s$ denoting the saturation field.
(b) Normalized magnetization calculated at zero temperature by LSWT (black line) and DMRG (red circles), with $M_s$ the saturation magnetization. Both calculations use $J_c/J_{\perp}=-0.2$, $g_c=3.75$, and $B\parallel c$.}
\label{fig:parameter-constraints}
\end{figure}

\textit{Field-polarized spectrum.}---For $B\parallel c$, the XXZ Hamiltonian is
\begin{equation}
H=\sum_{\langle ij\rangle}\left[
J_{\perp}(S_i^xS_j^x+S_i^yS_j^y)+J_cS_i^zS_j^z\right]
-h\sum_iS_i^z,
\end{equation}
where $h\equiv g_c\mu_BB$ and $S=1/2$.
Using the Holstein--Primakoff representation $S_i^z=S-a_i^\dagger a_i$ and $S_i^+\simeq\sqrt{2S}\,a_i$, the quadratic Hamiltonian about the fully polarized state becomes
\begin{equation}
H_2=(h-6SJ_c)\sum_i a_i^\dagger a_i
+SJ_{\perp}\sum_{\langle ij\rangle}(a_i^\dagger a_j+\mathrm{H.c.}).
\end{equation}
Fourier transformation therefore yields
\begin{equation}
\varepsilon_{\bm q}=h-6SJ_c+6SJ_{\perp} g_{\bm q},\qquad
g_{\bm q}\equiv\frac13\sum_{\ell=1}^3\cos(\bm q\cdot\bm\delta_\ell),
\label{eq:sm-polarized-dispersion}
\end{equation}
where $\bm\delta_\ell=(-1/2,-\sqrt{3}/2),(-1/2,\sqrt{3}/2),(1,0)$ are the three nearest-neighbor vectors, in units of the lattice constant $a$.
Because the XXZ Hamiltonian preserves the one-spin-flip subspace, Eq.~\eqref{eq:sm-polarized-dispersion} gives the exact zero-temperature one-magnon dispersion when the ground state is fully polarized.

With $J_{\perp}$ fixed by fitting the measured bandwidth of the $4$-T excitation spectrum, the calculated dispersion remains consistent with experiment as long as
\begin{equation}
g_c\mu_BB-6SJ_c=\mathrm{constant},
\label{eq:sm-spectrum-line}
\end{equation}
where the constant is determined by the experimental spectrum.
This relation gives the gray constraint line in main-text Fig.~2(d).

\textit{Saturation-field constraint.}---At the saturation field $B_s$, the minimum magnon energy in Eq.~\eqref{eq:sm-polarized-dispersion} reaches zero.
Since the magnon energy is lowest at the $K$ point, we obtain
\begin{equation}
g_c\mu_BB_s-3S(J_{\perp}+2J_c)=0.
\label{eq:sm-saturation-field}
\end{equation}
Substituting the experimentally estimated saturation field $B_s\simeq 0.25$~T (Ref.~\cite{colman}) into Eq.~\eqref{eq:sm-saturation-field} renders the green constraint line in main-text Fig.~2(d).

Figure~\ref{fig:parameter-constraints}(b) shows the zero-temperature magnetization calculated by LSWT and DMRG for our parameters. Although the magnetization curves differ slightly near the polarization QCP, the agreement between the saturation fields supports the use of LSWT and the resulting parameter constraint.

\textit{Susceptibility fitting loss.}---The susceptibility-fit landscape provides an additional constraint beyond the two straight lines.
Writing $\Delta\equiv J_c/J_{\perp}$, we define the susceptibility fitting objective as
\begin{equation}
\mathcal L_\chi(\Delta,g_c) \equiv\log_{10}\!\left[\frac{1}{N\chi_{\mathrm{exp,max}}^2} \sum_{i=1}^{N}\left(
\chi_{\mathrm{cal}}(T_i;\Delta,g_c)-\widetilde{\chi}_{\mathrm{exp}}(T_i)
\right)^2\right],
\label{eq:sm-chi-loss}
\end{equation}
where $\widetilde{\chi}_{\mathrm{exp}}(T_i)$ is obtained by linear interpolation of the experimental data onto the theoretically simulated temperature points, $N$ is their number, and $\chi_{\mathrm{exp,max}}\equiv \max_i\widetilde{\chi}_{\mathrm{exp}}(T_i)$ is the normalization factor. Only theoretical temperature points within the fitting window and the experimental data range are included. Smaller, more negative value of $\mathcal L_\chi(\Delta,g_c)$ indicates a better fit.

\clearpage
\section{Sample dependence of the experimental specific heat}
\label{sec:experimental-specific-heat}

Figure~\ref{fig:experimental-specific-heat}(a) shows the zero-field magnetic specific heat measured by four experimental groups, including our present work.
The datasets differ substantially around the low-temperature broad peak, particularly in the peak height, while becoming more consistent at higher temperatures.
These differences indicate appreciable sample dependence of the low-temperature specific heat, making the parameter fitting with a single dataset less conclusive.
In contrast, our data at $B=0.25$~T agree closely with those of Ref.~\cite{Cao2025}, including the position and height of the broad peak [Fig.~\ref{fig:experimental-specific-heat}(b)].
Therefore, we use the specific heat at $B=0.20$ and $0.25$~T, rather than at zero or very low fields, to validate the model parameters [main-text Fig.~2(e)].

\begin{figure}[H]
\centering
\includegraphics[width=0.85\textwidth]{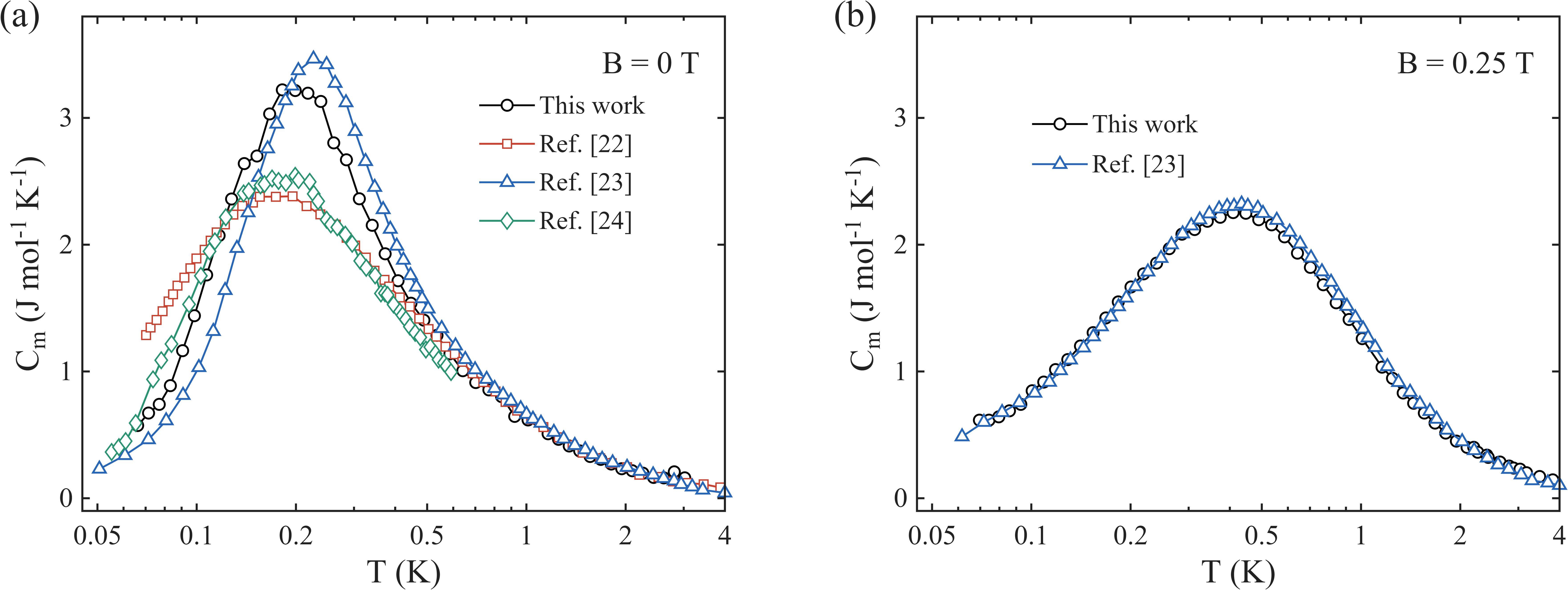}
\caption{Comparison of the magnetic specific heat $C_m$ measured by different experimental groups.
(a) Zero-field data from this work (black circles) and Refs.~\cite{Gao2026} (red squares), \cite{Cao2025} (blue triangles), and \cite{colman} (green diamonds).
(b) Data at $B=0.25$~T ($B\parallel c$) from this work and Ref.~\cite{Cao2025}, which are nearly indistinguishable.}
\label{fig:experimental-specific-heat}
\end{figure}

\clearpage
\section{Data convergence of tensor-network calculations}
\label{sec:D-convergence}

To assess the bond-dimension convergence of the calculations in the main text, we present results obtained with $D=300$, $500$, and $1000$, keeping the system size and exchange parameters fixed.
Here $D$ denotes the matrix-product-operator bond dimension for the finite-temperature tanTRG calculations and the matrix-product-state bond dimension for DMRG and zero-temperature real-time evolution.
Figure~\ref{fig:D-convergence} summarizes the convergence of the thermodynamic results in main-text Fig.~2 and the zero-temperature dynamical spectra in main-text Fig.~3.

\textit{Thermodynamic observables.}---As shown in Fig.~\ref{fig:D-convergence}(a), the susceptibility exhibits a small $D$ dependence at low temperatures, which becomes less pronounced as the temperature increases. The magnetization and specific-heat curves for different $D$ nearly coincide in Figs.~\ref{fig:D-convergence}(b) and \ref{fig:D-convergence}(c). Together, these results support the convergence of the thermodynamic calculations used in the main text.

\textit{Dynamical spectra.}---Figures~\ref{fig:D-convergence}(d)--\ref{fig:D-convergence}(i) show the $M$- and $K$-point energy cuts for different bond dimension $D$. The characteristic line shapes remain largely unchanged as $D$ increases for all $J_{\pm\pm}$ values considered here. In particular, the dominant peak positions remain stable, with only small changes in their intensities.
These results support the convergence of the zero-temperature dynamical spectra in the main text.

\begin{figure}[H]
\centering
\includegraphics[width=0.82\textwidth]{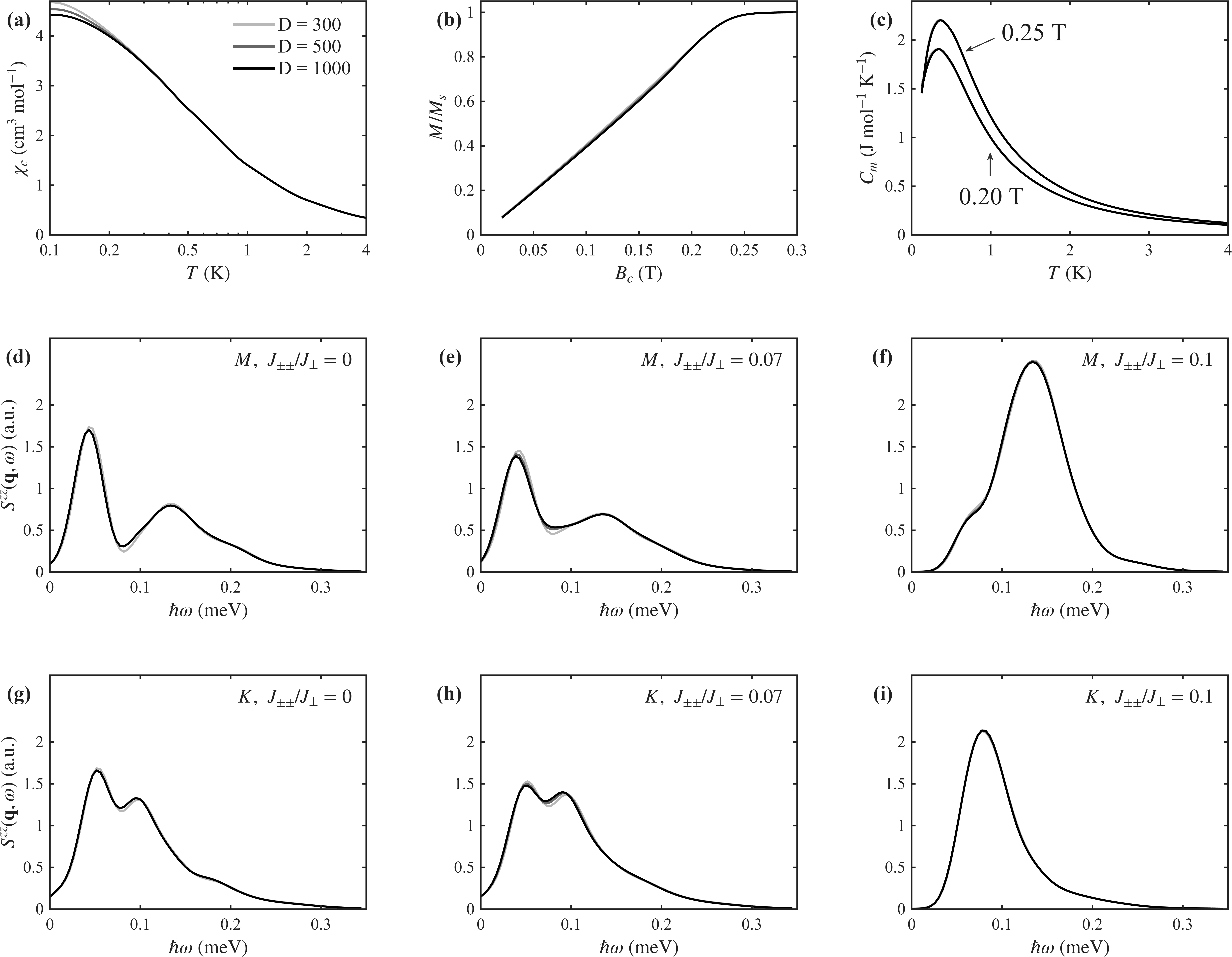}
\caption{Data convergence of the thermodynamics and zero-field spin dynamics calculations.
Light gray, dark gray, and black curves correspond to $D=300$, $500$, and $1000$, respectively, throughout all panels.
Calculations are performed on $6\times15$ cylinders with $J_{\perp}=0.65$~K and $J_c/J_{\perp}=-0.2$.
(a) Longitudinal susceptibility $\chi_c$ at $B=0.1$~T.
(b) Normalized magnetization $M/M_s$ at $T=0.03$~K.
(c) Magnetic specific heat $C_m$ at $B=0.20$ and $0.25$~T.
Panels (a)--(c) use $g_c=3.75$, $J_{\pm\pm}=0$, and $B\parallel c$.
(d)--(f) Energy cuts of $S^{zz}(\mathbf q,\omega)$ at $M$ for $J_{\pm\pm}/J_{\perp}=0$, $0.07$, and $0.1$, respectively; (g)--(i) show the corresponding cuts at $K$.
All dynamical curves are calculated at $T=0$ and the same energy resolution.}
\label{fig:D-convergence}
\end{figure}

\clearpage
\textit{Finite-temperature dynamical spectra.}---Figure~\ref{fig:FT-D-convergence} compares the spectra at $T=0.1$~K, computed with bond dimensions $D=1000$ and $2000$, respectively. Both the low-energy band and the broad higher-energy continuum remain largely unchanged as $D$ increases [Figs.~\ref{fig:FT-D-convergence}(a,b)]. The corresponding energy cuts at $M$ and $K$ [Figs.~\ref{fig:FT-D-convergence}(c,d)] show closely matching line shapes, retaining the low-energy peak and the higher-energy shoulder, with only small changes in their positions and intensities. These results support the bond-dimension convergence of the finite-temperature dynamical spectrum presented in main-text Fig.~3(b) at the resolution considered here.

\begin{figure}[H]
\centering
\includegraphics[width=0.75\textwidth]{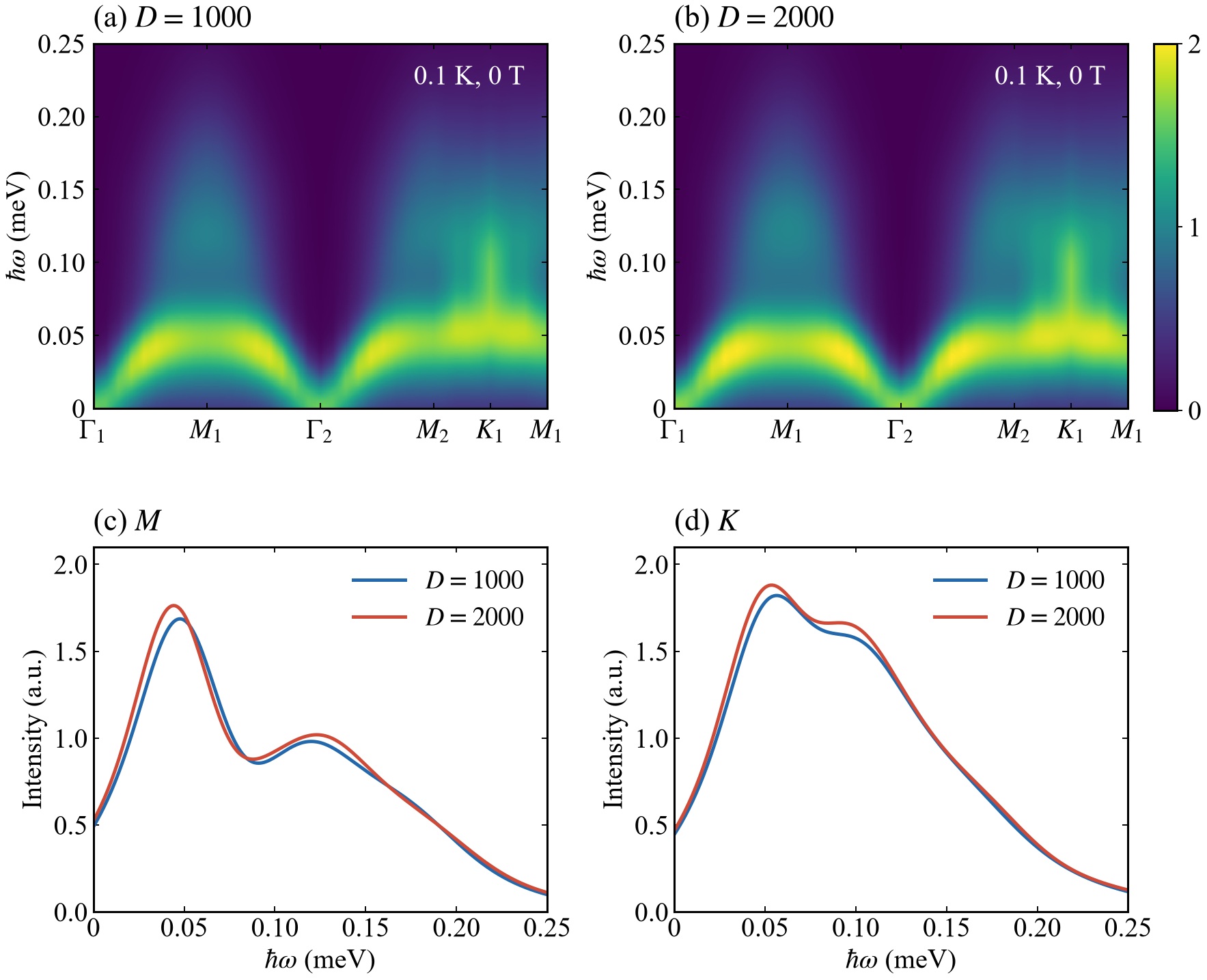}
\caption{Bond-dimension dependence of the finite-temperature spectrum at $T=0.1$~K and $B=0$, on a YC6$\times$15 cylinder.
The model parameters are $J_\perp=0.65$~K, $J_c/J_\perp=-0.2$, and $J_{\pm\pm}=0$.
(a,b) $S^{zz}(\mathbf q,\omega)$ calculated with bond dimensions $D=1000$ and $2000$ in real-time evolutions, respectively.
Both calculations share the same Parzen window and energy resolution ($\epsilon \simeq 0.037$~meV).
(c,d) Corresponding energy cuts at $M$ and $K$, respectively, for $D=1000$ (blue) and $2000$ (red).}
\label{fig:FT-D-convergence}
\end{figure}

\clearpage
\section{Supporting dynamical spectral data}
\label{sec:time-dependence}

The finite energy resolution in neutron experiments may broaden the measured spectral data. In our simulations, increasing the maximum evolution time $t_{\max}$ improves the energy resolution, with $\varepsilon\approx8/t_{\max}$ for the Parzen window used here. This distinguishes intrinsic spectral features from resolution-induced broadening. Figure~\ref{fig:time-dependence} shows the resolution dependence of the XXZ spectrum for our parameters; the spectra are obtained from the same real-time correlation functions using different time cutoffs and a fixed window-function shape.

\begin{figure}[H]
\centering
\includegraphics[width=0.85\textwidth]{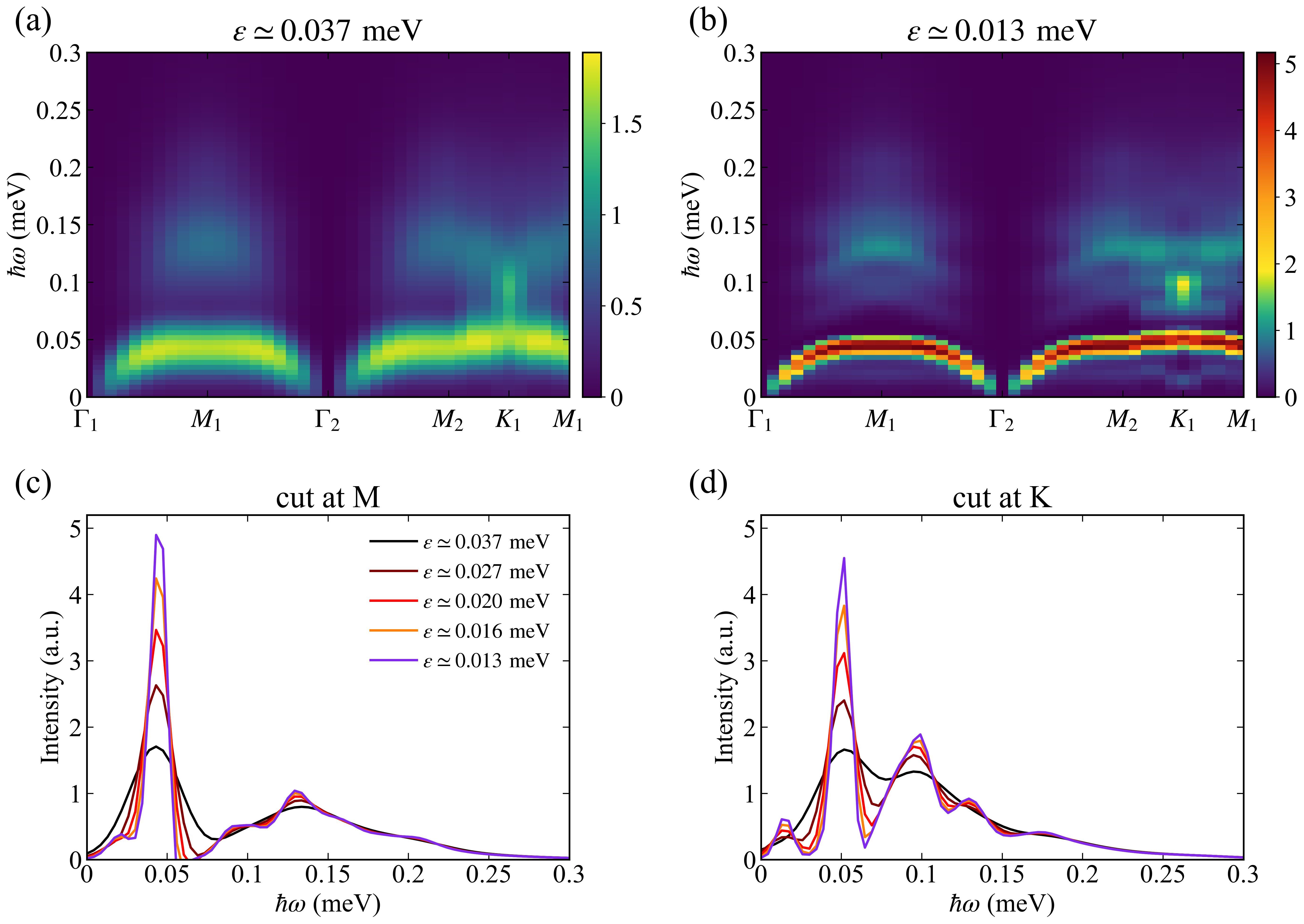}
\caption{Energy-resolution dependence of the zero-field dynamical structure factor $S^{zz}(\mathbf q,\omega)$ at $T=0$.
Model parameters are $J_{\perp}\simeq 0.65$~K, $J_c/J_{\perp}=-0.2$, and $J_{\pm\pm}=0$.
The calculations are performed on a $6\times15$ cylinder with retained bond dimension $D=1000$.
(a) Spectrum at an effective energy resolution $\varepsilon\simeq0.037$~meV.
(b) Spectrum at $\varepsilon\simeq0.013$~meV. The color scale in (b) extends that in (a) to accommodate the enhanced peak intensity; equal intensities have the same colors over their shared range.
(c,d) Energy cuts at $M$ and $K$, respectively, with the energy resolutions indicated in the legend of (c).}
\label{fig:time-dependence}
\end{figure}

At the lower energy resolution shown in Fig.~\ref{fig:time-dependence}(a), the low-energy response forms a broadened band.
Improving the energy resolution resolves a sharp, dispersive magnon branch of the $120^\circ$ state, as shown in Fig.~\ref{fig:time-dependence}(b).
In contrast, the higher-energy response remains broad even at the improved resolution.
The energy cuts at $M$ and $K$ points in Figs.~\ref{fig:time-dependence}(c) and \ref{fig:time-dependence}(d) further illustrate this contrast.
As the energy resolution $\varepsilon$ decreases, the low-energy peak becomes narrower and higher, whereas the higher-energy response remains distributed over a broad energy range despite the emergence of finer structure.
Based on these results, we propose that in CMAO a low-energy $120^\circ$ magnon branch coexists with an intrinsic higher-energy continuum whose broad spectral distribution cannot be attributed solely to finite-resolution broadening.

\clearpage
\section{Magnon damping via three-magnon decay}
\label{sec:three-magnon}

The energy cuts of the DMRG spectra at the $M$ and $K$ points for $J_{\pm\pm}/J_\perp=0$ and $0.07$ are shown in
Figs.~\ref{fig:D-convergence}(d,e,g,h). A small but finite $J_{\pm\pm}$ broadens the spectra at these momenta, suggesting enhanced
magnon damping. To explore the microscopic origin of this enhanced damping, we derive the
three-magnon decay vertex for the $120^\circ$ state.
The analysis first examines the symmetry and energy--momentum constraints in
the XXZ limit, before considering the effect of a finite $J_{\pm\pm}$.

\textit{Cubic interaction.}---The local axes are chosen as
$\bm e_i^z=(\cos\theta_i,\sin\theta_i,0)$,
$\bm e_i^y=(\sin\theta_i,-\cos\theta_i,0)$, and
$\bm e_i^x=(0,0,1)$, with $\theta_i=0,2\pi/3,4\pi/3$ on the three sublattices.
Here $\theta_i$ specifies the classical spin direction, and $\widetilde{\bm S}_i$ denotes the spin components in the local frame.
The cubic interaction originates from the longitudinal--transverse terms of the bond Hamiltonian,
\begin{align}
H_{ij}^{yz}={}&(D_{ij}+P_{ij})\widetilde S_i^y\widetilde S_j^z
 +(-D_{ij}+P_{ij})\widetilde S_i^z\widetilde S_j^y,\nonumber\\
D_{ij}={}&J_{\perp}\sin(\theta_i-\theta_j),\nonumber\\
P_{ij}={}&2J_{\pm\pm}\sin(\theta_i+\theta_j+\phi_{ij}),
\label{eq:sm-DP}
\end{align}
where $\phi_{ij}=0,\pm2\pi/3$ denotes the bond-dependent phase.
Substituting $\widetilde S_i^z=S-n_i$, with $n_i=a_i^\dagger a_i$, and
$\widetilde S_i^y=i\sqrt{S/2}(a_i^\dagger-a_i)+O(S^{-1/2})$ and collecting the surviving cubic terms gives
\begin{equation}
H_3=-i\sqrt{\frac S2}\sum_{\langle ij\rangle}\big[(D_{ij}+P_{ij})(a_i^\dagger-a_i)n_j +(-D_{ij}+P_{ij})n_i(a_j^\dagger-a_j)\big].
\label{eq:sm-H3real}
\end{equation}

The cubic interaction is transformed to momentum space using $a_{\bm Rm}=N_c^{-1/2}\sum_{\bm k}e^{i\bm k\cdot(\bm R+\bm r_m)}a_{\bm k m}$.
Here $\bm R$ labels magnetic unit cells, $m,n$ denote sublattices, $\bm r_m$ is the sublattice position within a unit cell, and $N_c$ is the number of magnetic unit cells.
Collecting the two longitudinal--transverse contributions from each bond gives
\begin{equation}
H_3=\frac{1}{\sqrt{N_c}}\sum_{\bm p,\bm q}\sum_{mn}
 \gamma_{\bm p}^{mn}a^\dagger_{\bm q m}a_{\bm q-\bm p,m}
 (a^\dagger_{-\bm p,n}-a_{\bm p n}),
\label{eq:sm-H3momentum}
\end{equation}
where
\begin{equation}
\gamma_{\bm p}^{mn}={}i\sqrt{\frac S2}\sum_{\bm\delta_{mn}} e^{i\bm p\cdot\bm\delta_{mn}}
 \big[J_{\perp}\sin(\theta_m-\theta_n) -2J_{\pm\pm} \sin(\theta_m+\theta_n+\phi_{\bm\delta_{mn}})\big].
\label{eq:sm-gamma}
\end{equation}
The sum includes all nearest-neighbor displacements $\bm\delta_{mn}$ from sublattice $m$ to $n$.

\textit{Decay vertex.}---The Bogoliubov transformation is
\begin{equation}
a_{\bm k m}=\sum_\lambda\big[
 u_{m\lambda}(\bm k)b_{\bm k\lambda}
 +v^*_{m\lambda}(-\bm k)b^\dagger_{-\bm k\lambda}\big].
\label{eq:sm-bogoliubov}
\end{equation}
For the process $3\to1+2$, each magnon is labeled by its momentum $\bm k_a$ and band index $\lambda_a$, with $a=1,2,3$. The momentum representatives are chosen to satisfy $\bm k_3=\bm k_1+\bm k_2$.
Write $b_a\equiv b_{\bm k_a\lambda_a}$, $u_{ma}=u_{m\lambda_a}(\bm k_a)$, $v_{ma}=v_{m\lambda_a}(\bm k_a)$, and $w_{ma}=u_{ma}-v_{ma}$, with $w_{m\lambda}(\bm k)=u_{m\lambda}(\bm k)-v_{m\lambda}(\bm k)$.
Substituting the Bogoliubov transformation into Eq.~\eqref{eq:sm-H3momentum} and collecting the terms describing the decay process $3\to1+2$ gives
\begin{equation}
H_3^{\rm dec}=\frac{1}{2\sqrt{N_c}}\sum_{123}
 V_{3;12}b_1^\dagger b_2^\dagger b_3+\mathrm{H.c.},
\label{eq:sm-Hdec}
\end{equation}
with the symmetric vertex
\begin{equation}
V_{3;12}=\sum_{mn}\big\{\gamma_{-\bm k_1}^{mn}w^*_{n1}
 (u^*_{m2}u_{m3}+v^*_{m2}v_{m3}) +\gamma_{-\bm k_2}^{mn}w^*_{n2}
 (u^*_{m1}u_{m3}+v^*_{m1}v_{m3}) -\gamma_{\bm k_3}^{mn}w_{n3}
 (u^*_{m1}v^*_{m2}+u^*_{m2}v^*_{m1})\big\}.
\label{eq:sm-vertex}
\end{equation}
Here $\sum_{123}$ runs over two independent momenta and all three band indices.

At zero temperature, the leading-order decay contribution to the retarded self-energy is
\begin{equation}
\Sigma^{R,\rm dec}_{\lambda_3}(\bm k_3,E)
=\frac{1}{2N_c}\sum_{\bm k_1,\lambda_1,\lambda_2}
 \frac{|V_{3;12}|^2}{E-\varepsilon_1-\varepsilon_2+i0^+}.
\end{equation}
Here $E$ is the energy argument of the self-energy, and $\varepsilon_a=\varepsilon_{\lambda_a}(\bm k_a)$ denotes the energy of magnon $a$.
The magnon damping rate is then obtained from the on-shell imaginary part of the self-energy,
\begin{align}
\Gamma_{\lambda_3}(\bm k_3)
&=-\operatorname{Im}\Sigma^{R,\rm dec}_{\lambda_3}(\bm k_3,\varepsilon_3)
\nonumber\\
&=\frac{\pi}{2N_c}\sum_{\bm k_1,\lambda_1,\lambda_2}
 |V_{3;12}|^2\delta(\varepsilon_3-\varepsilon_1-\varepsilon_2).
\label{eq:sm-damping}
\end{align}
Therefore, finite magnon damping requires both a nonzero decay vertex and available phase space satisfying energy--momentum conservation.

\textit{Symmetry selection rule.}---We now examine the symmetry constraints on magnon decay.
At $J_{\pm\pm}=0$, the $120^\circ$ state is invariant under a
primitive lattice translation combined with a compensating U(1)
spin rotation.  Under this symmetry, the three folded branches carry
symmetry labels $s=0,1,2$,
corresponding to unfolded momenta $\bm\kappa=\bm k+s\bm Q$, where
$\bm Q$ is the ordering vector and $3\bm Q$ is a crystallographic
reciprocal vector.  Conservation of the corresponding unfolded momentum,
with $\bm k_3=\bm k_1+\bm k_2$, then requires
\begin{equation}
s_3=s_1+s_2\pmod 3.
\label{eq:sm-selection}
\end{equation}
Thus, a nonzero decay vertex, and hence a finite magnon damping rate,
is symmetry-allowed only when Eq.~\eqref{eq:sm-selection} is satisfied.
However, a finite $J_{\pm\pm}$
breaks the U(1) spin-rotation symmetry and hence this combined
symmetry.  The decay vertices can then be generically nonzero
regardless of whether the channels satisfy the modulo-three selection
rule in Eq.~\eqref{eq:sm-selection}.
Thus, by lifting the symmetry restriction on the decay vertex,
$J_{\pm\pm}$ can generate additional damping channels without a substantial
change in the dispersion.

\textit{Energy--momentum constraints on magnon decay.}---The following analysis focuses on the decay channels of the magnon branch contributing to $S^{zz}$ in the XXZ limit.
The three branches are labeled by the sublattice vectors
$(1,e^{i\eta_\lambda},e^{2i\eta_\lambda})/\sqrt{3}$, where
$\eta_\lambda=(-2\pi/3,2\pi/3,0)$ for $\lambda=1,2,3$.
The branch contributing to $S^{zz}$ has $\lambda_3=3$ and $\eta_3=0$, corresponding to the symmetry label $s_3=0$ in Eq.~\eqref{eq:sm-selection}.
The corresponding magnon dispersions are
\begin{align}
\varepsilon_\lambda(\bm q)&=3SJ_{\perp}
 \sqrt{[1-g_\lambda(\bm q)][1+2\Delta g_\lambda(\bm q)]},\nonumber\\
g_\lambda(\bm q)&=\frac13\sum_{\ell=1}^3
 \cos(\bm q\cdot\bm\delta_\ell+\eta_\lambda),
\label{eq:sm-fixed-bands}
\end{align}
with $\Delta\equiv J_c/J_{\perp}$.

For each outgoing band pair, the energy mismatch is defined as
\begin{equation}
\Delta\varepsilon_{\lambda_1\lambda_2}(\bm k,\bm q)
=\varepsilon_{\lambda_1}(\bm q)+\varepsilon_{\lambda_2}(\bm k-\bm q)
 -\varepsilon_{\lambda_3=3}(\bm k).
\label{eq:sm-mismatch}
\end{equation}
This definition enforces momentum conservation, while energy conservation requires $\Delta\varepsilon=0$.
Figure~\ref{fig:decay-kinematics} shows the minimum and maximum of $\Delta\varepsilon_{\lambda_1\lambda_2}(\bm k,\bm q)$ over $\bm q$ for each $\bm k$, with $\bm q$ spanning a full reciprocal primitive cell.
When the minimum is negative and the maximum is positive, continuity guarantees the existence of momenta $\bm q$ satisfying energy--momentum conservation.
Since exchanging the two final magnons leaves these bounds unchanged, only the six band pairs with $\lambda_1\leq\lambda_2$ are shown.

\begin{figure}[H]
\centering
\includegraphics[width=0.85\textwidth]{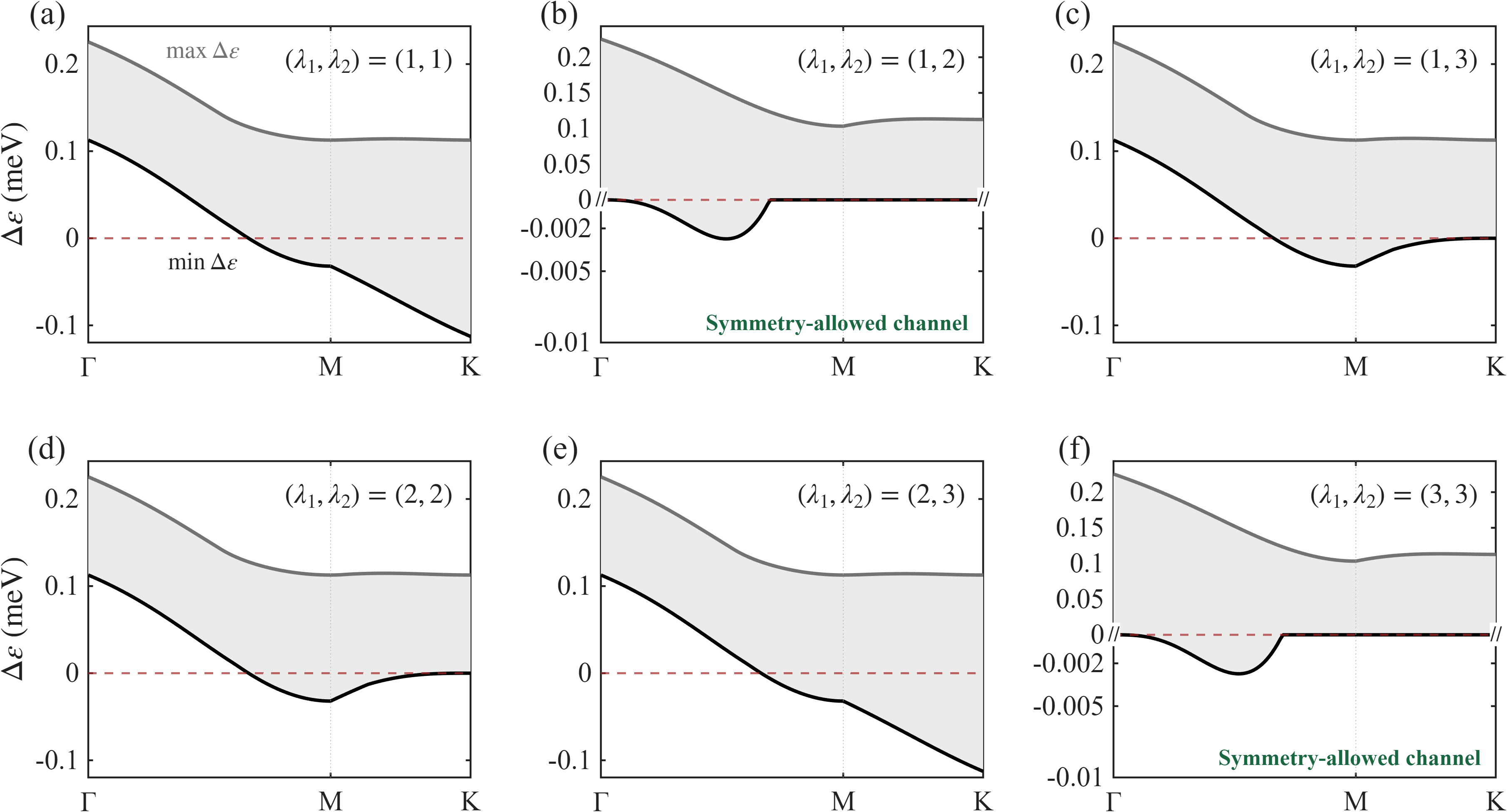}
\caption{Energy--momentum constraints on magnon decay in the $120^\circ$ state for $S=1/2$, $J_{\perp}=0.65$~K, $J_c/J_{\perp}=-0.2$, and $J_{\pm\pm}=0$.
The initial magnon has momentum $\bm k$ along $\Gamma$--$M$--$K$ and belongs to the phase-zero branch $\lambda_3=3$, which contributes to $S^{zz}$.
Panels (a)--(f) correspond to the final-state band pairs $(\lambda_1,\lambda_2)=(1,1)$, $(1,2)$, $(1,3)$, $(2,2)$, $(2,3)$, and $(3,3)$, respectively.
The band indices $\lambda=1,2,3$ correspond to the sublattice phases $\eta_\lambda=-2\pi/3,2\pi/3,0$, respectively.
Black and gray curves show the minimum and maximum of $\Delta\varepsilon_{\lambda_1\lambda_2}(\bm k,\bm q)$ over $\bm q$ at each $\bm k$, respectively, with the interval between them shaded.
The dashed red line marks the energy-conservation condition $\Delta\varepsilon=0$.
Only the channels in (b) and (f) satisfy the symmetry selection rule in Eq.~\eqref{eq:sm-selection}.}
\label{fig:decay-kinematics}
\end{figure}

For the phase-zero branch, the selection rule in Eq.~\eqref{eq:sm-selection} permits only decay channels satisfying $s_1+s_2=0\pmod 3$. With the band labeling defined above, these channels are $(\lambda_1,\lambda_2)=(1,2)$ and $(3,3)$, shown in Figs.~\ref{fig:decay-kinematics}(b) and \ref{fig:decay-kinematics}(f), respectively.
In these channels, decay into two finite-energy magnons satisfies energy--momentum conservation over part of $\Gamma$--$M$, but is kinematically forbidden along $M$--$K$.
By contrast, the other four channels satisfy energy--momentum conservation in the interior of $M$--$K$, but their vertices vanish by the selection rule.

The damping rates calculated from Eq.~\eqref{eq:sm-damping} are shown in Fig.~\ref{fig:sm-damping}. The damping vanishes along $M$--$K$ at $J_{\pm\pm}=0$, but becomes nonzero at $J_{\pm\pm}/J_\perp=0.07$, consistent with the symmetry and kinematic analysis above.

\begin{figure}[H]
\centering
\includegraphics[width=0.45\columnwidth]{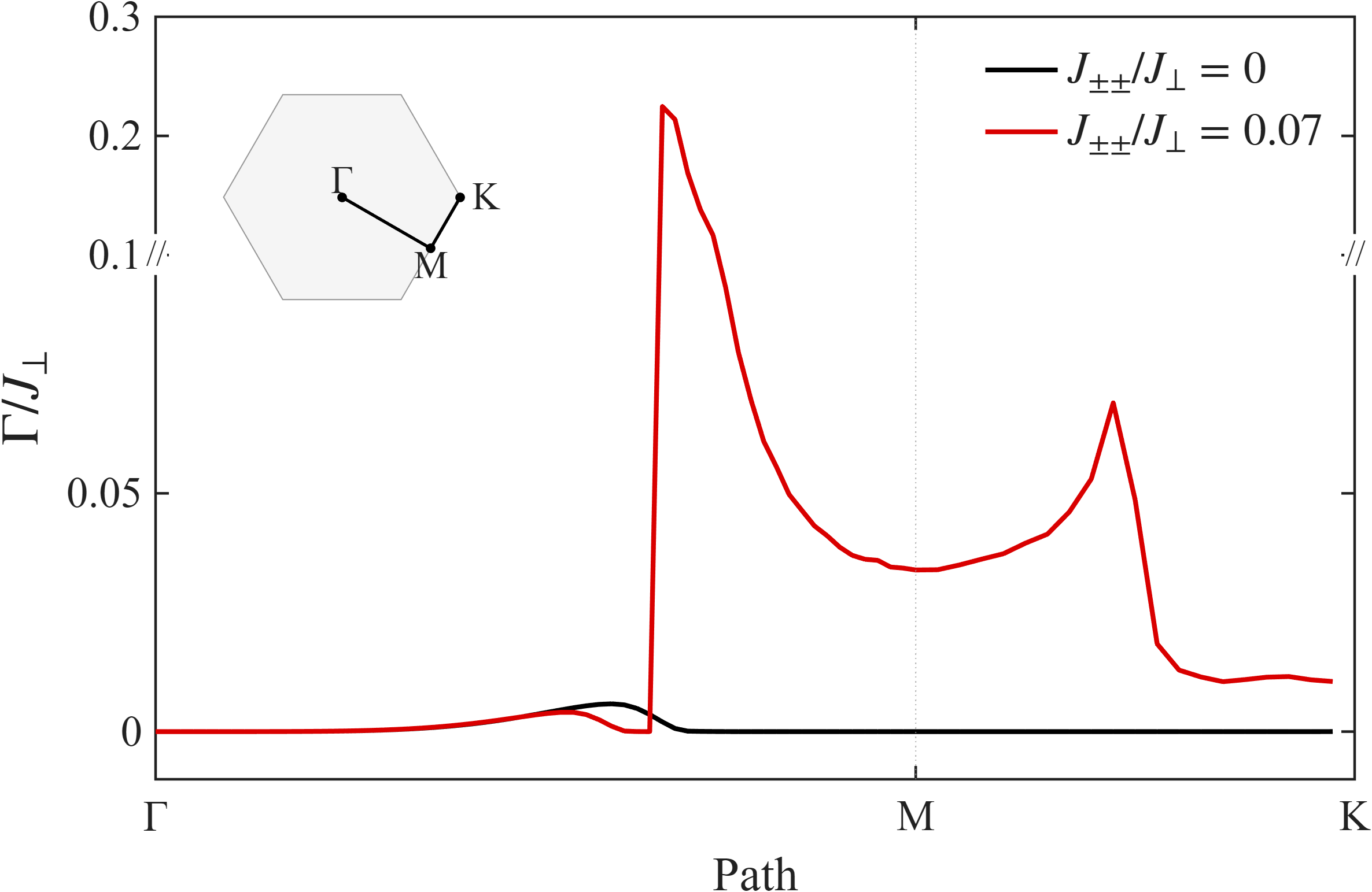}
\caption{Zero-temperature three-magnon damping rate along $\Gamma$--$M$--$K$
for the $120^\circ$ state at $J_c/J_\perp=-0.2$.
Black and red curves correspond to $J_{\pm\pm}/J_\perp=0$ and $0.07$, respectively.}
\label{fig:sm-damping}
\end{figure}

\clearpage
\section{Dipole--Triakontadipole Structure Factor}
\label{sec:dipole-triakontadipole-sf}

To examine correlations associated with the competing phases, we calculate the total static pseudospin structure factor, $S(\bm q)=\sum_{\alpha=x,y,z}S^{\alpha\alpha}(\bm q)$, as a function of temperature. In particular, $S(K)$ characterizes the $120^\circ$ correlations, whereas $S(M)$ characterizes the stripe correlations. Figure~\ref{fig:static-MK-total}(a) shows the temperature dependence of these two correlations. Upon warming within the low-temperature regime, the dominant $K$-point structure factor decreases, while the $M$-point structure factor initially increases. The temperature slices across the Brillouin zone in Figs.~\ref{fig:static-MK-total}(b,c) illustrate the thermal evolution of correlations near the $K$ and $M$ points. This temperature dependence suggests that low-energy excited states above the $120^\circ$ ground state contain stripe-like correlations.

\begin{figure}[H]
\centering
\includegraphics[width=0.9\textwidth]{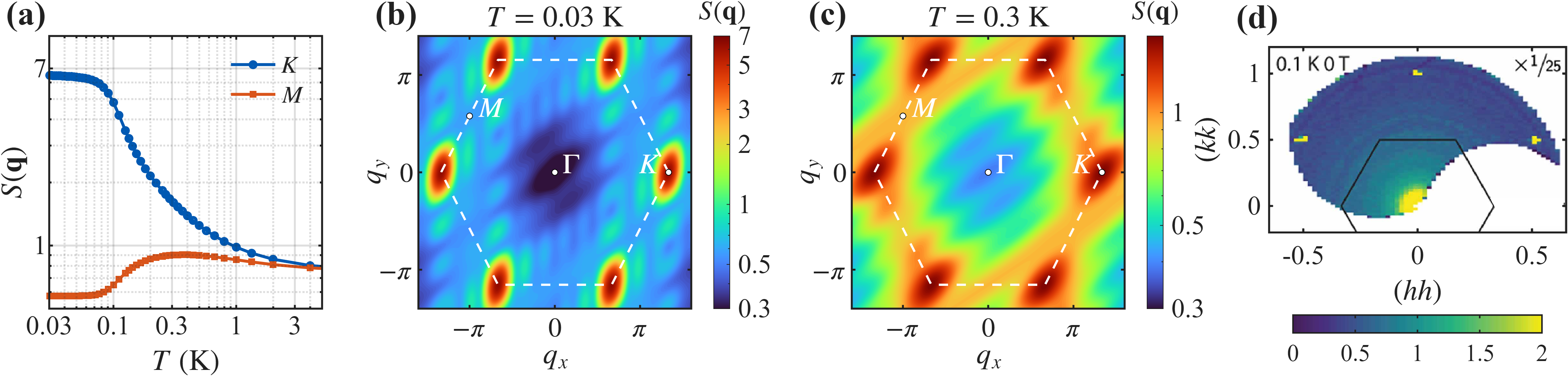}
\caption{Temperature dependence of the zero-field pseudospin structure factor in (a--c), calculated with CMAO parameters $J_\perp=0.65$~K, and $J_c/J_\perp=-0.2$, on a YC6$\times$15 cylinder with $D=1000$. The correlations are evaluated in the central $9\times6$-site region. (a) Temperature dependence at $K$ (blue circles) and $M$ (orange squares). The $M$ data are averaged over the three $M$ points. (b,c) Zero-field spin structure factors computed at $T\simeq0.03$ and $0.3$~K, respectively. White dashed hexagons mark the first Brillouin zone. (d) Experimental elastic neutron-scattering intensity in the $(hk)$ plane, reproduced from Fig.~3A of Ref.~\cite{Gao2026}. Black lines mark the first Brillouin zone. Note the experimental data show no magnetic Bragg peak at the $K$ point, whereas the corresponding numerical peaks in (a--c) are of multipolar origin and thus hidden from neutron probes.}
\label{fig:static-MK-total}
\end{figure}

It is worth noting that our results in Figs.~\ref{fig:static-MK-total}(b,c) are not contradicted by the absence of magnetic Bragg peaks at the $K$ points in the experimental elastic map [Fig.~\ref{fig:static-MK-total}(d)]. The dominant $120^\circ$ correlations at the $K$ points shown in Figs.~\ref{fig:static-MK-total}(b,c) arise from the transverse multipolar components of the pseudospin and are hidden from neutron scattering.

This behavior supports the competing-order picture discussed in the main text. A small $J_{\pm\pm}$ drives a first-order transition between the $120^\circ$ and stripe phases, where their energy branches cross. Near the transition on the $120^\circ$ side, stripe configurations have slightly higher energy than the ground state and can contribute short-range stripe correlations. Since a small change in $J_{\pm\pm}$ can induce a transition between the two phases, their low-energy responses differ substantially, whereas their higher-energy responses remain broadly similar. Indeed, the stripe spin-wave dispersion overlaid on main-text Fig.~3(e) follows the higher-energy continuum inside the $120^\circ$ phase. Together with the temperature dependence of the static correlations, this agreement supports interpreting the higher-energy continuum as precursor excitations of the nearby stripe phase.

\clearpage
\section{Further analysis of the spin excitation continuum}
\label{sec:continuum-analysis}

This section examines the size dependence of the zero-temperature spin excitation spectrum and the two-magnon excitation continuum. These results suggest the higher-energy continuum arises from precursor excitations of the nearby multipolar stripe phase.

\textit{Cylinder-size dependence.}---Figure~\ref{fig:continuum-size} compares the zero-temperature dynamical structure factor $S^{zz}(\mathbf q,\omega)$ on YC6$\times$15 and YC9$\times$18 cylinders. Both calculations use our parameters, $J_\perp=0.65$~K, $J_c/J_\perp=-0.2$, and $J_{\pm\pm}=0$, with retained bond dimension $D=1000$. The low-energy band and the broad higher-energy continuum remain essentially unchanged upon increasing the cylinder width [Figs.~\ref{fig:continuum-size}(a,b)]. The energy cuts at $M$ and $K$ [Figs.~\ref{fig:continuum-size}(c,d)] also show closely matching line shapes and intensities. Thus, the higher-energy continuum is robust against this increase in cylinder size at the resolution considered here.

\begin{figure}[H]
\centering
\includegraphics[width=0.75\textwidth]{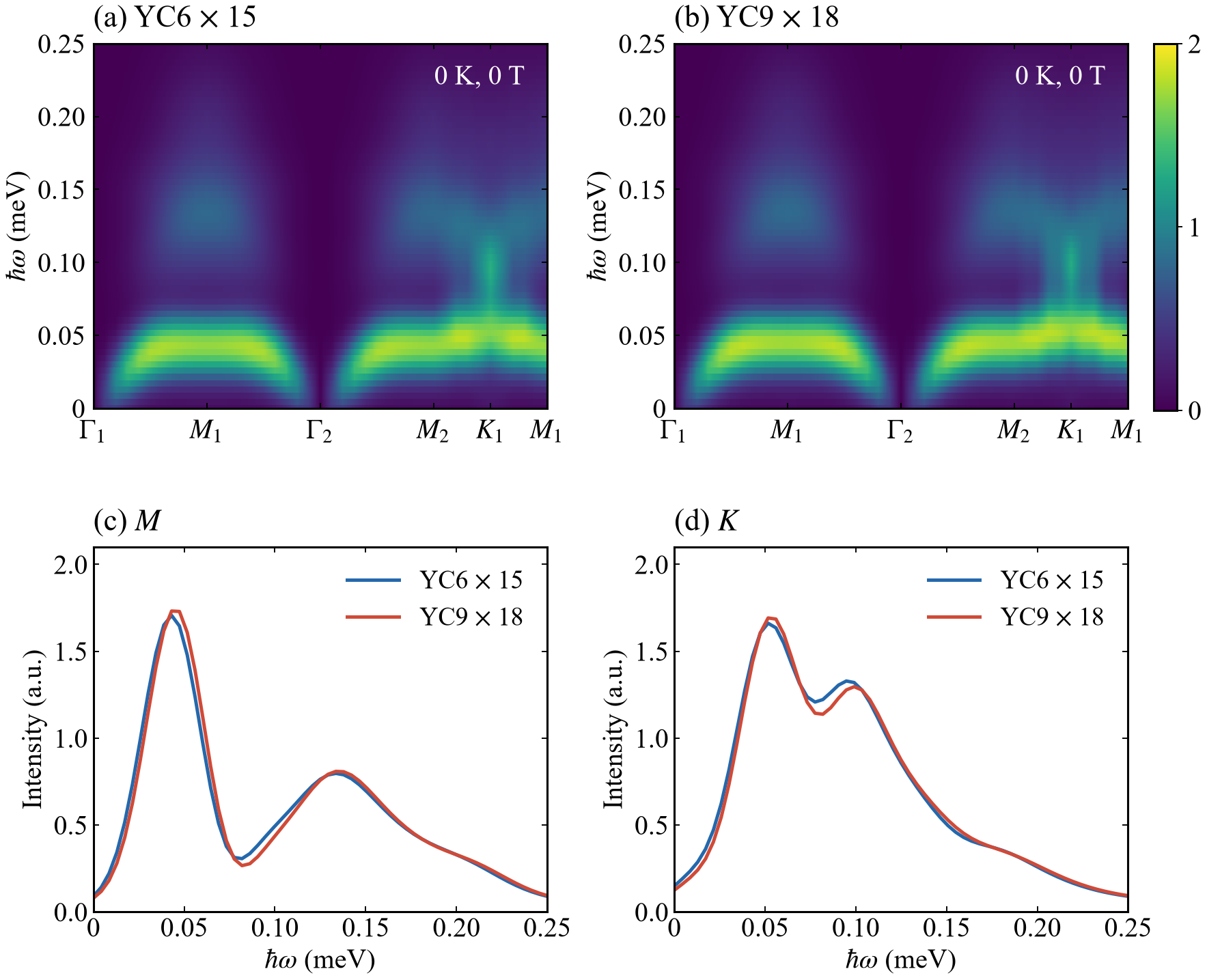}
\caption{Cylinder-size dependence of the zero-temperature spectrum at $B=0$, $J_\perp=0.65$~K, $J_c/J_\perp=-0.2$, and $J_{\pm\pm}=0$.
(a,b) $S^{zz}(\mathbf q,\omega)$ on YC6$\times$15 and YC9$\times$18 cylinders, respectively, with $D=1000$.
Both spectra use a Parzen window with  corresponding energy resolution $\varepsilon \simeq 0.037$~meV.
(c,d) Corresponding energy cuts at $M$ and $K$, respectively, for YC6$\times$15 (blue) and YC9$\times$18 (red) cylinders.}
\label{fig:continuum-size}
\end{figure}

\clearpage
\textit{Two-magnon continuum comparison.}---We also test whether the two-magnon continuum in the $120^\circ$ phase can reproduce the higher-energy continuum profile. For $J_{\pm\pm}=0$, the unfolded linear spin-wave dispersion is $\varepsilon_{\bm q}=3SJ_\perp\sqrt{(1-g_{\bm q})(1+2\Delta g_{\bm q})}$, with $g_{\bm q}$ given in Eq.~\eqref{eq:sm-polarized-dispersion}. The equal-weight two-magnon density of states is
\begin{equation}
\rho_2(\bm k,E)=\frac{1}{N_q}\sum_{\bm q}
\delta\!\left(E-\varepsilon_{\bm q}-\varepsilon_{\bm k-\bm q}\right).
\label{eq:two-magnon-DOS}
\end{equation}
The sum over $\bm q$ covers a full reciprocal primitive cell. The lower and upper boundaries are the minimum and maximum of $\varepsilon_{\bm q}+\varepsilon_{\bm k-\bm q}$ at fixed $\bm k$. Here, $\rho_2$ is an equal-weight count of two-magnon states, without neutron-scattering matrix elements or magnon-interaction corrections. The two-magnon states considered here belong to the phase-zero sector associated with the longitudinal response (see selection rule in Sec.~\ref{sec:three-magnon}).

\begin{figure}[H]
\centering
\includegraphics[width=0.85\textwidth]{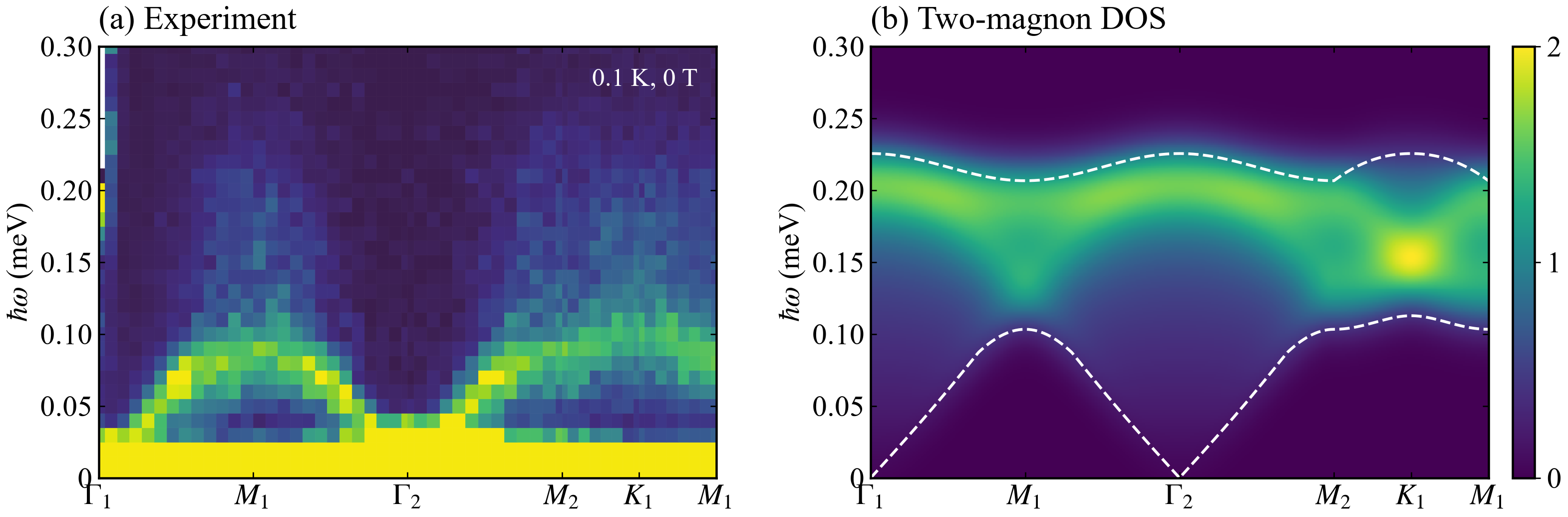}
\caption{(a) Experimental INS spectrum reproduced from Ref.~\cite{Gao2026}. (b) Equal-weight two-magnon density of states $\rho_2(\bm k,E)$ for $J_\perp=0.65$~K, $J_c/J_\perp=-0.2$, and $J_{\pm\pm}=0$, calculated from Eq.~\eqref{eq:two-magnon-DOS} and convolved with a Gaussian of full width at half maximum $0.04$~meV.
White dashed curves in (b) mark the boundaries of the two-magnon continuum.}
\label{fig:two-magnon-DOS}
\end{figure}

Compared with the experimental spectrum in Fig.~\ref{fig:two-magnon-DOS}(a), the two-magnon energy window [Fig.~\ref{fig:two-magnon-DOS}(b)] overlaps part of the higher-energy response, but its boundaries and DOS profile are clearly distinct from the experimental continuum. Therefore, two-magnon spectrum in this description is insufficient to explain the excitation continuum observed in CMAO experiments. 

\clearpage
\endgroup

\end{document}